# Further searches of the Higgs scalar sector at the ESS

Carlo Rubbia
GSSI, Gran Sasso Science Institute, L'Aquila, Italy
CERN, Geneva, Switzerland
carlo.rubbia@cern.ch

***Abstract.***

Recent decades have witnessed remarkable confirmations of the Standard Model (SM) describing the Electro-Weak and Strong Interactions. The Higgs boson was observed at CERN-LHC at 7-8 TeV and 13 TeV. The HL-LHC, a *major* luminosity upgrade has been recently approved.

The HL-LHC may be already an early "Higgs factory", however only capable to perform $H^o$ related measurements with rather large uncertainties because of persisting backgrounds and uncertainties. New projects using leptons rather than hadrons should be investigated

$H^o$ has a spin zero and its coupling is proportional to the square of the lepton mass, greatly enhancing the production from pairs of *muons*. A $\mu^+\mu^-$ Collider may operate at a much higher magnetic field respect to the $e^+e^-$ Collider and it has a smaller radius, easily fitting within one existing European site. However muons are unstable particles: they must be produced in sufficient amounts from pions of a proton beam, cooled and quickly accelerated to the required energies.

The scenario is here primarily concentrated to further developments of the *European Spallation Source (ESS)* under construction in the Lund site as the most intense future source of spallation neutrons.

As a initial part of the program, muon cooling should be experimentally demonstrated with the much cheaper and simpler *Initial Cooling Experiment.*

### *1.- Expectations and limits of the LHC.*

At the CERN-LHC, ATLAS [1] and CMS [2] teams have announced in 2012 a narrow line of high significance at a mass of about 125.5 GeV, compatible with the Higgs boson $H^0$. At the LHC, Higgs related events may be observed only once every $10^9$ collisions. In the two discovery experiments [1] and [2] several channels were actually recorded. These searches have been performed in the presence of very substantial backgrounds.

In the first LHC's run — at a collision energy of 7-8 TeV — decays of the Higgs boson involving pairs of electroweak bosons were observed. Between 2015 and 2018 at a collision energy of 13 TeV [3] ATLAS and CMS have further pursued several properties of the $H^0$. These results include new searches of the Higgs boson into pairs of muons and into pairs of charm quarks. Both ATLAS [64] and CMS [65] also measured previously unexplored properties of decays of the Higgs boson that involve electroweak bosons (the W, the Z and the photon) and compared these with the predictions of the Standard Model (SM). These studies will be continued from 2021 to 2023. The main Higgs decay channels for 300 $fb^{-1}$ of integrated LHC luminosity have the typical accuracy of about 10% – 15% at the single standard deviation level, (Figure 1) depending on the channel. In order to firmly confirm any new phenomenon, evidence of at least 5 standard deviations should be needed.

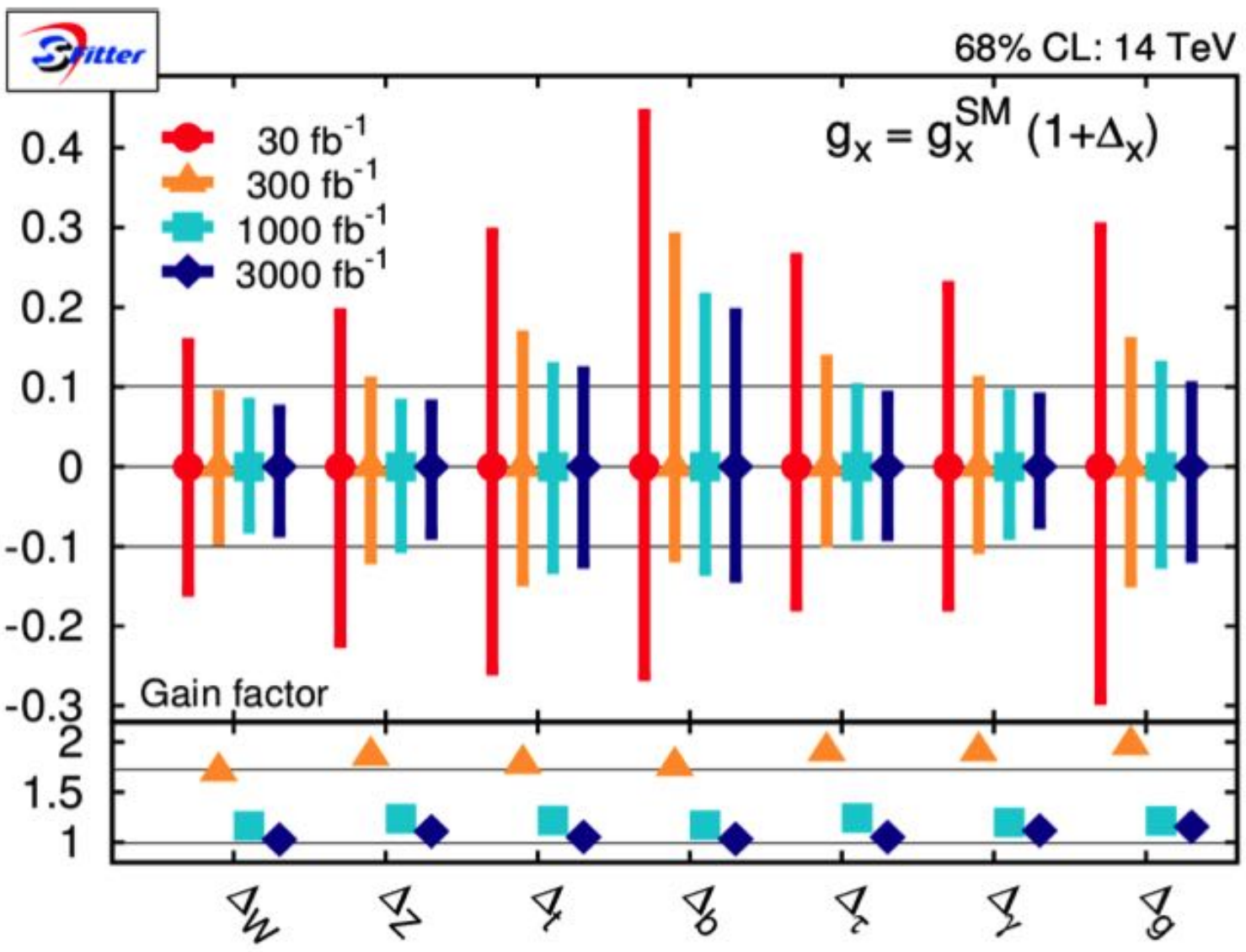

***Figure 1.-*** *Higgs estimates of the present and future LHC programs [3] of the ATLAS [1] and CMS [2] collaborations, accumulating increasing amounts of data, They are dominated by systematic errors according to their indicated predictions. These are 1 standard deviation values.*

The major luminosity upgrade [3] — the HL-LHC — has been recently approved, operational from 2026 onwards. Targets for the HL-LHC are of 250 $fb^{-1}$ / year and an integrated luminosity of 3000 $fb^{-1}$ by the mid 2030. Both ATLAS and CMS detectors will be upgraded to handle an average number of ≈ 140 pile-up background events at each bunch crossing, corresponding to a remarkable line density of 1.3 events per mm, an instantaneous luminosity of about $5 \times 10^{34}\ cm^{-2}s^{-1}$ for operation with 25 ns wide beams at $\sqrt{s}$ = 14 TeV and a visible cross section $\sigma_{vis}$ = 85 mb. The proton energy stored in the HL-LHC beams will be about 1000 MJ. The other two main detectors

ALICE [61] and LHCb [62] will be also upgraded to operate at instantaneous luminosities of up to $2 \times 10^{31}$ $cm^{-2}s^{-1}$ and $2 \times 10^{34}$ $cm^{-2}s^{-1}$.

The HL-LHC may even exceed its predicted design luminosity $L_{peak} = 5 \times 10^{34}$ $cm^{-2}s^{-1}$ and eventually reach $7.5 \times 10^{34}$ $cm^{-2}s^{-1}$, provided its practical rates and the upgraded detectors may accept an even higher pile-up. With a performance of 300 $fb^{-1}$/year, at least 4000 $fb^{-1}$ may be obtained by the year 2037, *the presently estimated termination date of the HL-LHC data taking*.

The HL-LHC — that should reach such a huge integrated sensitivity — may be already an early “Higgs factory”, however only capable to perform $H^{o}$ related measurements with rather large uncertainties, mainly because of the persisting backgrounds and uncertainties. For 3000 $fb^{-1}$ (Figure 1), uncertainties are expected to improve only by a factor of less than 2. The rate of rare channels such as [Zγ], [γγ] and invisible decays may be also measured, but with an ultimate single standard deviation precision of order 10 %.

HL-LHC plans are to further pursue the hadronic production of the $H^{o}$ related sector and the possible existence of SUSY particles. Super-symmetry (SUSY) is a theory that links gravity with the other fundamental forces of nature proposing a relationship between the two basic types of elementary particles: bosons, which have integer-valued spin and fermions, which have half-integer spin.

It had been argued by many theorists that *“new and additional uncovered physics”* must also be necessarily present at the TeV scale. SUSY theories [4] are considered as one of the most attractive extensions of the SM since they may protect the Higgs boson mass and stabilize the hierarchy between the electroweak and Planck scales. The Higgs sector may be extended to contain several additional Higgs fields. For instance, the Minimal Super-symmetric Standard Model (MSSM) predicts the existence of as many as five Higgs particles: two CP–even Higgs particles, a CP–odd A boson and two charged Higgs particles.

However, at a $H^{o}$ mass at $\sqrt{s}$ = 125.5 GeV as recently observed, additional Higgs particles — either charged or neutral, eventually associated to higher symmetries — may not be necessarily present in the LHC energy domain. The $H^{o}$ potential may develop instability [4] only around $10^{11}$ GeV and with a lifetime much longer than the age of the Universe. Taking into account theoretical and experimental errors, stability even up to the Planck scale cannot be excluded [4].

Therefore, there is no longer the inevitable need of additional related particles within the LHC range. So far LHC experiments have found no additional Higgs boson up to approximately 600 GeV [1] [2].

The future experimental determinations of the Higgs sector should closely follow the previous and well-known observations of the $Z^{o}$ and the W’s, in which the initial search and discovery with the p-pbar hadron Collider had been followed by the systematic studies with leptons at LEP. Also in the present case, new and more precise determinations are needed in order to place this new scalar $H^{o}$ particle more precisely within Particle Physics beyond the LHC. Like in the previous $Z^{o}$ case *new projects using leptons rather than hadrons should be investigated.*

### 2.- The Higgs sector.

According to the Standard Model (SM), the width of $H^O$ is only ≈ 4.2 MeV (Figure 2), to be compared for instance to the previous and much larger $Z^O$ width of 2.5 GeV. At $\sqrt{s}$ = 125.5 GeV the $H^O$ has several substantive decay branching fractions (Figure 3): (bb), 60%; (WW), 20%; (gg), 9%; (ττ), 6%; (ZZ), 3%; (cc), 3%. The channels (γγ) and (Zγ) with 0.2% are also substantive due to its high mass resolution and relatively low background.

Already at MURA in the fifties it was realised that compression of the beam phase-space may be required from the source to the collision point.

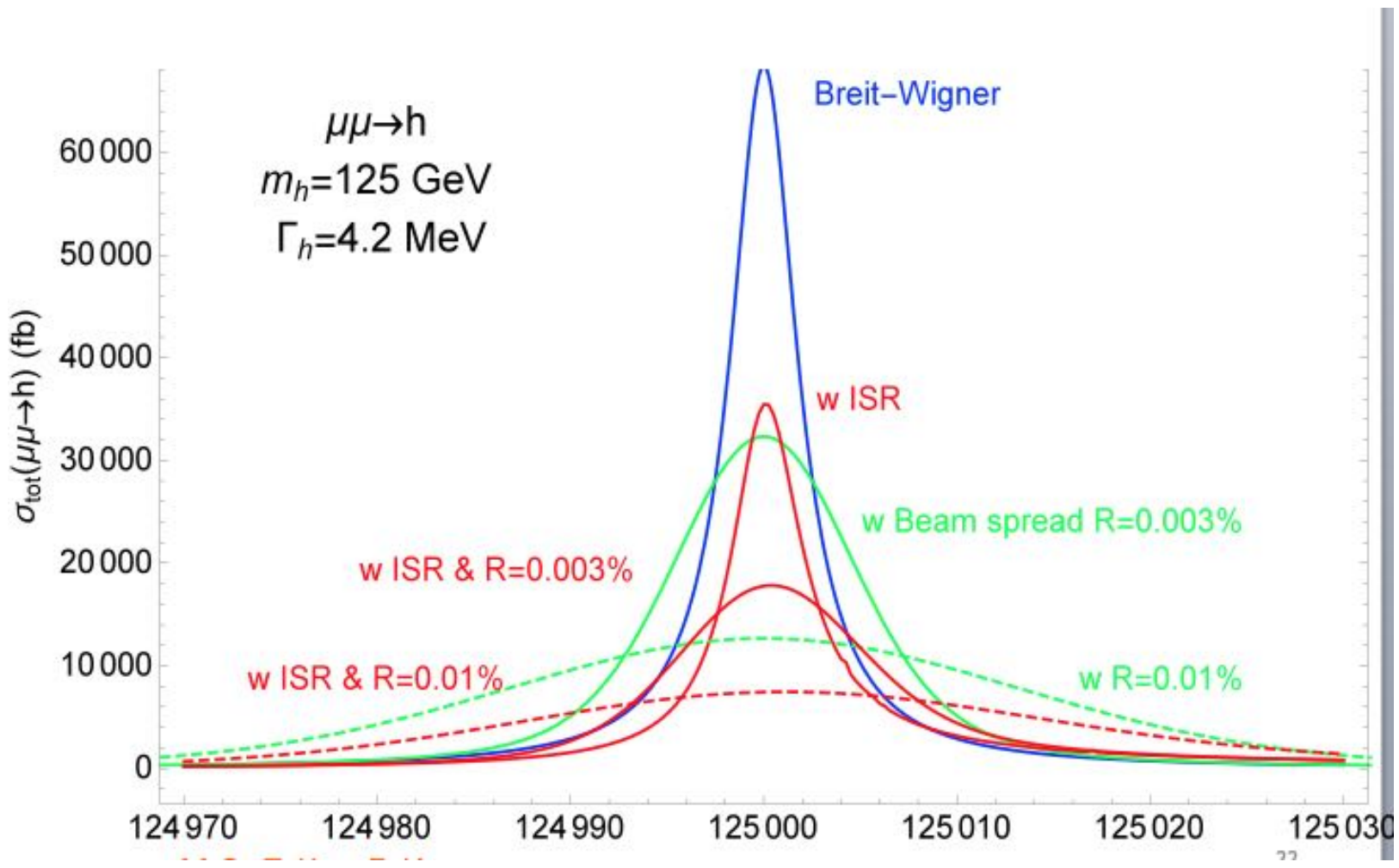

***Figure 2.-*** *According to the Standard Model (SM), the width of the $H^o$ is only ≈ 4.2 MeV, compared for instance to the much larger $Z^o$ width of 2.5 GeV.*

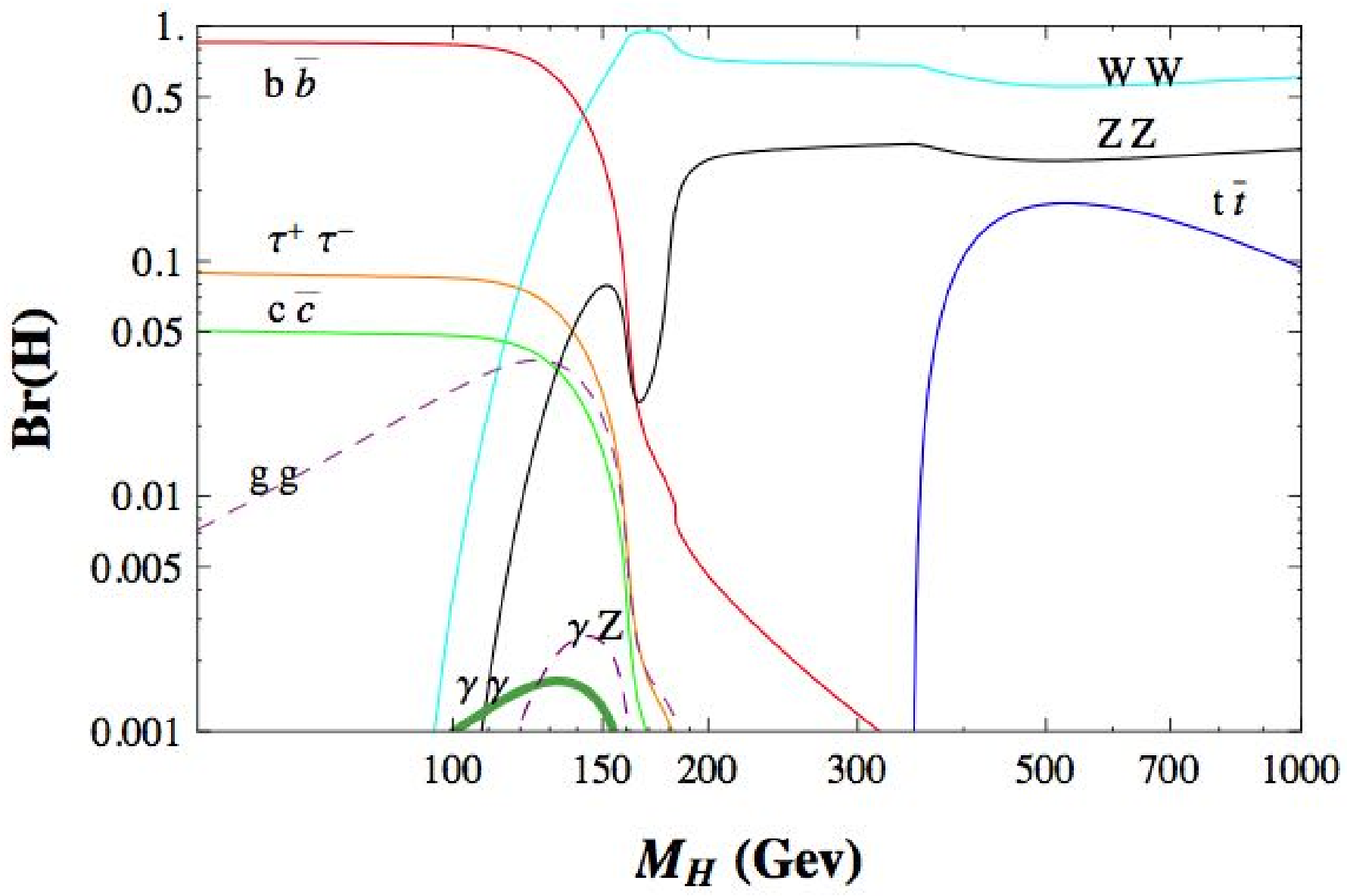

***Figure 3.-*** *Leading Higgs channels as a function of the Higgs mass.*

The Liouville theorem [54] states that whenever there is a Hamiltonian (i.e. a force derivable from a potential) the six dimensional phase space of the beam ($q_i$, are positions and $p_i$ conjugate momenta) is preserved, namely, dV/dt = 0, since the rate of change of volume has to be equal to the volume integral of its divergence. This is a very powerful constrain since *in an accelerator any combination of magnetic and electric fields is derivable from a Hamiltonian*. Therefore in order to introduce compression of a beam during acceleration we must provide for a dissipative non-Liouvillian drag force working against the particle speed and *not derivable from an Hamiltonian*.

Several non-Liouvillian alternatives are possible. The *"Ionization Cooling"* method [5] – here discussed for incoming muons – closely resembles the synchrotron compression of the relativistic electrons, with the multiple energy losses in a thin, low Z absorber substituting the classic synchrotron radiated photon emissions of $e^{\pm}$. The main feature of this method is that it may produce an extremely fast cooling, compared to other traditional methods, a necessity for the muon case.

As well known, Ionization Cooling may occur passing a beam periodically through a thin material to cause energy losses in the medium through effects such as ionization and excitation. The effect of these ionization losses is to reduce the beam momentum in all three dimensions. To maintain the reference momentum of the beam, momentum is restored by an accelerating RF cavity, but only in the longitudinal dimension.

Since the muon is an unstable particle, it must be generated from a beam striking a production target. Muons of both signs that then emerge from the decays of the target must be gathered, cooled and accelerated rapidly to higher energies, at which point they can finally collide in a compact storage ring and high field superconducting magnets.

As well known, the $H^{o}$ being a *scalar* (spin = 0) particle, it is characterized by a much stronger coupling when initiated from muons than from electron pairs. The energy losses of the muon are described by the Bethe-Bloch theory and the multiple-scattering heating by the Moliere theory. The derivations and discussions of the basic formulae of Ionization Cooling can be found in many places.

In order to put this work in its context, it is important to view the efforts in a historical perspective and to give proper credit to the many predecessors [5]. The concept of Ionization Cooling was initially proposed in the 60's and early 70's by Budker [6] and by Skrinsky [7], However, there was little substance to the concept until Skrinsky and Parkhomchuk [7] further developed the idea.

In the US, the initial ideas were presumably due to Neuffer [8] in 1979 and to Cline and Neuffer [9] in 1980. Neuffer [10] had then shown that the muon ionization cooling of the transverse emittance was limited by beam heating due to multiple Coulomb scattering. After the first outlines by Neuffer et al. in 1983 and the development of the solenoidal capture by Palmer (1984), a first Snowmass feasibility study has been organized in 1996 and a US collaboration has been formed in 1997 with DOE organization and funding [5].

As discussed for instance already in 1994 by Barletta and Sessler [11], muon pairs may be produced by two different classes of processes, (1) $\mu^+\mu^-$ pair electro-production and (2) production of pions and kaons from protons, subsequently decaying into muons. The claimed advantage of process (1) initiated from a high energy (45 GeV) $e^+$ accelerator on $e^-$ at rest is that the $\mu^+\mu^-$ pairs are produced near threshold where a tiny signal is concentrated — also in absence of an effective cooling — in a very short bunch before acceleration to the desired energy. However, at that time, the achievable luminosity (1) had been considered as vastly insufficient for any practical realization. Several other authors have later confirmed such a conclusion on the process (1).

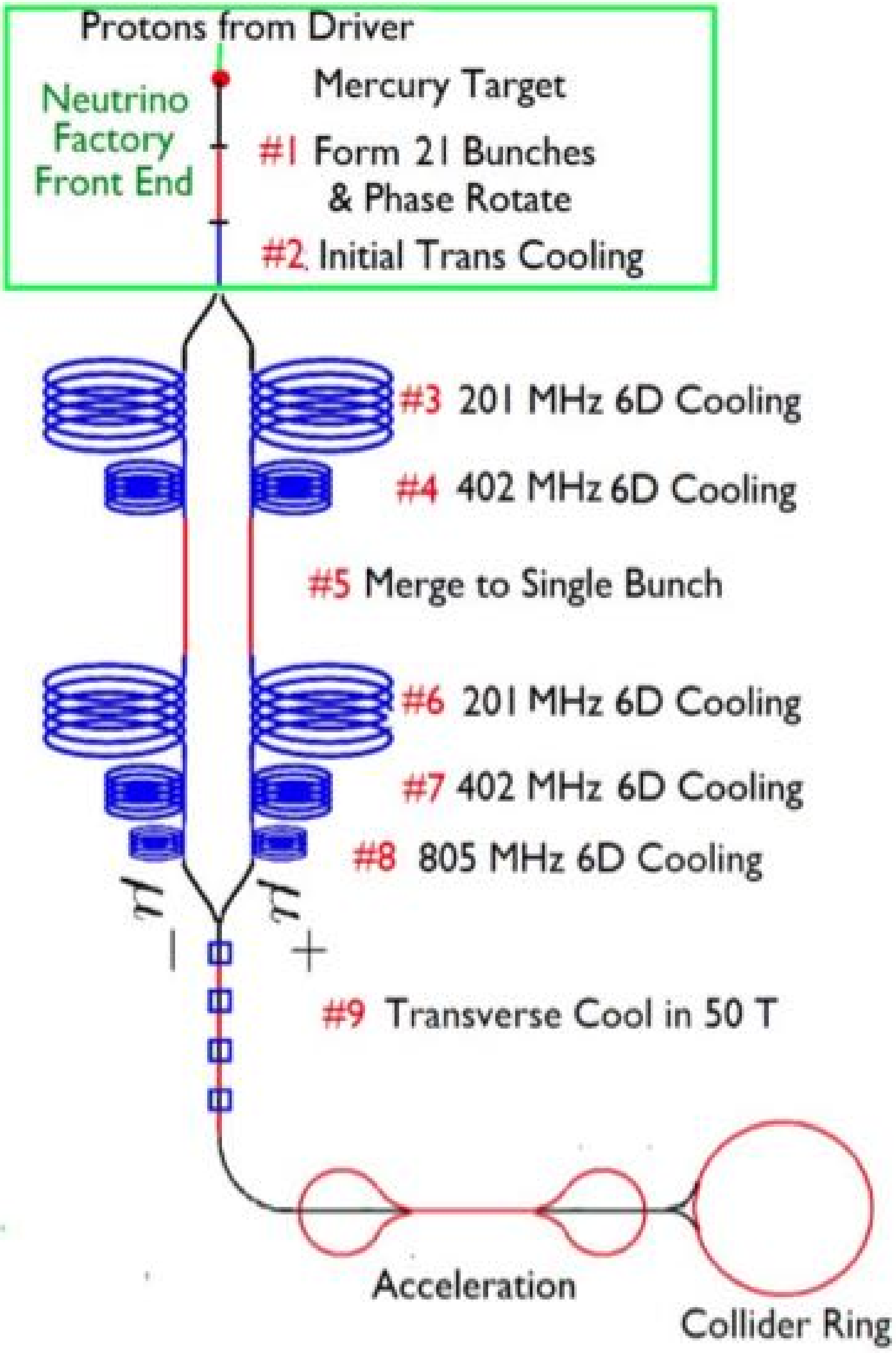

***Figure 4.-*** *Main components of a Muon Collider as described in COOL-2007 [13]; Phase (#1) is the phase rotation;(#2) initial cooling; (#3) 201Mhz 6D cool; (#4) 402 Mhz 6D cool; (#5) merge to single bunch; (#6) 201Mhz 6D cool; (#7) 402Mhz 6D cool; (#8) 805 Mhz 6D cool; (#9) transverse cool at 50 Tesla; acceleration and Collider ring.*

During the subsequent two decades, Neuffer, Palmer, Cline and many others in the US and elsewhere have greatly expanded Ionization Cooling [12] of process (2) with protons. These have been very important developments, but only very few verifying experimental tests have been performed.

Out of the many different descriptions, we shall focus on the COOL-2007 Conference report (Figure 4) of Palmer et al. [13] in which a complete scheme for capture, phase manipulation and cooling of the muons, every component of which has been simulated at a significant level. Their specific choice has been of a 1.5 TeV ($\mu^+ \mu^-$) Collider with 1.5 km circumference. Other final energies like the 0.125 TeV for the Higgs factory and 0.35 TeV for the Top physics have been considered.

A recent (2013) study LEMMA [14] of process (1) on electro-production (Figure 5) is presently being carried out at the INFN laboratory in Frascati, Italy initiated by a positron beam of 1.5 $\times 10^{18}$ $e^+$/s at an energy of about 45 GeV, slightly above the $\mu^+\mu^-$ 22.5 GeV production threshold in a 6 km circumference ring (the size of the CERN-SPS). At these energies, a main feature is the absence of cooling. The initial acceleration to 45 GeV of such a large number of positrons is an additional and very considerable task, for instance demanding a ≈ 3 km long Linear Accelerator, comparable in size to the old and well known of the Stanford Linear Accelerator (SLAC) in the US. In absence of cooling, the performances of the LEMMA scheme are insufficient to study the $\Gamma$ width, since the $H^0$ at $\sqrt{s}$ = 125.5 GeV requires a remarkable energy resolution, $R = \Gamma/\sqrt{s} = 3.2 \times 10^{-5}$ ($\Gamma \approx 4.2$ MeV). Their project has been concentrated on a *"Very High Energy"* option, eventually well above 10 TeV. A 6.0 TeV Collider has been described in detail.

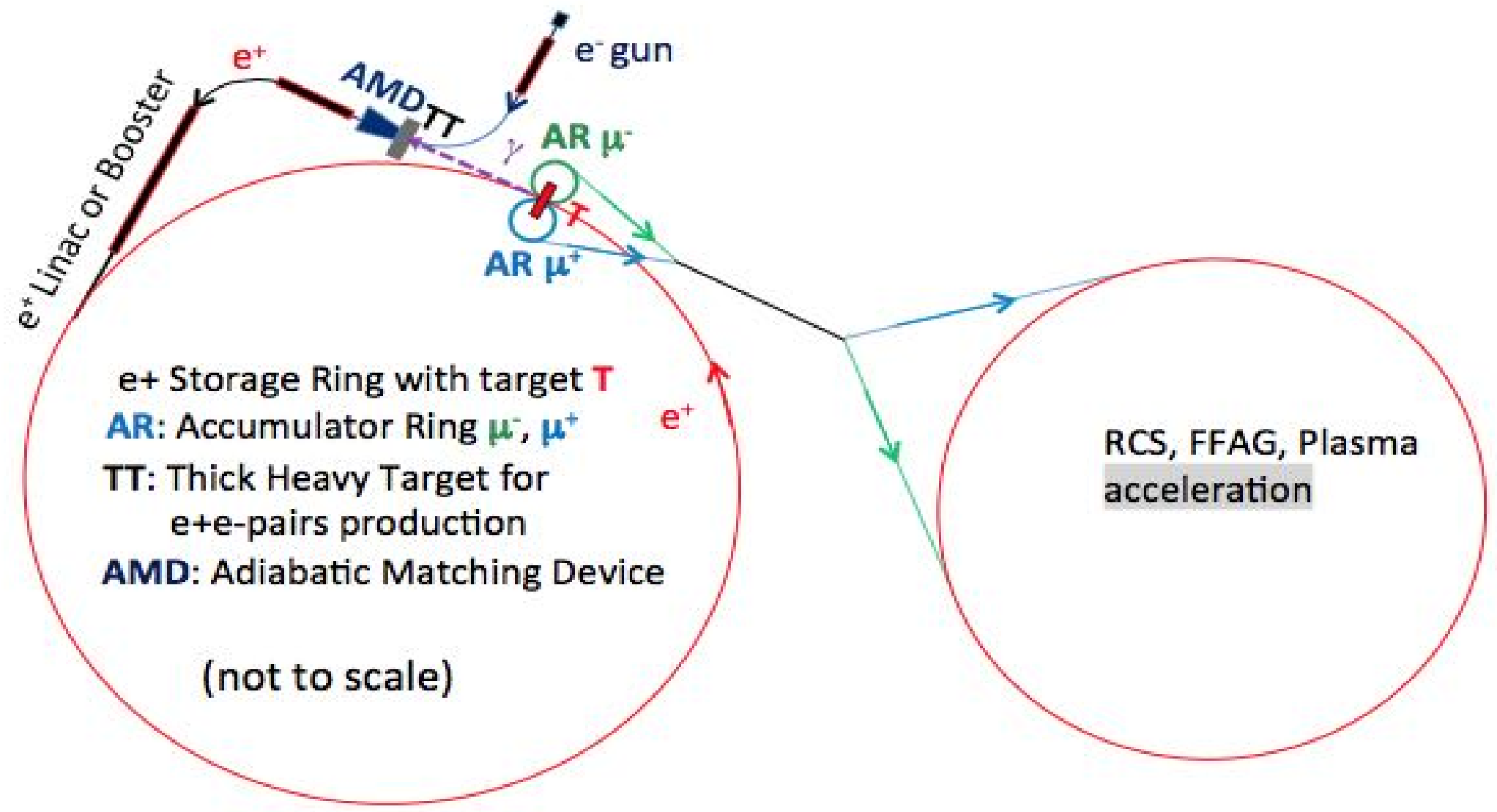

***Figure 5.***- *Low EMittance Muon Accelerator (LEMMA) [14] is based on muon pair production by a positron beam impinging on electrons at rest in a target, constrained into a very small longitudinal and transverse phase space regions. If successful, this approach would allow the exploration of the multi TeV region; a preliminary set of parameters for a low emittance 6 TeV muon collider is described.*

The physics of Ionization Cooling is straightforward; however its actual experimental realization with incoming muons has not been yet fully demonstrated. Further developments, like for instance the "*Initial Cooling experiment"* described in Paragraph 14, are needed and should provide valuable knowledge about the challenges of actually constructing an appropriate ionization cooled muon project. The presently described ionization cooled muon program will then be compared with the several other proposals based on $e^+e^-$ collisions.

### 3.- Future $e^+e^-$ proposals at CERN and elsewhere.

These new $e^+e^-$ developments are also possible in a revisited CERN-LEP tunnel (LEP3), where the Higgs signal could be further studied [15] [16] with the reaction $e^+ + e^- \rightarrow H^0 + Z$ presumably up to a $\sqrt{s} \approx 240$ GeV. The LEP3 Collider may use most of the already existing infrastructures, including the tunnel, cryogenics, injection and the two general-purpose LHC experiments ATLAS and CMS.

Two new $e^+e^-$ rings should be introduced inside the present CERN 27 km tunnel: (1) a low emittance Collider ring and (2) a separate accelerator complex injecting periodically electrons and positrons to top up the beams.

This new LEP3 Collider should have a luminosity of $10^{34}$ $cm^{-2}$ $s^{-1}$ (about 100 x LEP), an energy loss/turn of ≈ 7 GeV with a luminosity lifetime of a few minutes. According to the Standard Model (SM), a cross section of 200 fb and of the order of 2 x $10^4$ $H^0$ events may be collected in each of two experiments for an effective $10^7$ s/year, offering very precise measurements of the $H^0$ mass, cross-section and decay modes, including invisible ones. The total wall plug power of the LEP3 complex would be between 200 and 300 MWatt.

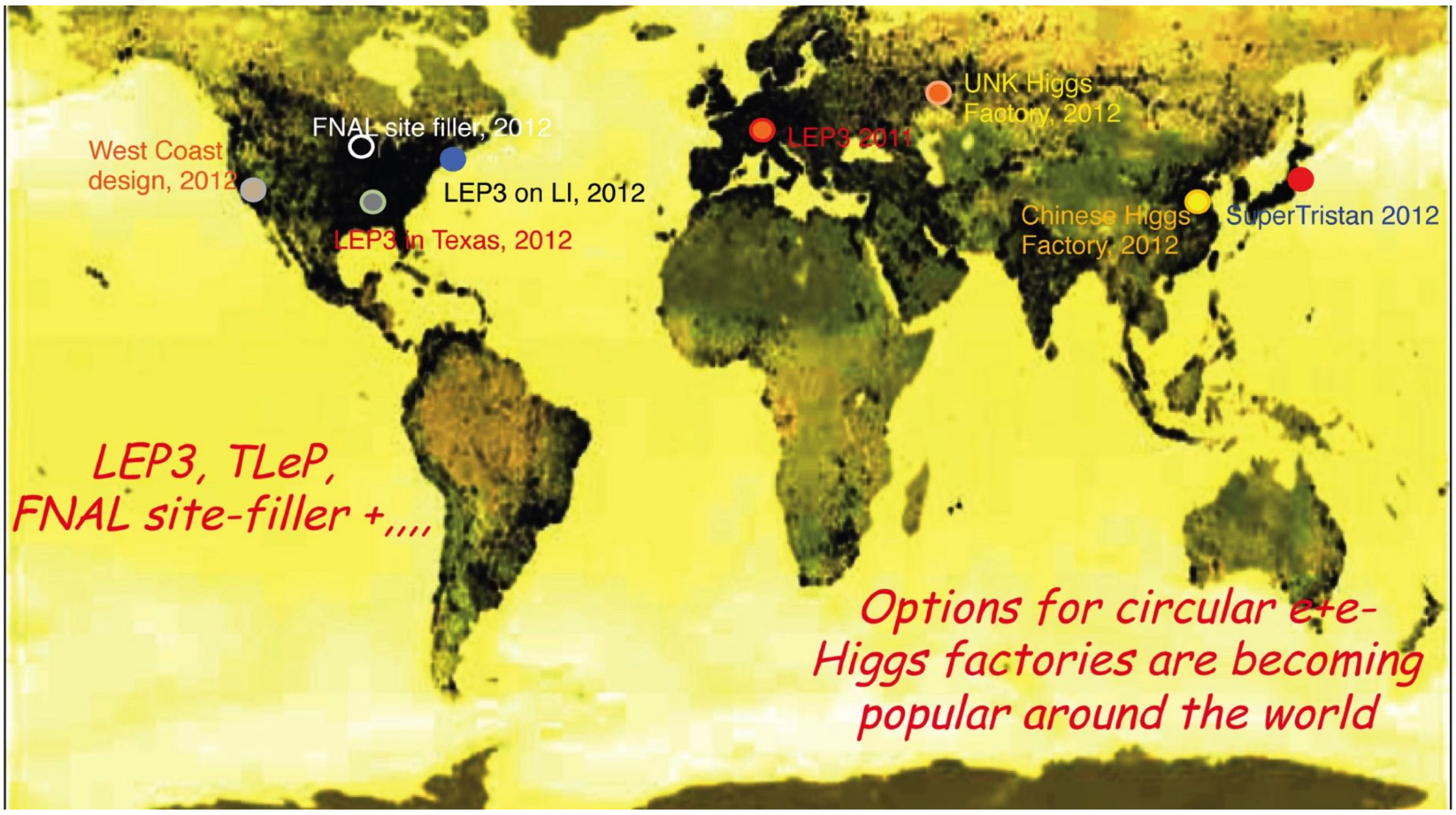

***Figure 6.-*** *Several other new projects based on circular rings of $e^+e^-$ described for different locations world wide.*

Several other new projects based on $e^+e^-$ have also been described for different locations (Figure 6). They all require huge new sites, vastly exceeding the dimensions of the LHC:

(a) *Conventional Collider Rings,* limited in $\sqrt{s}$ by the practical dimensions of the ring [17], requiring a luminosity near the beam-strahlung limit (about several hundred times x LEP2) and an extremely small vertical emittance with a beam crossing size the order of 0.07 $\mu$m (it has been 3 $\mu$m at LEP2);

*(b)* *Linear Collider (ILC)* [18] eventually up to $\sqrt{s} \approx 1$ TeV and a length of ≈ 50 km, a major new technology being developed. Two bunches of 5 nanometers (0.005 $\mu$m!) in height, each with 2 x $10^{10}$ particles are colliding 14’000 times per second. The 16’000 Niobium accelerating cavities are operating at 2 K with an accelerating gradient of ≈ 30 MV/m. Two damping rings are also required, each one with a circumference close to the one of the CERN-SPS.

Two typical examples of $e^+e^-$ circular rings for alternative (a) are quoted: FCC-ee [17] from CERN with circumference of 100 km (3.7 x LEP), and CEPC [19] from China with circumference of 100 km (arXiv:1811.10545) with an expectation of about $10^6$ Higgs in 10 years of operation, primarily from $e^+ + e^- \rightarrow H^0 + Z$.

Such a large luminosity and a synchrotron beam power of the order of 100 MWatt (corresponding to an electricity consumption of 500 MWatt) are realized with a very large number of $e^+e^-$ bunches (2 x 50 bunches for CEPC) and a very small vertical betatron focal radius $\beta \approx 1$ mm (it was 50 mm for LEP2).

In many instances and in particular in China [19], the $e^+e^-$ phase is eventually to be followed by a subsequent phase of *proton-proton collisions* with very high fields (≥ 20 Tesla) magnets, thus approaching 50 to 100 TeV, about 4 to 8 times the present energy of the LHC.

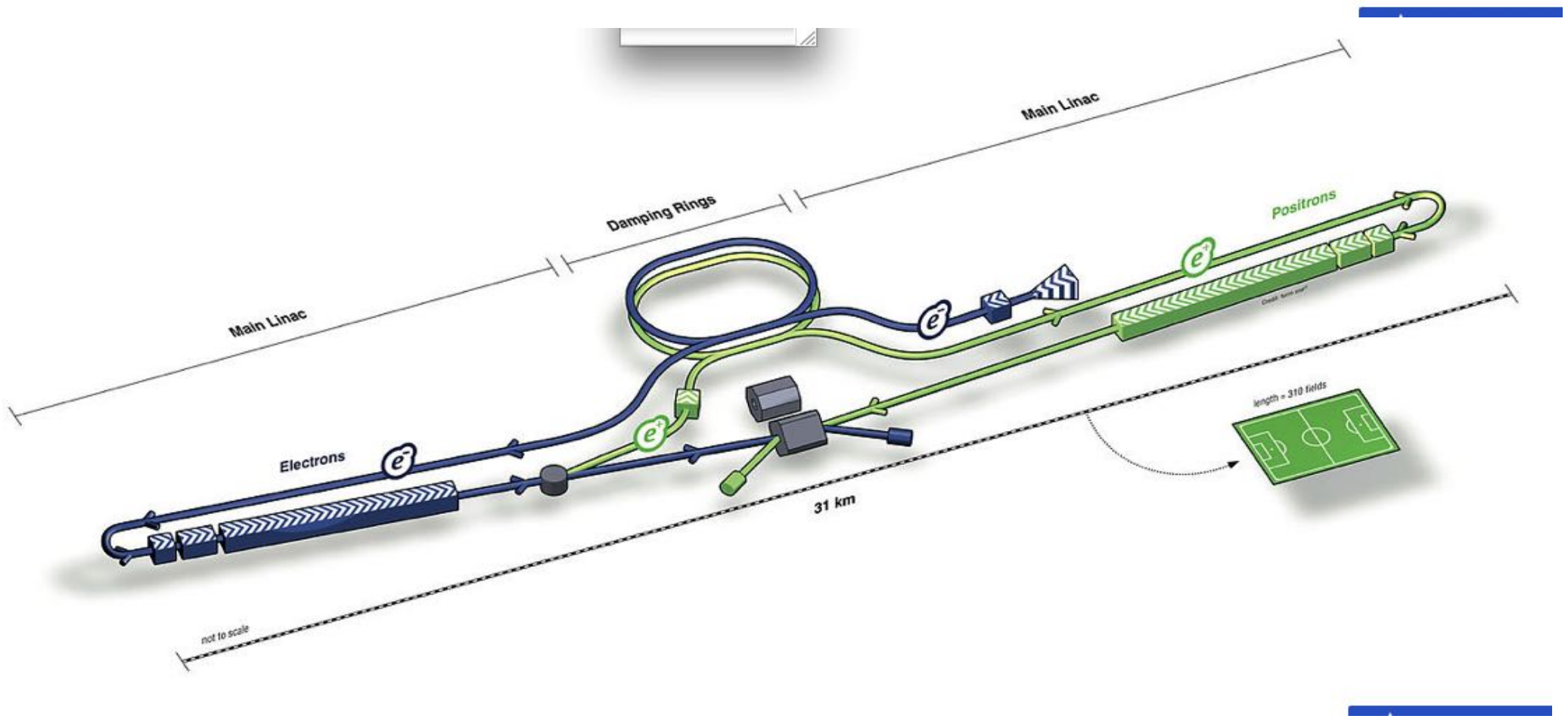

***Figure 7***.- *The International Linear Collider (ILC) as a proposed linear particle accelerator. It is planned to have initially a collision energy of 500 GeV, with a later upgrade to 1000 GeV (1 TeV) in the Kitakami highland in the Iwate prefecture of northern Japan has been the focus of ILC design efforts. Japan is willing to contribute half of the costs. However the Government has recently put the program on an indefinite hold.*

In order to further develop the $e^+e^-$ alternative (b) for a future ILC Collider (Figure *7*), a dedicated experimental facility, the ATF2 (*Accelerator Test Facility*) has been constructed in Japan at KEK. The main R&D research focus of ATF2 is the production, measurement and control of a ultra-low emittance beam. The ATF project has been initiated in 1993 by a number of Japanese universities as a collaboration of accelerator physicists, high energy physicists and engineers. By now the collaboration has been further extended and it includes significant participation from SLAC and LBNL of US, PAL of Korea, IHEP of China, BINP of Protvino, Russia, DESY of Germany and of CERN.

Nearly 300 laboratories and universities around the world are presently involved in the ILC development program [18]: more than 700 people are working on the accelerator design and another 900 people on detectors. The design is coordinated for the accelerator by the Global Design Effort and for physics and detector by the World Wide Study. According to some European accounting, the estimated cost for the 500 GeV ILC option is ≈ 8 Billion $ (2012).

A relatively short-term approval, financing and actual construction of any of these gigantic new $e^+e^-$ options, ether (a) a Ring or (b) a Linear Collider is rather unlikely. Therefore other alternatives are here taken into consideration.

### *4.- The Muon Collider as a future $H^o$ Facility.*

The $\mu^+\mu^-$ Collider [5] may be preferable to the previously described huge $e^+e^-$ options because of its much smaller dimension and cost and since it may easily fit within one of the already existing European sites. However it requires the success of a substantial R&D in order to convincingly produce the adequate compression in 6D phase space of the muon beams.

Two different configurations of a $\mu^+\mu^-$ Collider are here described namely (1) the √s =125.5 GeV Higgs mass s-channel resonance to study the many $H^o$ decay modes with L ≈ $10^{32}$ $cm^{-2}$ $s^{-1}$ and very small backgrounds and (2) a higher energy Collider with L ≈ $10^{34}$ $cm^{-2}$ $s^{-1}$ to study other main $H^o$ related processes of the scalar sector.

Alternative (1) requires a $\mu^+$ $\mu^-$ Collider ring at the √s = $H^o$ mass with a tiny *60 m radius* for a later described ≈ 7 Tesla magnetic field and alternative (2) a Collider of ≈ *220 m radius* at ≈ 7 Tesla to explore $\mu^+$ $\mu^-$ collisions up to about √s ≈ 700 GeV or even beyond.

In analogy to the well known previous production of the $Z^o$ at CERN-LEP, alternative (1) for the $H^o$ offers remarkable conditions of cleanliness. As already pointed out, the $\mu^+\mu^-$ initiated cross section is greatly enhanced with respect to the one initiated with $e^+e^-$, since the leptons pair coupling to a $H^o$, being a scalar, is proportional to the square of the lepton mass.

Because of the much more favorable experimental environment when compared to the LHC, muon Colliders may extend with greater accuracies the properties of the Higgs boson. The production (1) is especially important. The resonant Higgs factory could produce the Higgs boson in the s-channel and provide the most direct measurement of the Higgs boson total width and the Yukawa coupling to muons.

The narrow width of the Higgs boson, about 4.2 MeV according to the Standard Model (*Figure* 3) may be quantified by convoluting the Breit–Wigner resonance for the Higgs signal and the Gaussian distribution for the profile of (BES) beam energy spread [26]. The Initial State Radiation (ISR) of the QED effect is very relevant since it would further degrade the peak luminosity of the muon Collider. Effective $\mu^+\mu^-$ cross sections at the resonance √s = 125.5 GeV, are of 71 pb for Breit–Wigner resonance profile alone, of 37 pb for ISR alone, of 17 pb and 41 pb for BES alone for two choices R = 0.01 % and R = 0.003 % of the energy resolution, and of 10 pb and 22 pb for both the BES and ISR effects included.

For an $e^+e^-$ Collider [26] and its much smaller cross section these differences would be even stronger. Effective $e^+e^-$ cross sections at the resonance √s = 125.5 GeV, are of 1.7 fb for Breit–Wigner resonance profile alone, of 0.5 fb for ISR alone, of 0.12 fb and 0.41 fb for BES alone for two larger choices R = 0.04 % and R = 0.01 % of the energy resolution, and of 0.048 fb and 0.15 fb for both the BES and ISR effects included.

Signal and background effective cross sections at the resonance at a $\mu^+\mu^-$ Collider of alternative (1) for the two main channels $H^0 \rightarrow bb_{bar}$ and $H^0 \rightarrow WW^*$ are displayed in Figure 8 with a beam energy resolution R = 0.003 % with ISR effects taken into account and SM branching fractions of 58% and 21%. The $H^0 \rightarrow bb_{bar}$ cross section is respectively 12 pb (5.6 pb with BES and ISR effects included) with a background of 20 pb and the $H^0 \rightarrow WW^*$ cross section is respectively 4.6 pb (2.1 pb with BES and ISR effects included) with a background of 0.051 pb.

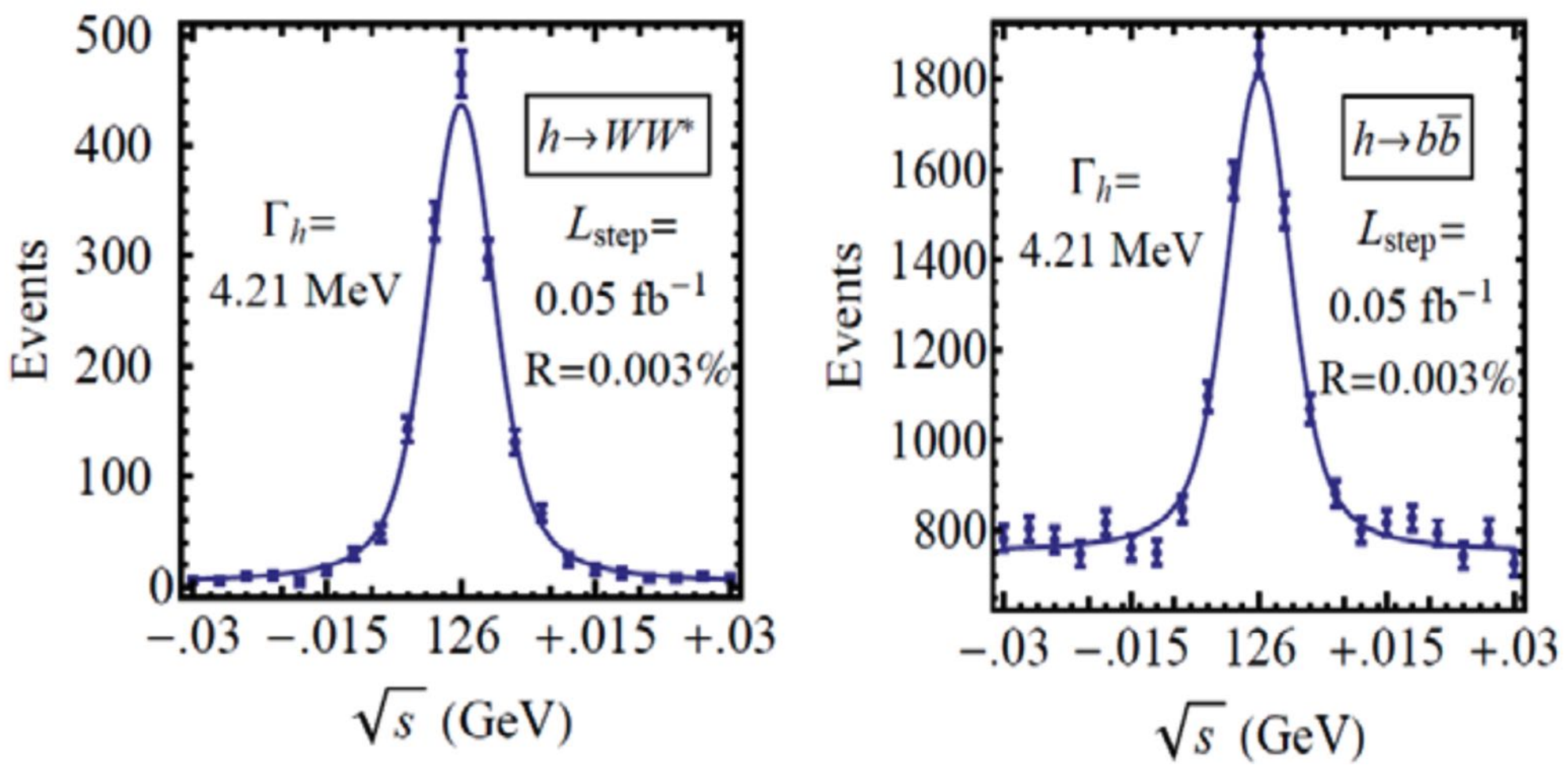

***Figure 8.-*** *Signal and background for H → bb and WW∗ at a energy resolution R = 0.003%. folded with a Gaussian energy spread Δ = 3.75 MeV and 0.05 fb$^{-1}$/step with detection efficiencies included.*

While the $\mu^+\mu^-$ produced $bb_{bar}$ event rate is larger by about a factor three, the latter process has a $H^0$ signal to background ratio of about 100:1. Its very narrow width and most of its decay channels may be compared with very high accuracy to the predictions of the SM. A remarkable and very demanding relative energy resolution R of the $H^0$ signal is however required since, as already pointed out, the expected $H^0$ width at $\sqrt{s}$ = 125.5 GeV corresponds to a relative width R = 4.2 MeV/125.5 GeV = 3.3 x $10^{-5}$.

No doubt, better future mass determinations of the $H^0$ will be performed already at the LHC. Accuracy of the order of ± 200 MeV may become reasonable. However as discussed later on, with $\mu^+\mu^-$, the actual value of its mass can be determined with a remarkable precision observing the (g – 2) precession frequency of polarized muon decays [20]. It may then be feasible to determine the experimental value of the mass of the $H^0$ particle to of the order of 100 keV.

The extensive studies with a $\mu^+\mu^-$ Collider at $\sqrt{s}$ = 125.5 GeV of alternative (1) are however not entirely sufficient in order to fully elucidate the underlying physics of the $H^0$. According to alternative (2), diagrams involving the production of single and double $H^0$ in higher energy collisions should be detailed up to energies of the order of several hundred GeV, well beyond the $\sqrt{s}$ = 125.5 GeV mark.

The main production cross sections of these Higgs related diagrams (Figure 9) are the so-called Higgs-strahlung diagram, the W-boson fusion process and the top-quark association. Double Higgs boson diagrams are generated mainly by the off-shell

Higgs-strahlung and by the W-boson fusion processes. The further study of all these diagrams requires new and substantial energies up to √s ≈ 0.5 TeV, or even beyond.

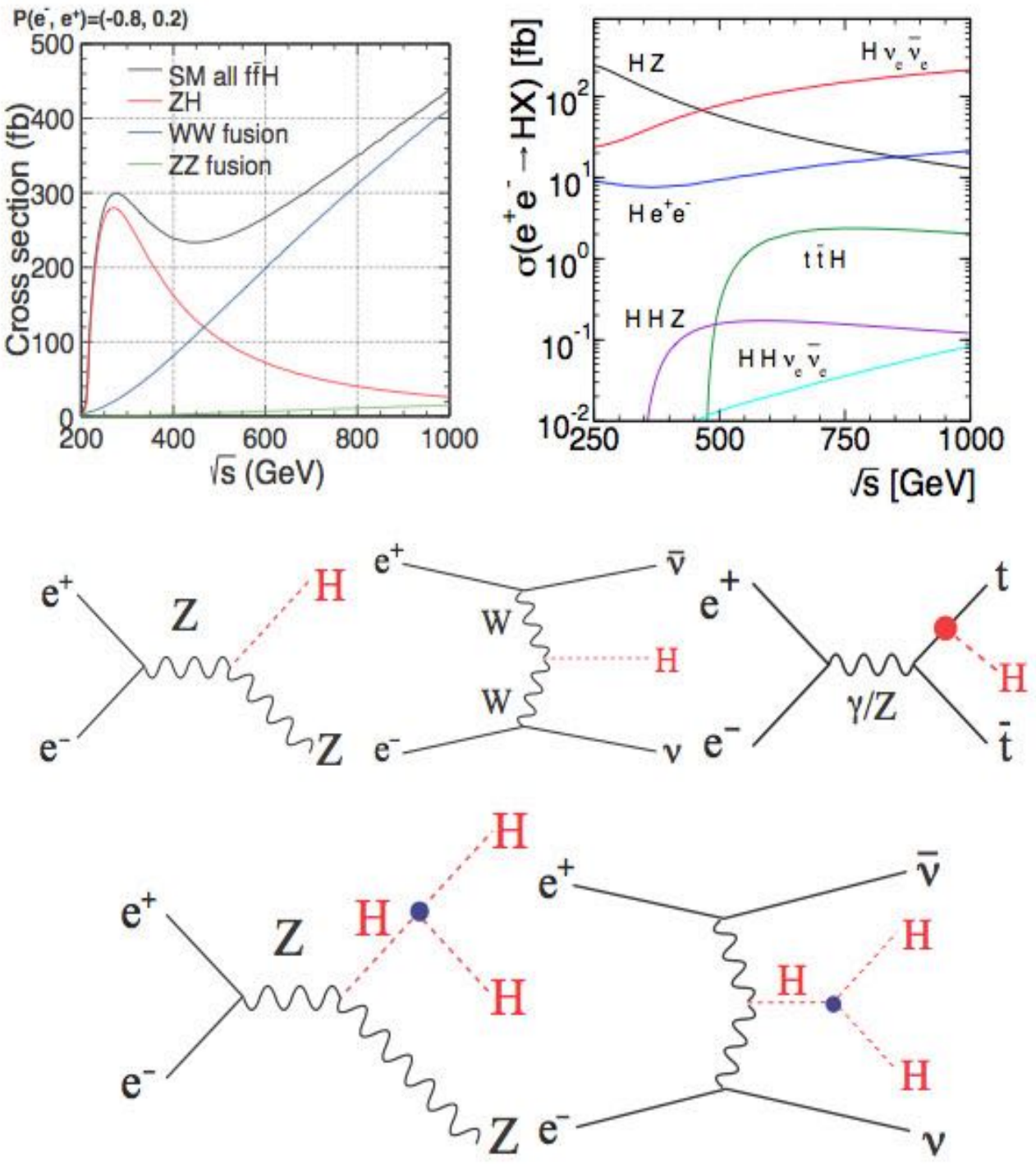

***Figure 9.-*** *Production cross sections from $e^+$-$e^-$ or $\mu^+$-$\mu^-$ -> Higgs + X at √s ≤ 1 GeV. On the top: The Higgs-strahlung diagram (Left), the W-boson fusion process (middle) and the top-quark association (right). On the bottom: double Higgs boson diagrams via off-shell Higgs-strahlung (left) and W-boson fusion (right).*

The measurements of the cross sections (2) of double-Higgs production and Higgs boson self-coupling projections with and without a triple-Higgs boson vertex have been already considered at the LHC. However due to the negative interference of these contributions, only a handful of events is expected in the final state HHZ even after 3000 $fb^{-1}$ of data. As of today, it is thus unclear if a fully meaningful measurement of the Higgs boson self-couplings (2) may be possible in the future with the LHC.

The predicted $H^o$ cross sections (2) are of the order of 2 x $10^{-37}$ $cm^2$ (200 fb). According to alternative (2), a precision determination in the √s domain up to several hundred GeV requires a ($\mu^+$ $\mu^-$) luminosity of about $10^{34}$ $cm^{-2}s^{-1}$. A convenient ($\mu^+\mu^-$ radius for a guide field of the order of 7 Tesla would represent only 5% of the LHC.

## *5.- The European Spallation Source.*

Most of the world's neutron sources were built decades ago, and although the uses and demand for neutrons have increased throughout the years, few new sources have been built. To fill that need new, improved neutron sources are necessary.

The present discussion for a $\mu^+\mu^-$ Collider will be primarily concentrated on a optimal scenario offered by further developments of the *European Spallation Source (ESS)* [21] under construction in the Lund's site.

The ESS linear accelerator (Figure 10) is a new major user facility where researchers from Academia and Industry will investigate scientific questions initially using neutron beams. These methods provide insights about the molecular building blocks of matter not easily available by other means. They are used for both basic and applied research.

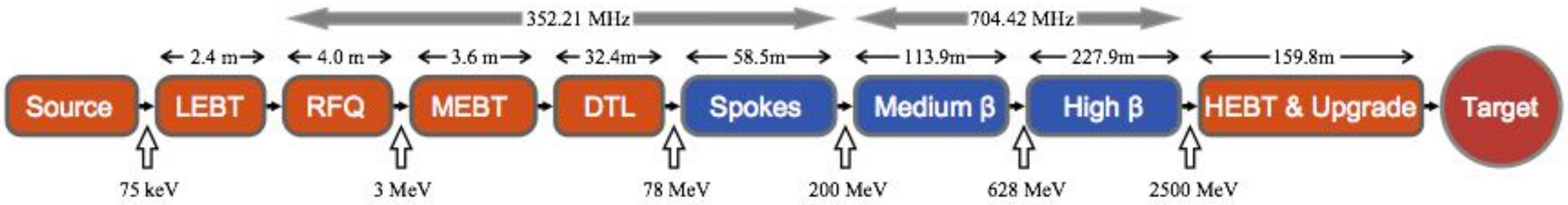

***Figure10***.- *Schematic drawing of the 2.0 GeV proton linear accelerator of the European Spallation Source in Lund [21]. Orange items are normal conducting while blue items are superconducting.*

ESS AB is a shareholding company under Swedish law in which Sweden holds approximately 75% of the shares and Denmark the remaining 25%. Its Steering Committee currently includes 17 partner countries: Sweden, Denmark, Norway, Latvia, Lithuania, Estonia, Iceland, Poland, Germany, France, the United Kingdom, the Netherlands, Hungary, the Czech Republic, Switzerland, Spain, and Italy. The 17 partner countries support ESS both through financial contributions and through in-kind contributions of materials, equipment and expert scientific and technical services.

ESS is supported by the European Strategy Forum for Research Infrastructures (ESFRI). ESFRI has concluded that a new type of neutron source, a very high intensity long-pulse source capable of dramatic gains, would be needed to address new science challenges.

ESS will provide major advantages in large domains of Science. The construction of the ESS facility for the spallation neutron production has started in September 2014. The first proton beam with low energy and intensity is already expected by 2019. The ESS will produce its first neutrons in 2019 and reach full accelerator potential in 2024 as the world's first fourth-generation neutron source.

The design phase for the spallation related instruments will continue past 2020, when the final instrument concepts are slated for selection. This staged approach will permit ESS to remain engaged with the European user community and to choose instrument designs that are state-of-the-art and scientifically relevant when they enter user operation.

The layout of the ESS site is shown in Figure 11. According to the presently approved initial design, the LINAC will make possible acceleration at 14 Hz of 2.86 ms long and 62.5 mA proton pulses to 2.0 GVolt kinetic energy (p = 2.784 GeV/c,□$\beta_{\pi}$ = 0.9476, $\gamma_{\pi}$ = 3.131), corresponding to 1.1 x $10^{15}$ protons/pulse and to an average beam power of nearly 5.0 MWatt. The duty cycle of the accelerator complex is 4%. The RF is 352.2 MHz in the first low energy spoke cavity section of the ESS LINAC and 704.4 MHz in the second high–energy elliptical cavity section. Available additional space (≈ 160 m) may permit to bring the energy up to 3.5 GeV with an extension of the LINAC.

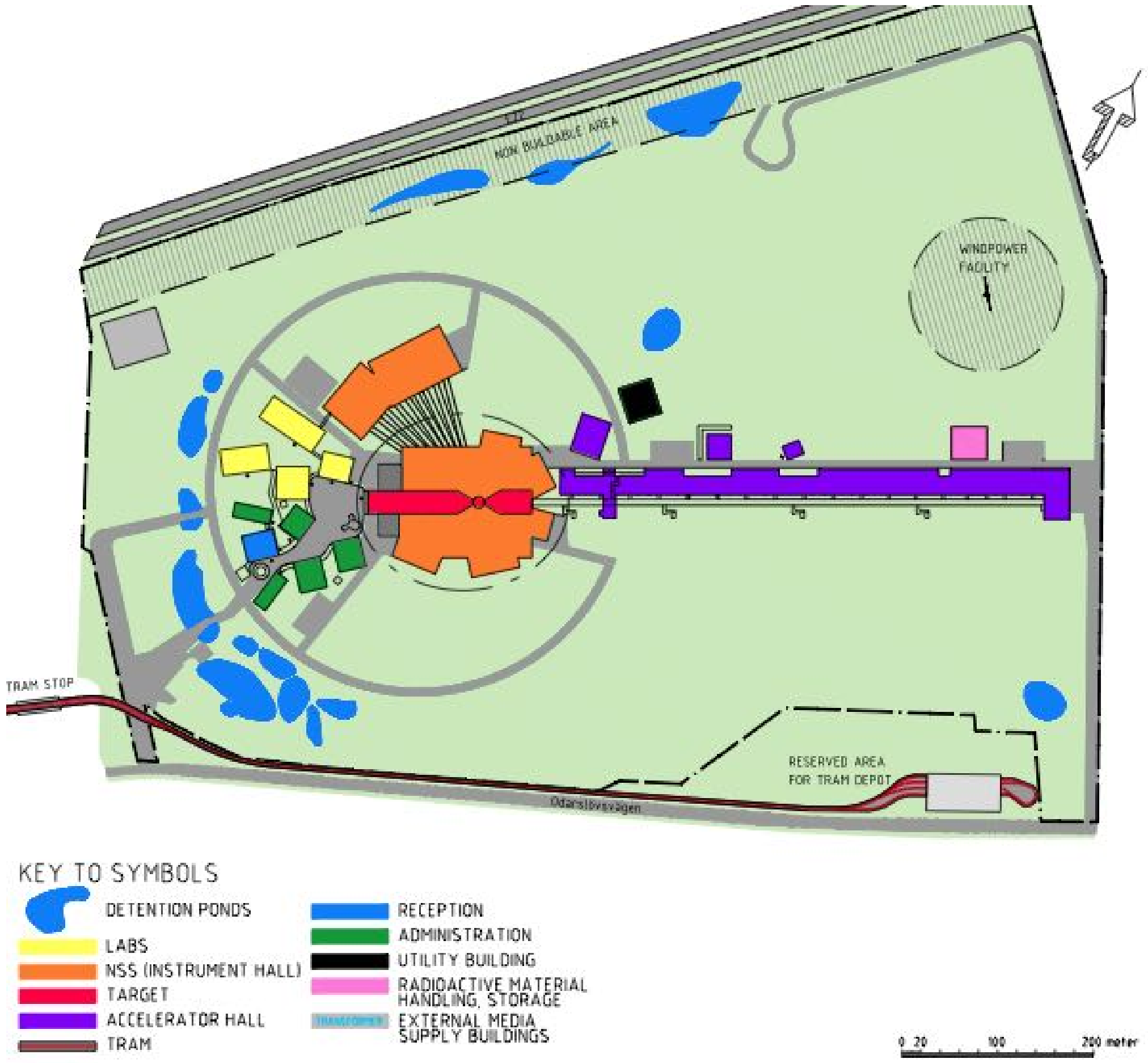

***Figure 11.-*** *The European Spallation Source, now in construction in Lund and 5 MWatt at 14 Hz of 2.86 ms and 62.5 mA proton pulses at 2.0 GeV kinetic energy (p = 2.784 GeV/c, $\beta_\pi = 0.9476$, $\gamma_\pi = 3.131$), corresponding to the remarkable intensity of 1.1 x $10^{15}$ protons/ pulse or 1.5 x $10^{16}$ p/sec. The duty cycle is 4%. The RF is 352.2 MHz in the first section and 704.4 MHz in the high–energy elliptical cavity section. Available additional space (≈ 160 m) may permit to bring the energy up to 3.5 GeV with an extension of the LINAC structure.*

It will deliver its protons to a solid, rotating tungsten target that will in turn generate neutrons for an initial suite of seven neutron scattering research instruments. ESS will later reach its full design specifications with a suite of 22 research instruments.

A spallation target is a very prolific source of neutrons. When for instance a high-energy proton hits the nucleus of a mercury atom, many neutrons are "spalled" I,e, thrown off, with for instance the order of 50-60 neutrons for an incident 2 GeV proton. The proton beam may be hitting a Helium cooled, 4 ton tungsten spallation target. Neutrons are guided out of the target vessel into beam guides that lead directly to instrument stations. Neutrons from the target must be moderated to room temperature or colder. Neutrons are slowed down by passing them through cells filled with water (to produce room-temperature neutrons) or through containers of liquid hydrogen at a temperature of 20 K (to produce cold neutrons). Neutrons are moderated adjacent to the target, before being transported through neutron guides to a large number of experimental instruments.

ESS should be compared with the other main neutron sources worldwide. In the US, the highly successful SNS Spallation Neutron Source [22] was completed at Oak Ridge in April 2006. The first three instruments began commissioning and were available to the scientific community in August 2007. As of 2017, 20 instruments have been completed and SNS is hosting about 1'400 researchers each year. A linear accelerator (LINAC), accelerates the $H^-$ beam from 2.5 MeV to 1 GeV. The ring structure bunches and intensifies the ion beam for delivery onto the mercury target to produce the pulsed neutron beams. The ions pass through a foil which strips off each ion's two electrons, converting it to a proton.

A major SNS upgrade is proceeding in the US involving two major projects: a proton power upgrade of the existing accelerator structure (SNS-PPU) and construction of a second target station complex with an initial suite of beam lines (SNS-STS). The SNS-PPU project scope consists of doubling the proton beam power capability from 1.4 MWatt to 2.8 MWatt, and upgrading the STS target systems to accommodate beam power up to 2 MWatt. The power doubling is achieved by increasing the proton beam energy by 38% and peak beam current by 45%.

The Asia-Pacific research infrastructure is also going through a transition, driven by the region's rapidly growing and strengthening materials research. China will build an ISIS-level facility with the China Spallation Neutron Source (CSNS), while the Australian Nuclear Science and Technology Organization (ANSTO) and the High-Flux Advanced Neutron Application Reactor (HANARO) in South Korea are achieving high impact, productivity, and regional capacity.

One of the most advanced source and linchpin of the region is the Material and Life Sciences Facility (MLF) at the Japan Proton Accelerator Research Complex (J-PARC) [23]. As this highly advanced source reaches design performance at 1 MW, it will rival SNS. In addition, neutron beams at both J-PARC and ESS will have low repetition rates and broad energy bandwidths that are well matched to emerging challenges in complex hierarchical materials.

Within such wide international framework, the ESS initiative holds the promise of future spallation neutron intensities that will be a factor of six more intense per megawatt of proton beam power than any contemporary existing or planned facility. In addition the beam power of ESS is larger than any other similar installation elsewhere. For instance the typical experimental average spallation neutron flux is $1.6 \times 10^{15}$ $cm^{-2}$ $s^{-1}$ compared with $1 \times 10^{14}$ $cm^{-2}$ $s^{-1}$ of SNS at Oak Ridge [22].

In 2007, SNS was entered into the Guinness Book of World Records as the most powerful, pulsed spallation source. It will be vastly overcome by the future ESS facility.

Unlike the Spallation Neutron Source (SNS) in the US [22] and J-Park [23] in Japan, ESS does not use for the pulsed spallation source a compressor ring. In view of its very large power, the ESS macro-pulses of 2.86 ms are sent directly to the target from the LINAC complex rather than with the ~1 µs duration which is for instance used in the ≈ 1 MWatt SNS compressor.

The intent of ESS is to deliver the same kind of impact that the "Institut Laue-Langevin" (ILL) the international research centre at the leading edge of neutron science and technology had achieved during the last 40 years —representing a game-changing level of performance in cold neutrons [24]. Situated on the Polygone Scientifique in Grenoble, France, ILL is today one of the world centres for research using neutrons. Founded in 1967 and honouring the physicists Max von Laue and Paul Langevin, the ILL provides one of the most intense neutron sources in the world and the most intense continuous neutron flux in the world in the moderator region: $1.5 \times 10^{15}$ neutrons per second per $cm^2$, with a thermal power of 58.3 MW.

The "Institut Laue-Langevin" (ILL) makes its facilities and expertise available as a service to visiting scientists. Every year, some 1400 researchers from over 40 countries visit the ILL. More than 800 experiments selected by a scientific review committee are performed annually. Research focuses primarily on fundamental science in a variety of fields: condensed matter physics, chemistry, biology, nuclear physics and materials science, etc.

Although ESS may eventually lead to a progressive phase-out of ILL, at least one other major reactor source, such as the Forschungs-Neutronenquelle Heinz Maier-Leibnitz (FRMII) reactor in Germany [25], remains essential in providing sufficient beam line capacity and development capabilities in order to meet the European needs.

Much of the fleet of highly successful medium-flux facilities elsewhere may be eventually retired as their performance becomes no longer competitive against the new capabilities.

## 6.- Future developments of ESS.

The further extension of the *European Spallation Source (ESS)* [21] may offer unique opportunities primarily because of its unique very large beam power and its fast repetition rate. The ESS may become a highlight project for further studies of the Higgs in Europe since in other existing locations – like for instance at CERN – the development of adequately intense proton beams is presently no longer under active consideration.

Major new developments of the ESS program are possible as further additions to the already approved direct injection to the target generating the intense source of spallation neutrons. The LINAC repetition rate could be doubled from 14 Hz to 28 Hz with additional 5 MWatt protons to new facilities, bringing the duty cycle of the LINAC from 4% to 8%. In these future upgrades, although the initial ESS proton kinetic energy will be the relatively low value of 2 GeV, the LINAC could be eventually extended to ≈ 3.5 GeV within the already available space.

A main feature for the further applications of the ESS is the addition of a chain of storage rings handling the proton beam: a first ring called the *"accumulator"* to collect the LINAC pulse and eventually also a second ring called the *"compressor"* to further steer the beam according to some specific requirements of the target.

The transition from the LINAC to the accumulator ring should be performed by a multi-turn injection using negative $H^-$ ions. Negative [p+2e-] ions — i.e. three simultaneous m.i.p.'s, namely two electrons and a proton — are accelerated by the LINAC, stripped of their two electrons at the entrance of the accumulator ring, either with a thin absorbing foil or of an appropriate LASER beam.

The use of negative ions is more demanding. Already during their path inside the LINAC, negative $H^-$ ions may lose some of their electrons in collisions with residual gas, blackbody photons or the so-called intra-beam stripping, first observed at LEAR and the primary source of beam losses at SNS [22].

Both the longitudinal and the transverse emittances of the resulting proton beam are strongly compressed when accumulated in the ring. As already pointed out, the beam conversion of ($H^-$) to protons is not derivable from an Hamiltonian, like it would be generally the case of any combination of electric and magnetic fields and where the 6D beam emittance is instead conserved.

The "*proton power*" (namely a number of protons inversely proportional to its energy) is almost independent of the proton energy in the interval between 8 and 20 GeV and it is about a factor two lower for 2 GeV protons. Therefore the secondary pion beams are primarily dependent on *proton power* rather than on the actually chosen *proton energy*. Because of its proton energy and power, the ESS represents therefore an almost ideal solution for the production of the huge intensity of O($10^{13}$) $\mu^{\pm}$/pulse.

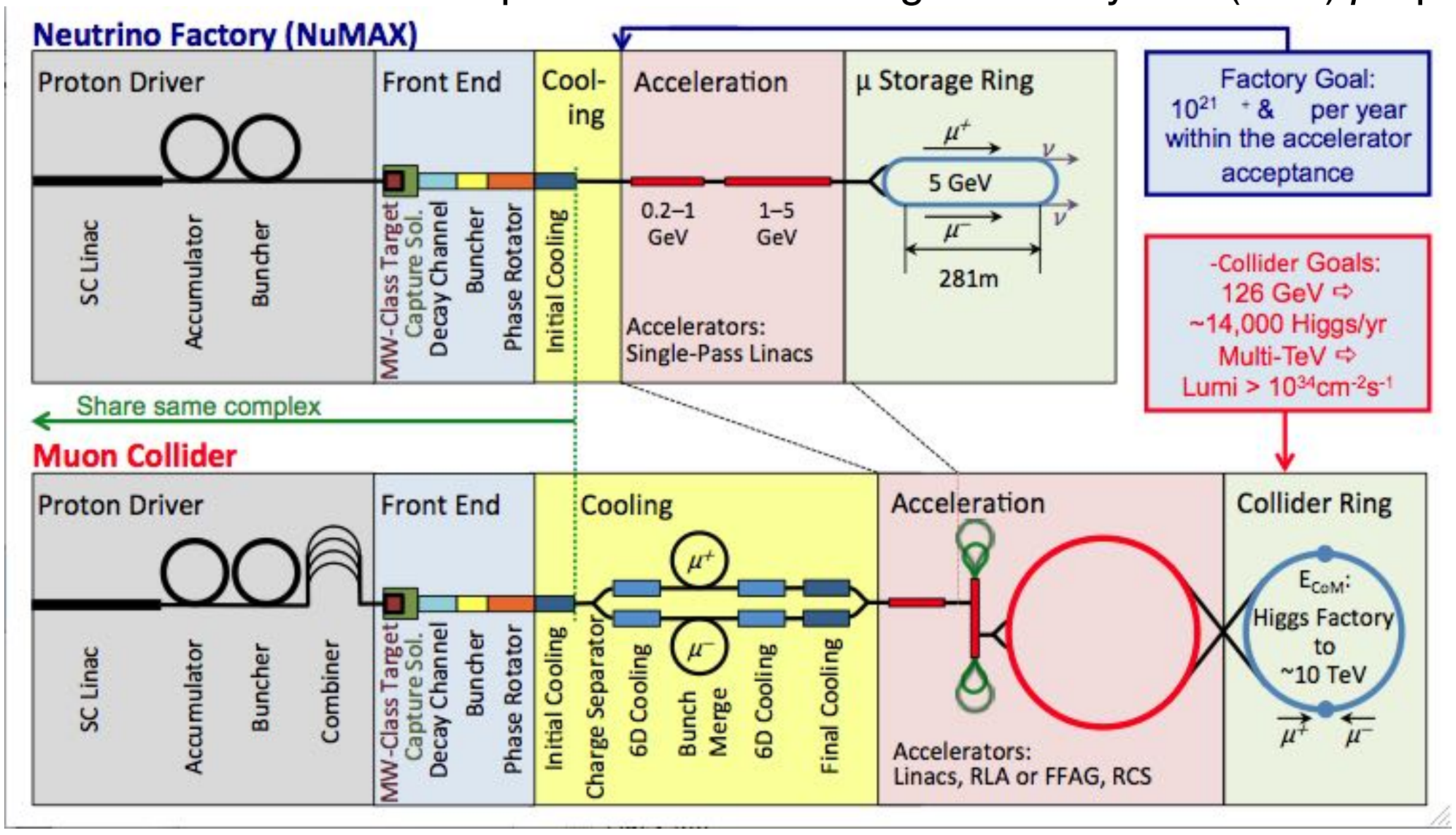

***Figure 12.-*** *The ESSνSB and the ESSμSB, are dedicated to the neutrino and muon programmes. High intensity bunches from the negative H- source are converted into protons, producing secondary particles (mostly π±). The π decay to μ's and the μ's are then captured, bunched, cooled and accelerated in a storage ring to produce an appropriate rate of high energy collisions.*

Two major experimental projects, the **ESSνSB** and the **ESSμSB,** dedicated respectively to the neutrino and muon programmes have been considered (Figure 12). In both instances high intensity bunches from the negative H- source are converted into protons, producing secondary particles (mostly $\pi^{\pm}$). The π decay to μ's and the μ's are then captured, bunched, cooled and accelerated in a storage ring to produce an appropriate rate of high energy collisions.

## *7.- The ESSνSB project.*

The ESSνSB project is described first and it is based on a *high intensity neutrino beam* (Figure 13). It has been widely discussed in several comprehensive papers [28]. This next-generation ESS neutrino observatory needs a huge, megaton-sized Water Cherenkov neutrino detector installed at ≈1000 m underground in a mine of a new laboratory. The best performance is obtained at a distance between 300 km to 600 km, depending on the future proton ESS energy in the interval from 2 GeV to 3.5 GeV. Two locations corresponding to existing mines have been proposed, one at 360 km and the other one at 540 km from Lund.

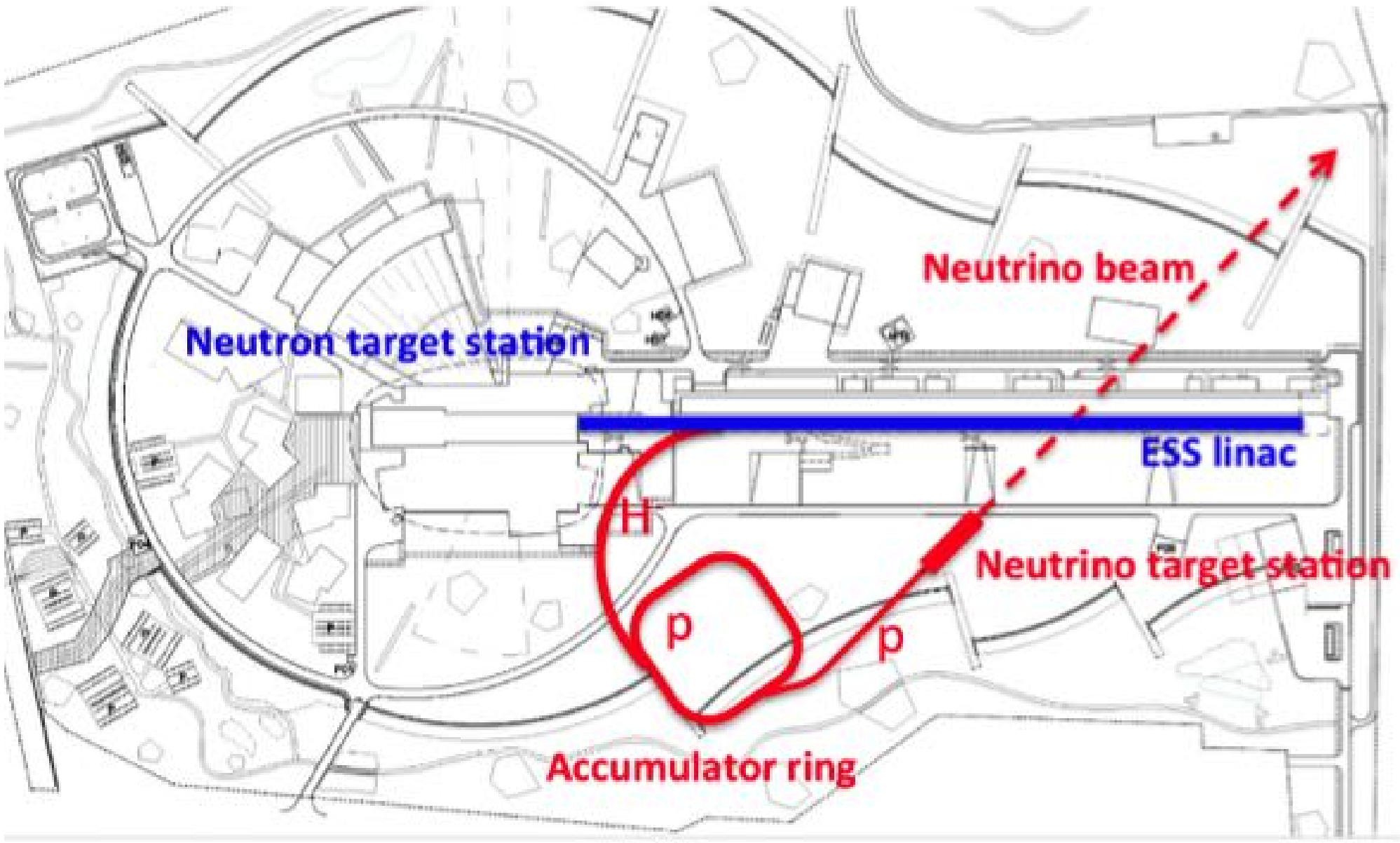

***Figure 13.-*** *Layout of the ESSvSB project. The repetition rate of the LINAC could be doubled from 14 Hz to 28 Hz with the additional 5 MW protons to new facilities. To this effect, ($H^-$) ions, instead of protons ($H^+$), are accelerated by the LINAC. The ($H^-$) ions will then be stripped of their two electrons. The here described ESSvSB (neutrino beam) and ESSμSB (muon collider) projects are both based to the intermediate proton accumulation with storage rings. Protons are accumulated in a ring structure and ejected to the neutrino target station.,*

Taking advantage of the low neutrino energy and the very high intensity of the ESS proton beam, the 2nd oscillation maximum (Figure 14) has been preferred [27].

A previous similar study for a neutrino facility had been described at CERN based on the proposed design of a Superconducting Proton Linac (SPL) Super Beam, in order to realize MEMPHYS [29] a large Water Cherenkov detector in the Fréjus tunnel at 130 km distance.

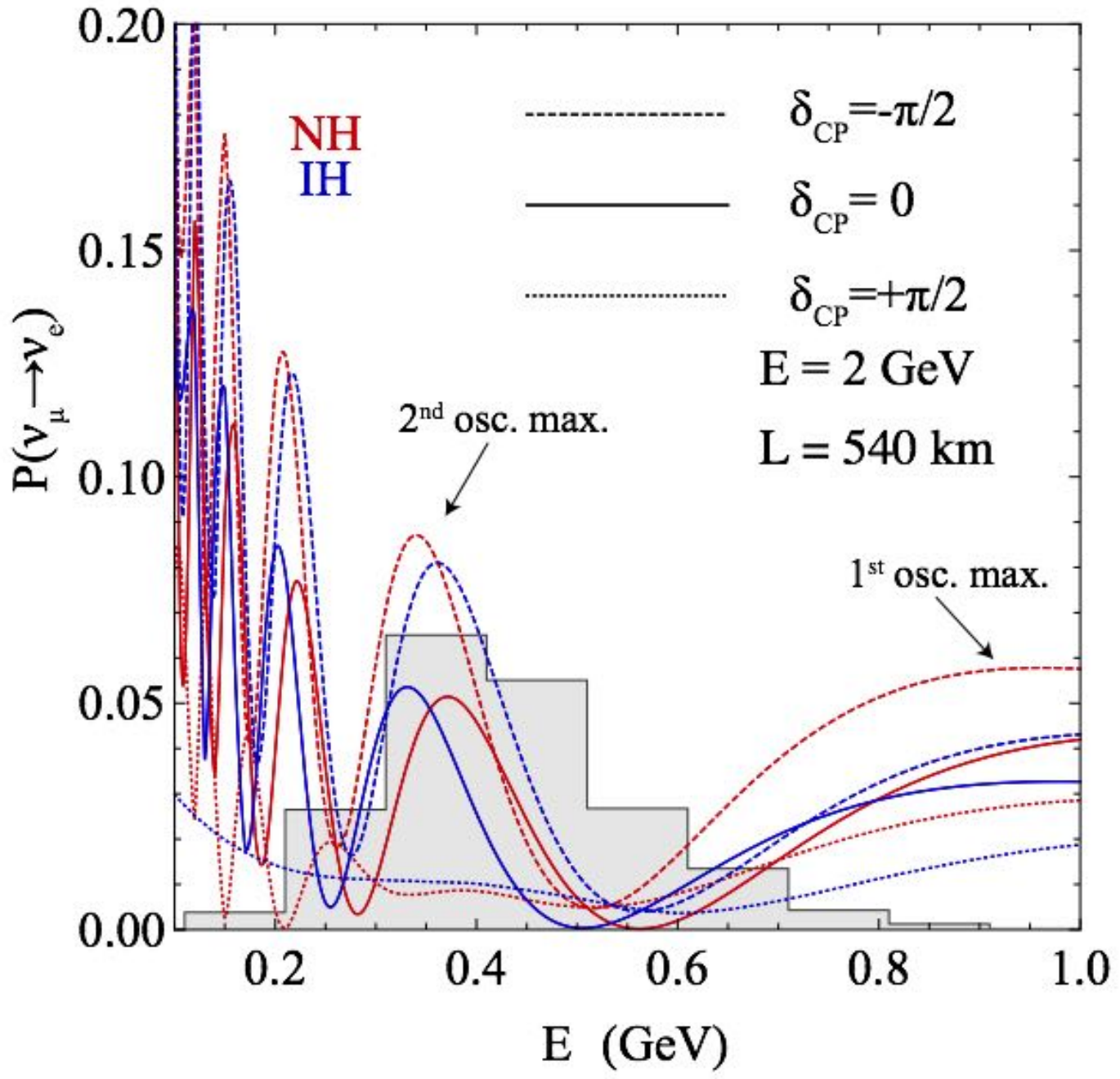

***Figure 14.-*** *In view of the low neutrino energy and the very high intensity of the ESS proton beam, the far more preferable 2nd oscillation maximum has been shown for neutrinos and for various indicative values of the* $\delta_{CP}$ *phase.*

The CERN-SPL [30] design (Figure 15) was made of a 160 MeV normal conducting sequence of four accelerating structures of Linac (RFQ, DTL, CCDTL and PIMS) followed by a 2.0 K superconducting LINAC producing a $H^-$ beam of about 5 GeV kinetic energy. In the full HP-HPL version of the CERN design the repetition rate is brought to 50 Hz, the power to 4 MWatt with a beam pulse length of 400 $\mu$s and to 1.0 x $10^{14}$ [$H^-$]/pulse.

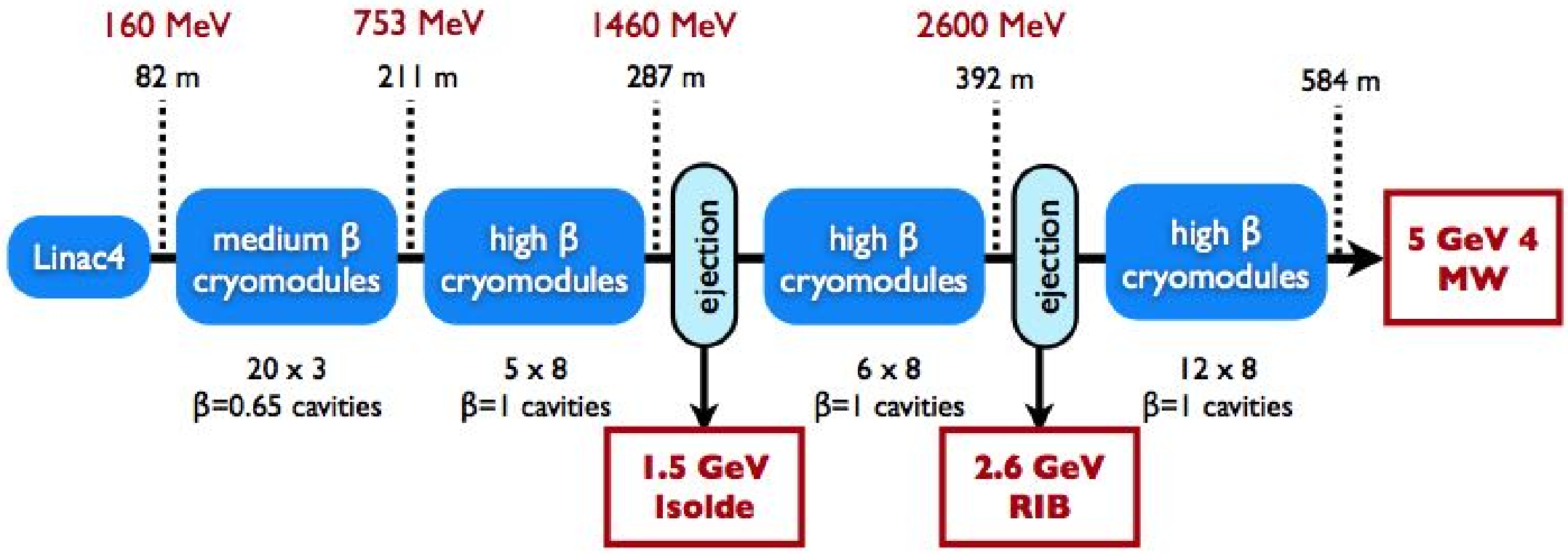

***Figure 15.-*** *The CERN-SPL design [27] is made of a 160 MeV normal conducting sequence of four accelerating structures of Linac (RFQ, DTL, CCDTL and PIMS) followed by a 2.0 K superconducting Linac producing a* $H^-$ *beam with about 5 GeV kinetic energy. In the full HP-HPL version of the CERN design the repetition rate is brought to 50 Hz, the power to 4 MWatt with a beam pulse length of 400* $\mu$*s and to 1.0 x* $10^{14}$ $H^-$*/pulse.*

In analogy with the ESS, the CERN-SPL design [31] was also based on the combination of two successive rings in succession, the accumulator and the compressor. The CERN rings were designed for an approximate circumference of 200 m (0.6 $\mu$s). The length of individual staggered proton pulses was 120 ns before compressor. The final one bunch accumulation was very challenging not only because of high bunch intensity but also its long accumulation of 1920 bunches.

A previous version with 4 MW beam power had also been described at CERN [32] with a 2.2 GeV proton LINAC operated at 75 Hz, recovering the RF cavities left over from the shutdown of LEP, followed by accumulator and compressor rings installed inside the previously existing (but very large) ISR tunnel (150 m radius). The duration of the pulses was 2.2 ms and the intensity 1.5 x $10^{14}$ $H^-$/pulse. The 40 and 80 MHz RF systems were chosen to capture, cool and phase-rotate pions and muons, downstream to the target. The lattices used the relatively large values of the max. betatron functions $\beta_t$ = 25 m because of the 1 km circumference of the ISR ring. The half width and the half height of vacuum pipe were 9 cm.

Following the decision taken in 2010 to base the high-luminosity LHC on an upgrade of the existing PSB and PS, the CERN program based on the proposed design of the high intensity neutrino beam with the Superconducting Proton Linac (SPL) has been cancelled (Figure 15).

The comprehensive analysis performed at CERN has however shown no evidence for instabilities. *In view of its close similarities, this represents also a valid criterion for the predictions for the ESS project.*

The future ESSvSB neutrino project is not without strong competitions. Several other next-generation neutrino detectors are also addressing similar subjects: fundamental properties of the neutrino like mass hierarchy, mixing angles, the CP phase, the low-energy neutrino astronomy with solar, atmospheric and supernova neutrinos and major improvements in the search for nucleon decay.

In the US [33], the Fermilab laboratory in Batavia is proceeding with a neutrino related improvement program [34], foreseen to initiate about in 2019. Proton Improvement Plan-II (PIP-II) is a new superconducting LINAC of 25 mA, 4 x $10^{12}$ p/p, 15 c/s and 400 MeV protons. Augmented by improvements to the existing Booster, Recycler, and Main Injector complex it will deliver at least 1 MWatt of protons from the Main Injector at energies between 60 and 120 GeV to neutrinos at the Sanford Underground Research Laboratory in Lead, South Dakota with a Liquid Argon detector (LAr-TPC) in excess of 40 kton.

Operated as an international mega-science project, the “Deep Underground Neutrino Experiment” (DUNE) is the future main experiment for neutrino science and proton decay studies in the US [35]. DUNE will consist of two simultaneous neutrino facilities placed in the US most intense neutrino beam. One detector facility (SBN) will record particle interactions near the source of the beam, at the Fermi National Accelerator Laboratory in Batavia, Illinois. A second, much larger detector will be installed more than one km underground at the Sanford Underground Research Laboratory in Lead, South Dakota — 1,300 km downstream of the source. The current design of the far detector is of four modules of instrumented liquid Argon LAr-TPC with fiducial volumes of 10 kilotons each, extending the originally developed technology invented in Italy for the ICARUS program. ICARUS has now been transported at Fermilab and it will initiate soon the SBN experiment with three LAr-TPC detectors at different distances with the Booster beam.

The first two DUNE 10 kiloton modules are expected to be complete by 2024, with the beam operational by 2026. The other modules are planned to be operational by 2027. The ground breaking for the excavation and construction at Sanford Lab has taken place on July 2017, with researchers hoping to apply for funding for the detectors in 2019.

Two prototype detectors each with about 1/15 of the final LAr-TPC single volumes, the proto-DUNE Single and Double Phase detectors are presently under construction at CERN.

China offers a comparative future alternative at the Jiangmen Underground Neutrino Observatory (JUNO) [36] in Kaiping, with an acrylic sphere filled with 20'000 tons of liquid scintillator in order to detect reactor antineutrinos to perform high precision neutrino oscillation measurements.

The detector will study neutrinos from two nearby groups of nuclear reactors, the most powerful power plants in the world, being built at around 50 km from the experiment. JUNO will require a 80 m high and 50 m diameter experimental hall located 700 m underground and will use ~20'000, (20′′diameter) and ~25'000 (3′′diameter) pm tubes covering approximately 80% of the surface of the detector to detect the scintillation light that is created when a neutrino hits a hydrogen atom. This project has been approved: ground breaking has begun in 2015 and operation for 2022.

In Japan, the Hyper-Kamiokande detector [23] is located in the Tochibora mine of the Kamioka Mining and Smelting Company, near Kamioka town in the Gifu Prefecture. The Hyper-K detector is an underground water Cherenkov detector with a total (fiducial) mass of 0.99 (0.56) million metric tons. Located 650 m underground in Kamioka, it consists of a tank of pure water some 74 m in diameter and 60 m tall, using 40,000 photomultiplier tubes that each have a diameter of 50 cm and producing collisions of neutrinos with water molecules. One of the main goals is the study of CP asymmetry in the lepton sector using accelerator neutrino and anti-neutrino beams from KEK (Tsukuba). Presently $1.24\times10^{14}$ protons per pulse at 30 GeV and an average beam power of 240 kWatt have been achieved. The design power will be brought to 750 kWatt in forthcoming years and conceptual studies on how to realize a 1-2 MWatt power are underway.

Another neutrino underground detector RENO-50 [36] in Korea will consist of 18,000 tons of ultra-low- radioactivity liquid scintillator and 15,000 high quantum efficiency 20" photomultiplier tubes, located at roughly 50 km away from the Hanbit nuclear power plant in Korea. Reactor neutrino experiments represent therefore a complementary alternative to accelerators.

Currently Daya Bay [37] in China is running. Double Chooz [38] in France and RENO [39] in Korea have been cancelled. Hyper-K and DUNE may eventually commence data taking after 2025. These cosmological observations will vastly advance our understanding of neutrino physics.

Hyper-K in Japan and DUNE in the US are the main next-generation neutrino experiments aiming to detect CP violation. While the US project is further along in terms of construction, the technology is much less established than that at Hyper-K, so physicists in Japan are confident – if the project is fully approved soon – that they can measure it first.

Therefore our main consideration will be focussed to the ESS$\mu$SB project

### 8.-. The ESSμSB project: the proton beam.

The ESSμSB proposal is based on the production, accumulation, cooling and acceleration of *intense muon beams for a future facility to study the Higgs related scalar sector* in optimal background conditions.

The requirements of production, accumulation and cooling of the intense muon beams to study the Higgs related scalar sector with the ESSμSB project are much more stringent than of the ones of the ESSvSB neutrino project since now it will be necessary to compress the protons hitting the production target to a duration of few nanoseconds.

In order to handle the *proton beam* from the ESS-LINAC, two small storage rings are required (Figure 16): the *"accumulator"* to collect the pulse from the LINAC and the *"compressor"* in order to reduce the duration of the protons pulses to a few ns before hitting the π-μ production target.

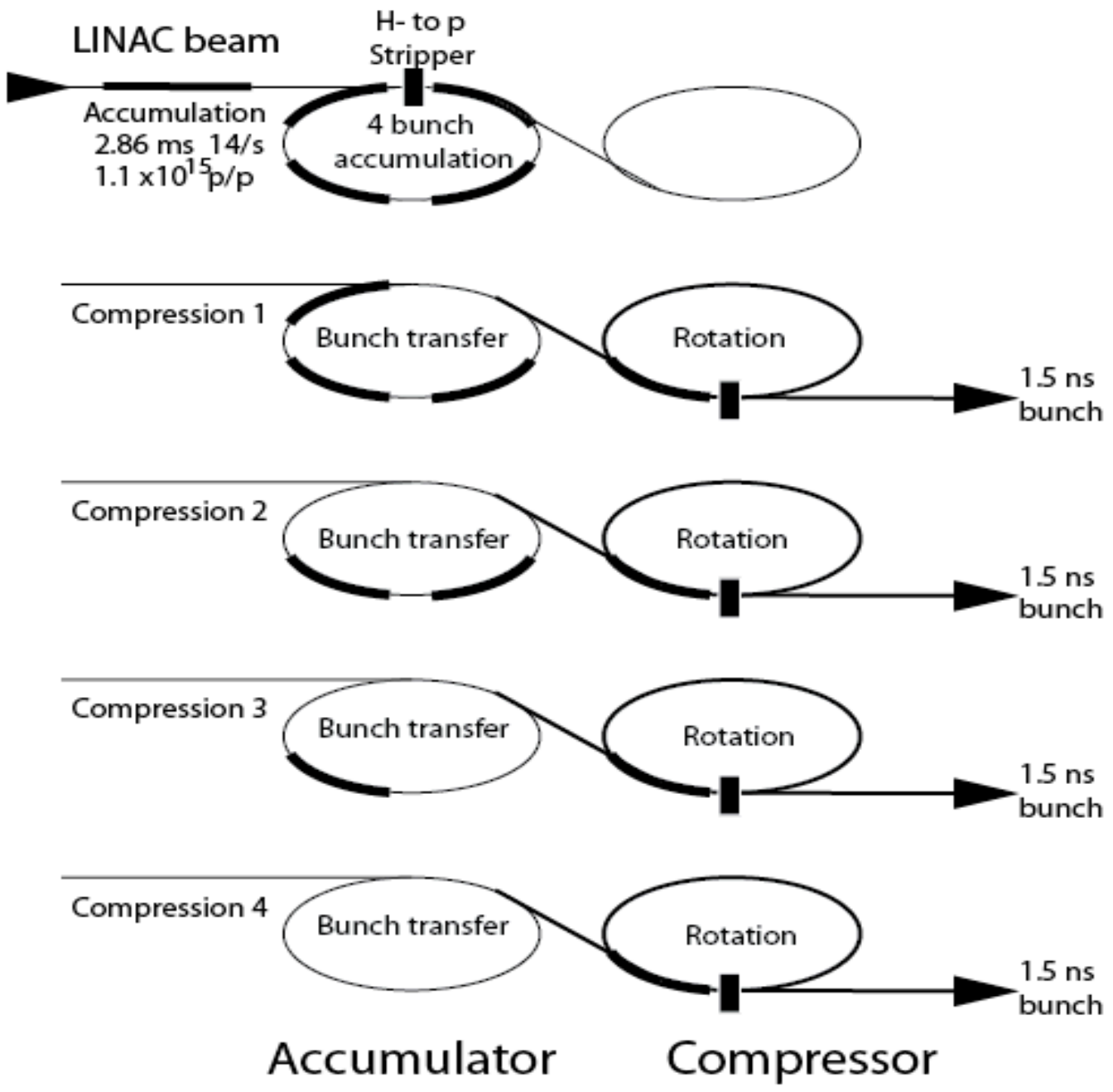

***Figure 16.-*** *The LINAC beam with 5 MWatt pulses is firstly accumulated in a thin stripper. A pair of proton rings subdivide the beam into four pulses extract the proton beam at 56 Hz, i.e. with bunches every 17.8 ms and 2.5 x $10^{14}$ p/bunch.*

In view of its relatively modest energy and the large number of protons/pulse when compared to other similar projects elsewhere, the ESS requires specific work in order to extrapolate many parameters, including longitudinal and transverse beam emittances, lattices for the accumulator and the compressor, RF, collimators, injection and extraction systems and so on.

In order to achieve an adequate final rate, the accumulator ring should be able to collect additional 5 MWatt pulses from the LINAC during the alternate 14 pulses/sec dedicated to ESS$\mu$SB (4% partial duty cycle). This system is much more complex than the one for the ESSvSB. It is therefore proposed to subdivide the extraction (Figure16) to 4 x 14 = 56 Hz, namely with four 1.25 MW separate batches each with ≈ 2.5 x $10^{14}$ protons, corresponding to one dedicated burst from the accumulator to the compressor every 17.8 ms.

Two alternative stripping options are described. The first option of *thin stripping carbon foils* [40], already chosen for instance for the SNS –US program, may be employed for a charge exchange injection. The melting point of carbon is 3925 K. The lifetime of a carbon stripping foil heating in vacuum decreases sharply when the foil temperature exceeds 2,000 K. For instance, the calculated lifetime of a typical 300 $\mu$g/cm$^2$ carbon foil against evaporation 56 Hz is 400 hrs at 2618 K, about 1 h at 3215 K and 60 seconds at 3810 K. The corresponding local heat densities deposited in the carbon foils are respectively 5.0, 9.6 and 19.1 Watt/mm$^2$, however dependent on the procedure followed.

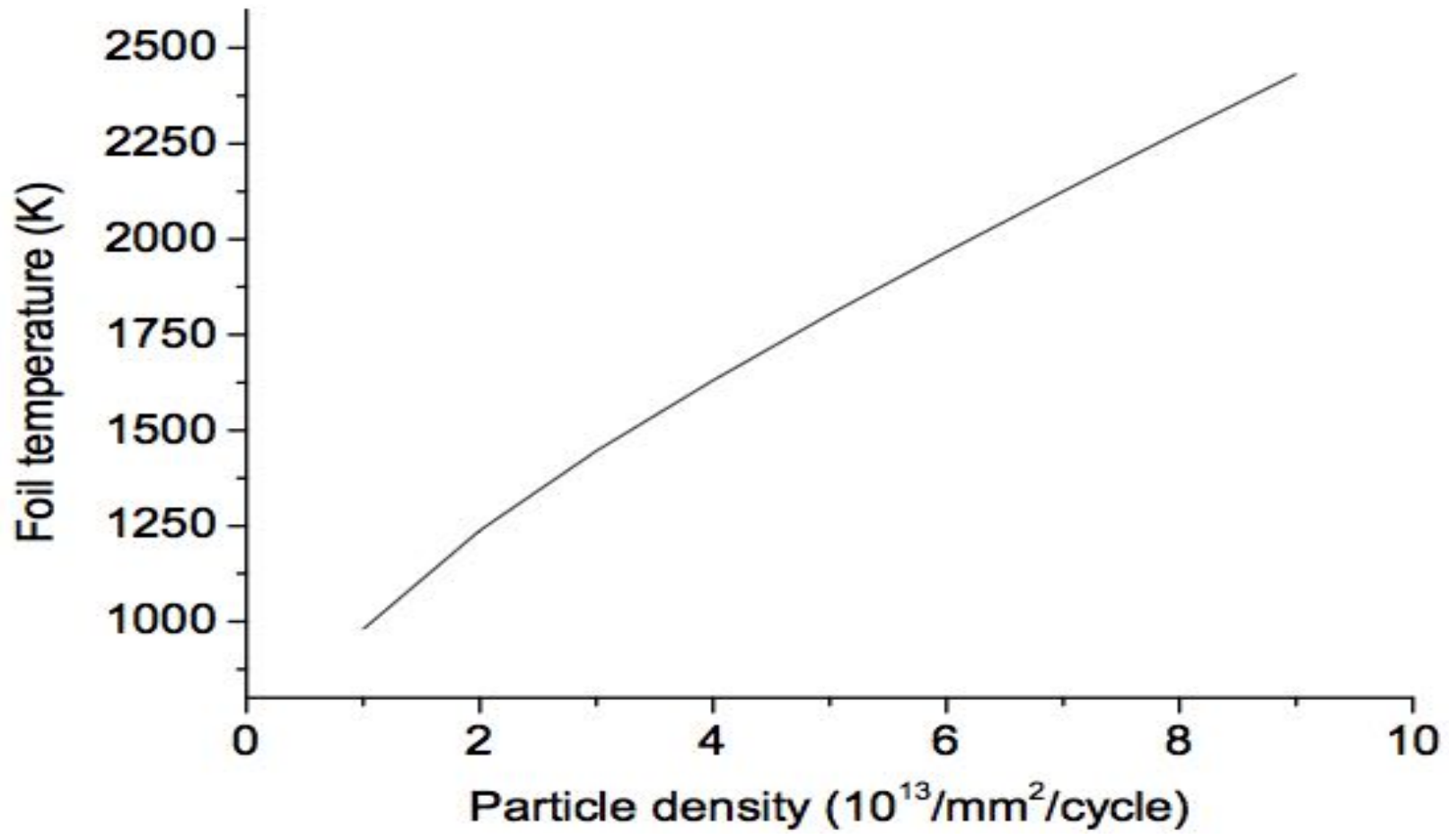

***Figure 17.-*** *Foil stripper temperature vs. particle density [22]. To compute the foil temperature, the carbon foil thickness of 300 $\mu$g/cm$^2$, an emissivity of 0.7 and a repetition rate of 50 Hz have been chosen. The feasibility of the injection has been studied using the tracking code ORBIT. In order to reduce the particle density on the foil, the two dimensional painting in both horizontal and vertical planes is used.*

An adequate converter could be a ≈ 300 $\mu$g/cm$^2$ thick carbon foil (Figure 17). The thin conversion foil will be necessarily traversed both by the incoming $H^-$ beam and by the recurrent traversals in the storage ring of the already accumulated protons. The Stephan-Boltzmann increase of the temperature during the pulse of the LINAC should be well below 2500 K, corresponding to an equivalent peak particle density of about $10^{14}$ (H)/mm$^2$ .

A more conservative criterion has been given here in order to limit the particle density to ≈ 6.0 x $10^{13}$ particles/mm$^2$/cycle and to keep the foil temperature well under 2,500 K with a safety margin of ≈ 500 K. At this rate, the "painting" of a full LINAC pulse with $10^{15}$ protons should require active foils for about 20 mm$^2$.

The multiple traversals of the protons during accumulation will also produce a so far not yet well estimated number of neutral ions $p + e^- \rightarrow H^0$, due to the background of foil stripping electrons neutralizing some of the protons from the foil. These generated neutral hydrogen atoms will be collinear with the direction of the proton beam and may require an efficient cooled dump.

With 300 $\mu$g carbon stripping foils, the first ESS protons must perform in the accumulator as many as 5958 traversals and the energy loss will be of 4.67 MeV lower than the last arriving protons, introducing an energy spread which must be compensated. The accumulated average energy loss radiated through the foils is 2.33 MeV representing slightly more than 1/1000 of the LINAC proton power (i.e. 3.1 kWatt/pulse).

The second option [41] is based on stripping with the help of *a LASER beam* (Figure 18).

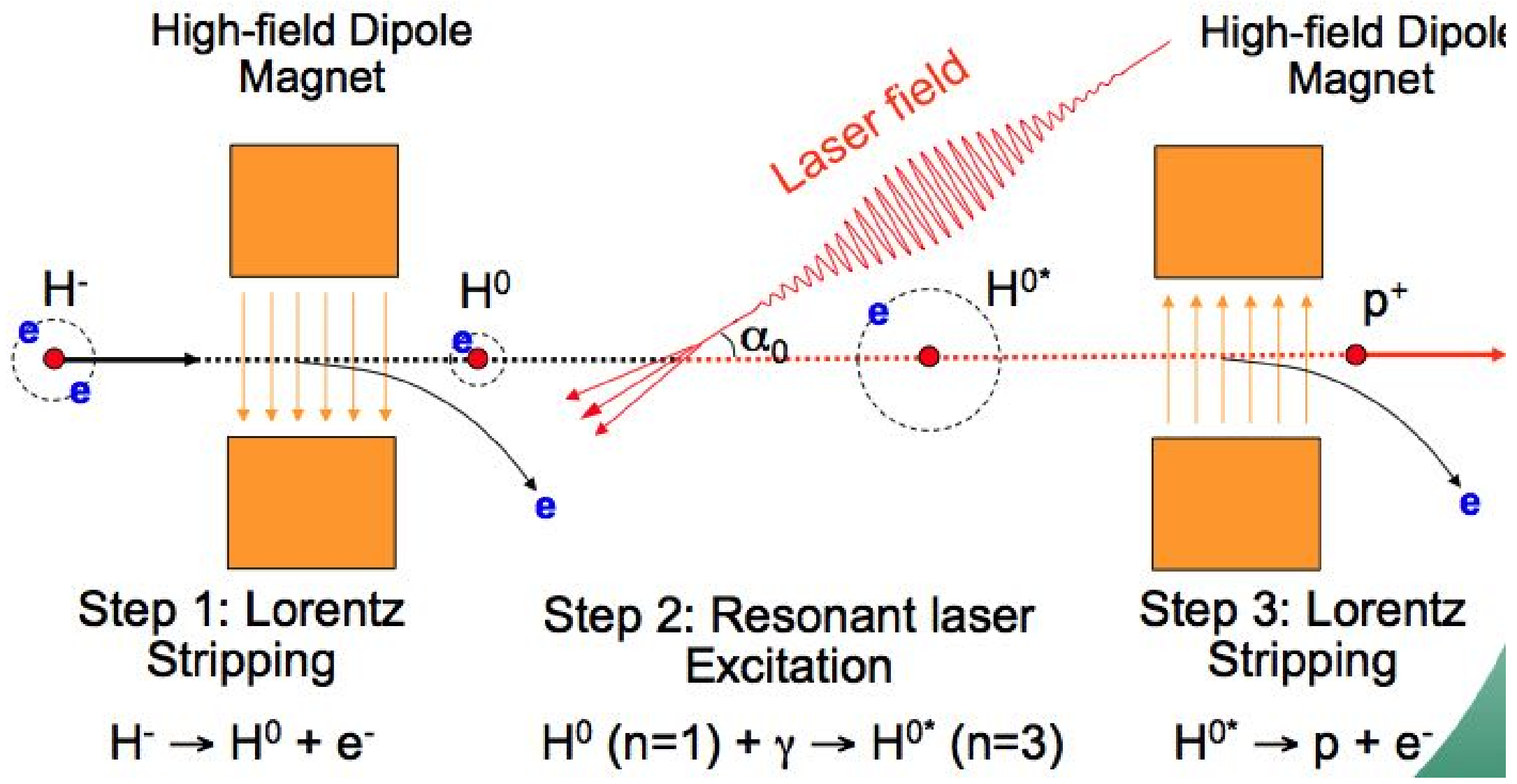

***Figure 18.-*** *General description of the LASER driven stripping method. Negative hydrogen atoms are stripped into neutral with a high magnetic field, excited with a resonant Laser to (n = 3) and finally stripped into protons again with a high magnetic field. An efficiency of 90% has been achieved*

The $H^-$ ion beam has two electrons, one tightly bound (binding energy of 13.6 eV), another loosely bound (binding energy of 0.75 eV). The beam from the ESS may strip off first electron with the help of a high field dipole magnet the through the process of Lorentz stripping ($H^-$ to $H^O$). The second inner electron is much too tightly bound to be easily stripped off. With a conventional magnet, a LASER is used to excite the electron to a higher quantum state ($H^O$ to $H^O$*) with a smaller binding energy. While in the excited state, the $H^O$* is passed through a second dipole magnet of comparable strength to the first, which strips off the remaining electron to produce a proton ($H^O$* to p). Due to the absence of material in the path of the beam, this technique does not suffer from the same performance limitations as foil-based injection and it may be scalable to high beam densities.

The idea of LASER stripping was presented over three decades ago [42]. The technique was limited by the Doppler broadening of the resonant excitation frequency of the LASER and of available LASER technology. This limitation was overcome in a 2006 proof of principle experiment that utilized a diverging LASER beam to introduce a frequency sweep in the rest frame of the LASER. The experiment demonstrated successful (> 90%) stripping of a 6 ns, 1 GeV H- beam using a 10 MWatt UV LASER [43].

Unfortunately, a direct scaling of this experiment to the full ESS duty factors implies unrealistically large LASER powers. To advance this concept one must find ways to reduce the LASER power requirement to a more achievable level. Therefore LASER stripping may not yet be today a proven technology and there is still R&D required before it may be fully implemented for such a high-intensity machine.

To conclude, mainly in view of the large power of the ESS, both the foil stripping and the LASER methods from incoming $H^-$ at the entrance of the accumulator *require future developments and a more conclusive proof.*

The *accumulator ring* (Figure 16) may have Q-values of 7.37 (H) and 5.77 (V), well above the proton transition energy. The ring period is 660 ns at $\beta_P = 0.9476$, to be filled by each dedicated pulse of the main ESS LINAC with 4 pulses each 120 ns long, equally distributed with four empty gaps of ≈ 45 ns for preparation to the subsequent extractions to the compressor. In order to fill the 1.1 x $10^{15}$ protons from the LINAC to the accumulator as many as 5958 turns are required, reduced eventually to 3273 turns for a 3.5 GeV operation. They are to be compared to the 1920 turns of the previously described CERN-SPL design [30] [31]. Because of the additional empty gaps, the total duration of each related LINAC pulse will have to be extended from 2.86 ms to 3.93 ms. An overall collection efficiency of ≈ 90% has been assumed.

In conclusion, four 120 ns long batches – each with 2.5 x $10^{14}$ protons – are simultaneously stored from the LINAC to the accumulator at the rate of 14 ev/s before being sequentially transferred as four batches to the compressor and where with the help of RF cavities a bunch rotation to a few ns width takes place.

The *compressor* ring may have Q-values of 4.21(H) and 2.74(V) and a RF voltage of 1.7 MVolt in order to ensure the final compression by bunch rotation. The compressor has large slippage factor and RF voltage for a fast phase rotation. The transverse emittance depends from the competing requirements of injection, foil heating, aperture, space charge and beam size on target, while the longitudinal parameters are set by phase rotation and required bunch length on target.

At the end of the compressor process, the Laslett [54] incoherent-space-charge transverse tune shift of the proton beam quantifies the severity of the effect. A reasonable maximum value of the space-charge tune shift at the center of a round Gaussian beam – in view of its short duration – is $\Delta\nu_{sc} \approx 0.4$, determining the acceptable transverse invariant emittances $\varepsilon_{V,H}$ and consequently the transverse size of the proton beam of the compressor before being ejected to the pion producing target:

$$\varepsilon_{V,H} = \frac{N r_p}{2\beta\gamma^2 \Delta\nu_{SC} b} \qquad \textbf{[1]}$$

Notice that there is no dependence of the radius of the ring. The so-called electromagnetic radius of the proton is $r_p$ =1.535 x $10^{-18}$ m, N = 2.5 x $10^{14}$, $\beta$ and $\gamma$ are the usual Lorentz kinematical factors at 2 GeV/c (i.e. $\beta\gamma^2$ = 9.285), b ≈ 2 ns / 120 ns = 1/60 is the bunching factor b of the initial 120 ns pulse rotated to 2 ns at the exit target point of the compressor.

For the value $\Delta\nu_{sc} \approx 0.4$, $\varepsilon_{V,H}$ = 3.1 x $10^{-3}$ ($\pi$) m rad corresponds to a r.m.s. radius $\sigma$ = sqrt($\varepsilon_{V,H}\,\beta^*/\gamma$) = 0.0315 x sqrt($\beta^*$) m. For $\beta^*$ = 0.3 m at the pion producing target, we find the large value $\sigma$ = 1.72 cm, not surprising in view of the relatively low proton energy and the large number of 2.5 x $10^{14}$ protons/bunch. It is expected to be able generate such a transverse proton beam dimensions and a specified value of the Laslett tune shift.

The behaviour of the compressor in units of three degrees is shown in Figure 19 for (a) the momentum distribution vs. the RF phase angle before and after the bunch rotation of the proton beam and (b) beam rotated to $\sigma$ = 1.98 ns r.m.s. RF phase angles.

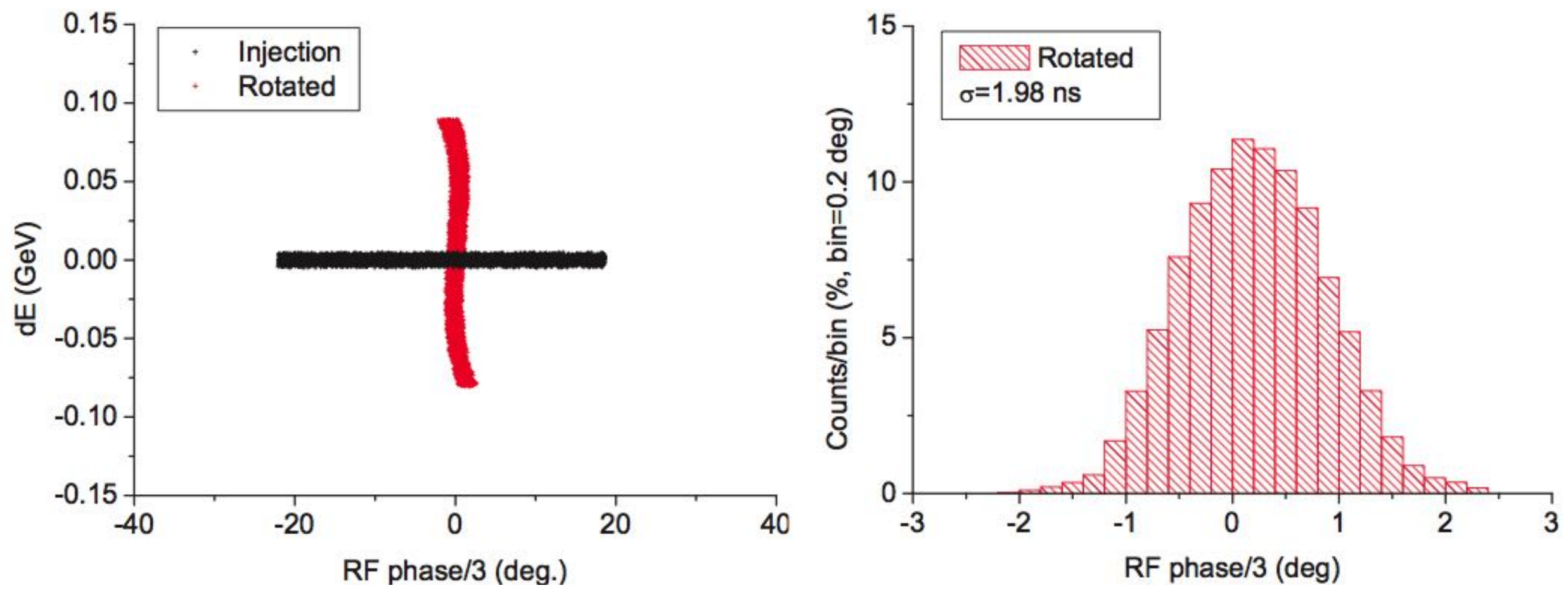

***Figure 19.-*** *(a) Momentum distribution vs the RF phase angle of the compressor before and after the bunch rotation of the proton beam and (b) beam rotated to s = 1.98 ns r.m.s. RF phase angles are given in units of three degrees.*

Bringing the energy to 3.5 GeV and with the same 5 MWatt power will be $\varepsilon_{V,H}$ = 0.90 x $10^{-3}$ ($\pi$) m rad and $\sigma$ = 7.54 mm at $\beta^*$ = 0.3 m. Therefore a higher proton energy could be beneficial for the realization of the appropriate target.

Alternatively, the final compression by bunch rotation could be extended from ≈ 2 ns r.m.s. to a larger value of for instance 6 ns with $\sigma$ = 5.7 mm at 2.0 GeV or $\sigma$ = 2.5 mm at 3.5 GeV, maintaining the chosen value $\Delta\nu_{sc}$ = 0.4 and extending the lengths of the subsequent transport systems.

### *9.- The ESSμSB project: the muon production.*

Pions produced from the *secondary production target* are quickly decaying into *μ*'s with a lifetime ct = 7.8 m, corresponding for instance to an average pion decay length of 8.94 m and $\beta$ = 0.753 at 160 MeV/c and of 16.77 m and $\beta$ = 0.907 at 300 MeV/c. Muons have instead ct = 659.1 m, corresponding for instance at 250 MeV/c to the much longer average decay length of 1.56 km. The pion energy spectrum at 2.0 GeV/c and 5.0 GeV/c are shown in Figure 20 for a heavy Z target.

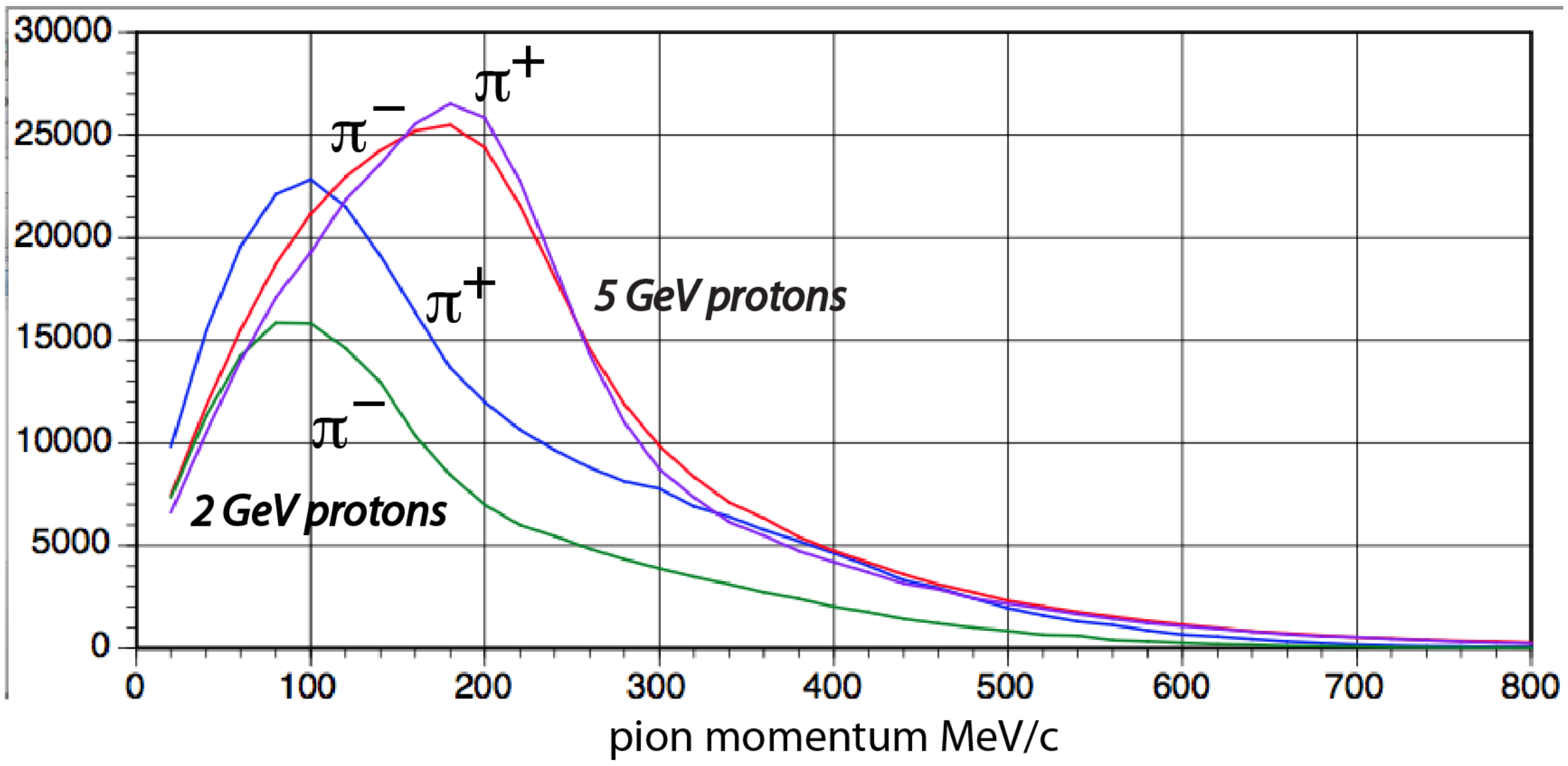

***Figure 20.-*** *Charged pion momentum spectra for 2 GeV and 5 GeV protons on a 30 cm long heavy Z (mercury) target.*

The production of secondary particles after the target will be performed in several phases to be described in the following paragraphs in more detail: a first phase with a *Front End* transport System (Figure 21 and Figure 22) of a substantial linear length with both polarities combined to compress with the help of RF cavities the momentum spread to about ±10%, followed in paragraph 10 (Figure 25) by a second phase with an extensive use of *Muon Cooling* and a strong 6D compression and accumulation of two rings, one for each sign, into single bunches. Finally in order to choose a ratio of the final transverse and longitudinal components as appropriate, a *Louvillian rotation* between transverse and longitudinal components may be further optimized — depending on the energy choice of the Collider ring — with a very high B ≈ 50 Tesla and sets of solenoids (Figure 25).

At the beginning of the Front End (Figure 22) secondary particles of both signs are sent to an axially symmetric target horn and a 30 cm long heavy Z (mercury) target immersed in a high field solenoid at a longitudinal $B_o$ = 20 T field. The proton beam is oriented with respect to the target with an angle of the order of ≈ 200 mrad in order to separate out the bulk of the surviving proton beam from the production of the secondaries. Such a large proton crossing angle requires an adequate target diameter, 0.2 rad x 30 cm = 6 cm, even without adding the contributions of $\sigma$ due to proton transverse dimensions. A target with a relatively large transverse size is therefore necessary.

The geometry of the horn follows closely the MERIT/CERN experiment [44], which has successfully injected a Hg-jet into a 15-T solenoid.

A GEANT4 simulation at the ESS has been used to estimate the charged pion productions of the 30 cm long heavy Z (mercury) target at the proton kinetic energy of 2.0 GeV . The 2.50 x $10^{14}$ protons/pulse at 56 Hz generate 6.72 x $10^{13}$ $\pi^+$/pulse and 4.15 x $10^{13}$ $\pi^-$/pulse for all angles, including also backwards directions to the proton beam direction. The calculations have been repeated also for 5.0 GeV and the same 5 MW proton power/pulse, giving 9.30 x $10^{13}$ $\pi^+$/pulse and 8.92 x $10^{13}$ $\pi^-$/pulse. As already pointed out (Figure 21), for a given proton power the pion yield is therefore not a very fast function of energy. If Liquid Mercury may not be allowed in Sweden, Titanium spheres may be an alternative.

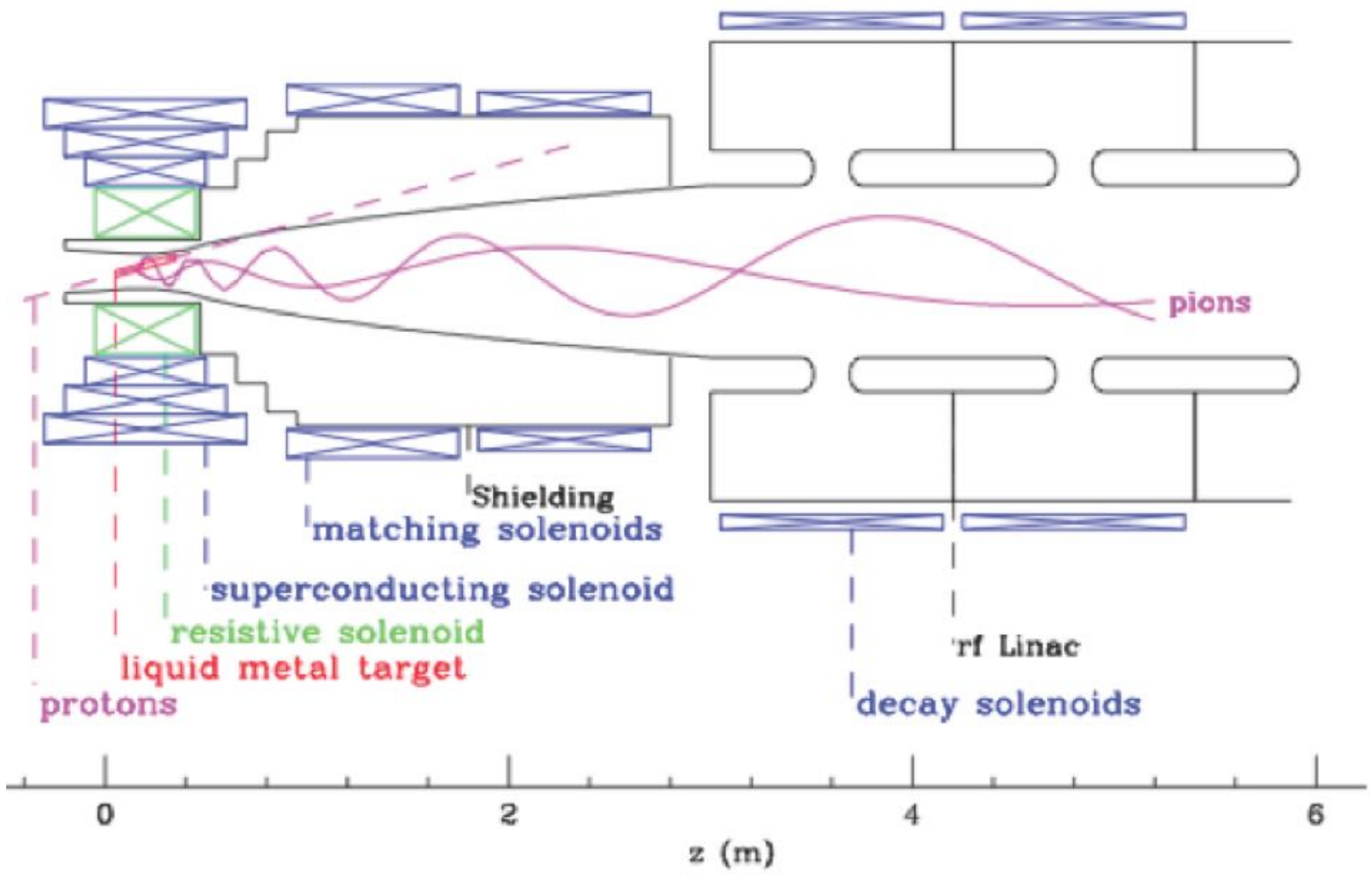

***Figure 21.-*** *Secondary particles of both signs are produced in a axially symmetric horn with a 30 cm target cm long heavy Z (mercury) target immersed in a high field solenoid at a longitudinal $B_o$ = 20 T field. The proton beam is oriented with respect to the target at an angle of 200 mrad to separate out he surviving protons from the secondaries. A purely transverse pre-cooling is initiated in a linear structure at a ≈ 2 T longitudinal magnetic field with alternate solenoids of an appropriately large diameter.*

Pion secondaries selected at the target with longitudinal momentum $p_L > 0$ oriented in the forward direction and with momenta between 50 MeV/c and 600 MeV/c are 2.97 x $10^{13}$ $\pi^+$/pulse and 1.91 x $10^{13}$ $\pi^-$/pulse. In the initial 20 Tesla field, the transverse momentum $p_T$ of pions produces a tight helix with a narrow magnetic radius equals to $p_T/60$ in MeV/c and cm.

The value of the magnetic field is progressively reduced in a linear longitudinal structure, to about 2.8 T (Figure 22). Following the Bush theorem, focusing is realized with secondaries in an axially symmetric solenoidal field. Secondaries of both signs are focused and they will be sign separated only at a later stage during the cooling process. By reducing the field, the rotational motion is converted into longitudinal motion, according to $p_\perp = p_\perp^{20T}\sqrt{B/B_o}$ and $p_T$ is reduced, with the longitudinal motion correspondingly incremented, since the total momentum is conserved in a magnetic field configuration. it is shown the momentum of positive pions near the horn position at 20 Tesla and after the helical evolution of the pions down to few Tesla.

After the horn we can conservatively assume an overall reduction of the muons to 80% due to various losses, arriving to 2.38 x $10^{13}$ $\mu^+$/pulse and 1.53 x $10^{13}$ $\mu^-$/pulse.

The detailed procedures of *drift and phase rotation* are followed according to the description of a complete scheme for a muon Collider given in COOL 2007 by the US team [45], but adapted to the ESS. This process is numbered from (#1) to (#9) of Figure 25.

Following Neuffer et al. [46] after the extraction from the solenoid, secondaries of both signs are momentum analysed with a magnetic structure called *"HFOFO Snake" or "chicane"* (Figure 26). The chicane is a pair of bent solenoids each about 6 m long and with a deflection angle of 15°. Higher momentum particles are not strongly deflected and are lost in or near the chicane and collimated shielded walls. Lower momentum particles are strongly focused by the solenoids and follow the chicane with little orbit distortion. After the decay into muons, some 30 m away, a 10 cm Beryllium absorber removes almost all of the remaining low energy protons and $e^+$ and $e^-$ electrons.

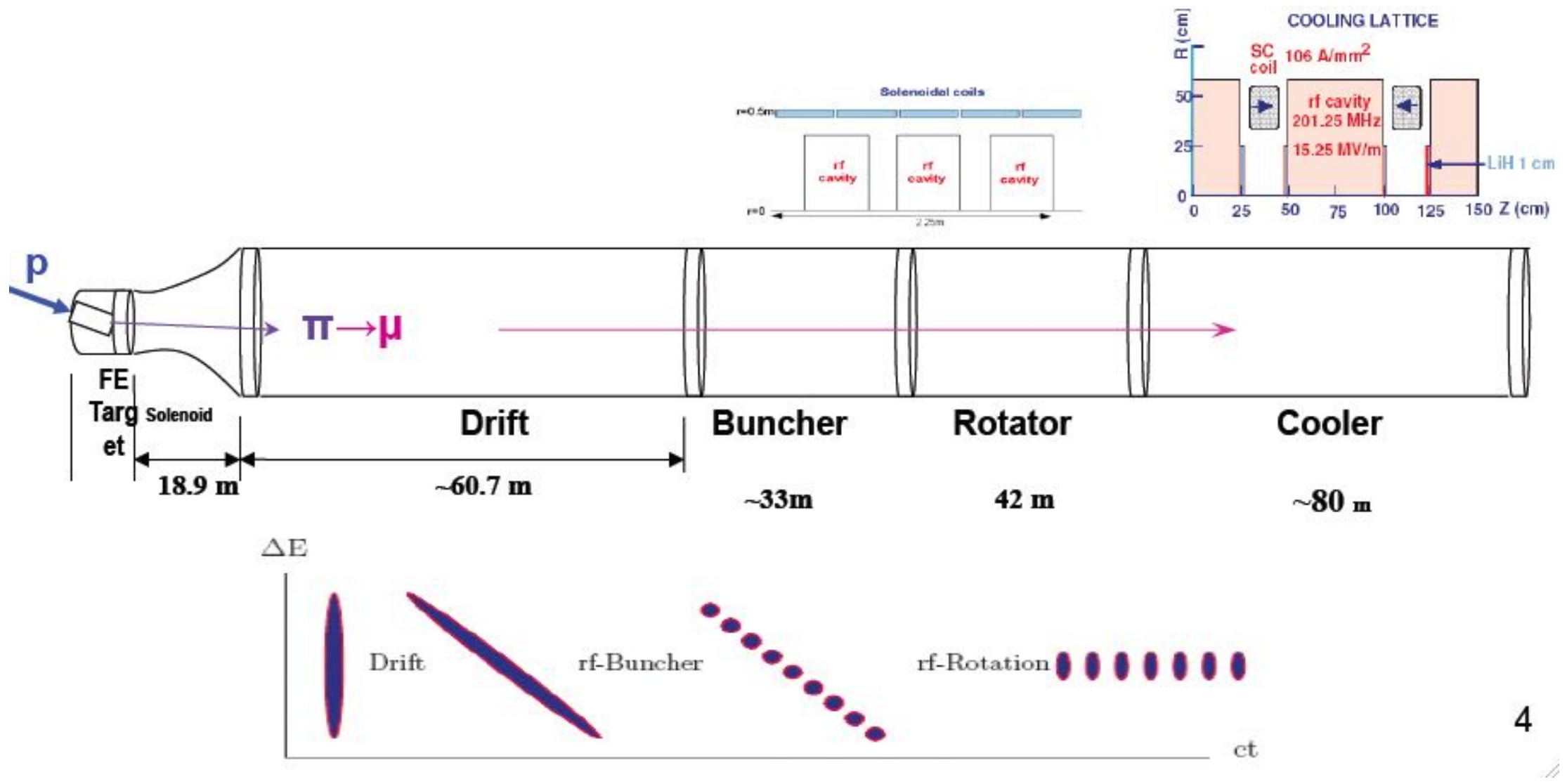

***Figure 22.-*** *The muon distribution with large energy spread and small bunch length stretches to an energy-position correlation in the Drift. The RF-Buncher forms the beam into a string of different energy bunches and the RF-Rotator moves bunches to equal energies, forming a string of bunches in which a trasverse Cooler is introduced in order to reduce the beam sizes to an acceptable value.*

After the target, muon bunches are followed by a drift space where a strong correlation develops between time and energy (Figure 25 ,#1),). The resulting momentum distribution for positive and negative muons is shown in Figure 21. At 2.0 Gev protons, in the muon interval between 100 MeV/c and 400 MeV/c there are about 70% of all events with 1.67 x $10^{13}$ $\mu^+$/pulse and 1.07 x $10^{13}$ $\mu^-$/pulse.

The momentum of the faster muons can be decreased and the slower muons increased. Two different methods may be used in order to provide a nearly non-distorting phase rotation, either (a) an appropriate length of Induction Linacs or (b) a chain of RF cavities with RF rotation.

Before entering in the cooling system, the buncher is followed by a rotator and a linear transverse pre-Cooler. A drift distance is inserted after the horn embedded in a over all longitudinal magnetic field of about 2 T. A RF driven buncher consists of many cavities with many different frequencies between 200 and 300 Mhz and a length of some 30 m. This is followed by a some 40 m long rotator with many different frequencies and cavities, compressing the muons to a series of bunches and to an average optimum cooling momentum of ≈ 250 MeV/c, near to the almost minimal dE/dx losses. Strings of both signs (Figure 23) are accumulated half-way at opposite values of the RF phase. The typical gradient of the cavities is 16 MV/m at the indicative frequencies, followed by 50 m long pre-Cooler.

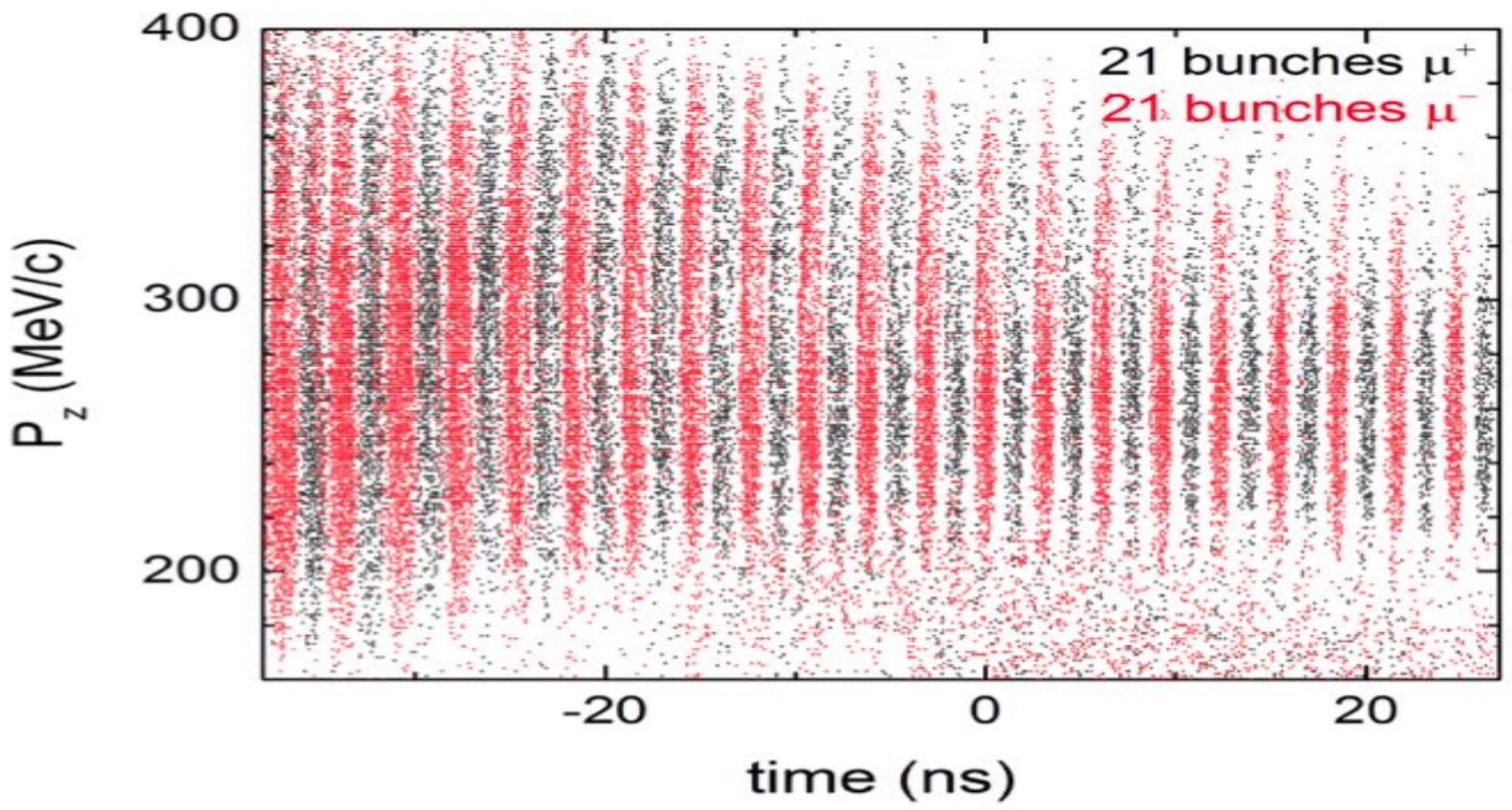

***Figure 23.-*** *Distributions of the positive and negative muons as a function of the longitudinal position, aligned in bunches, in position C at the end of the buncher. Strings of both signs are accumulated since half-way between each of the stable RF phases for one sign there is a stable phase for the opposite sign*

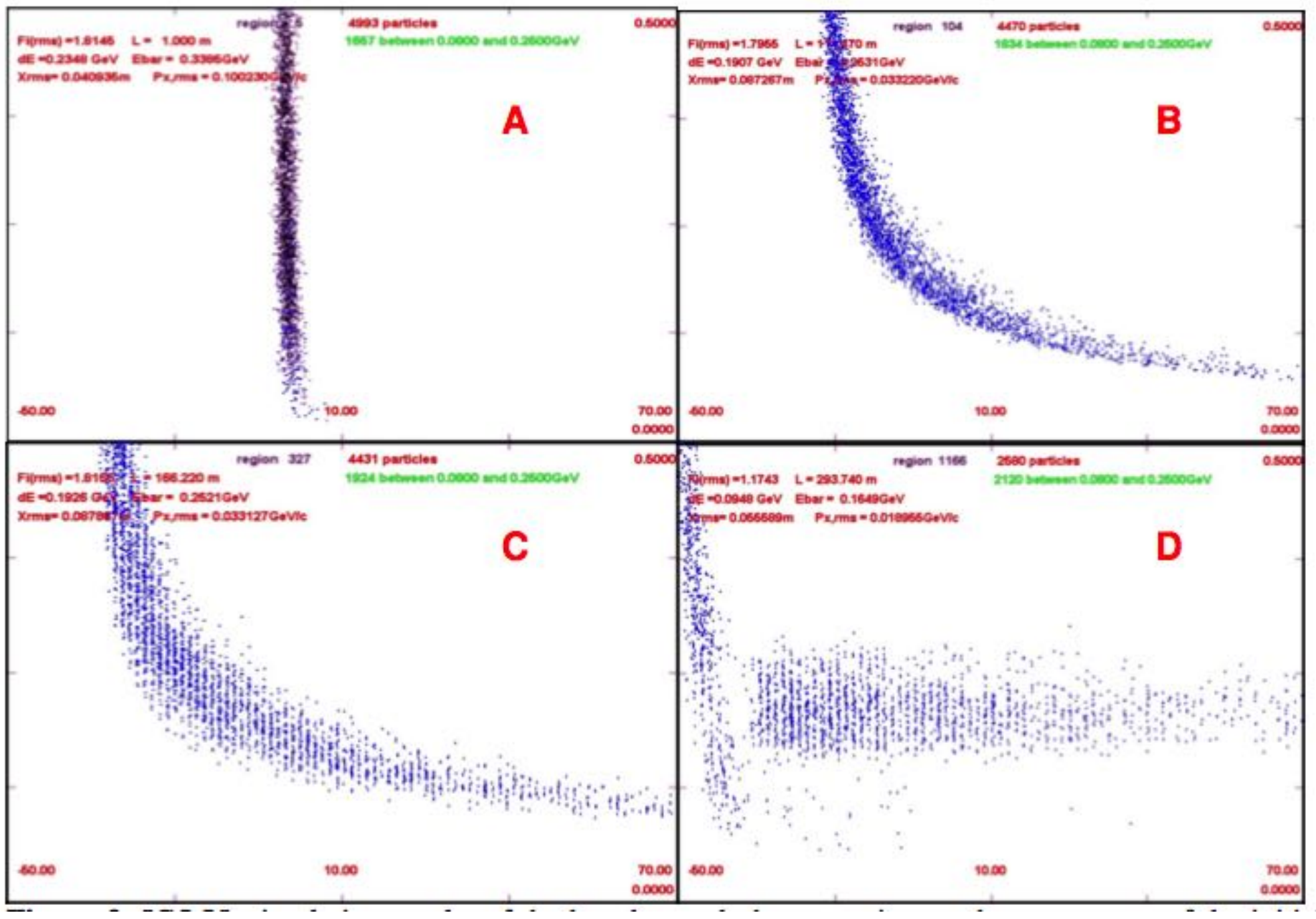

***Figure 24.-*** *ICOOL simulation results of the buncher and phase rotation. Each figure shows: A: π's and μ's as produced at the end of a 1.0m target. B: μ's at z=112m after a drift . C: μ's at z = 166m, end of the buncher. The beam has been formed into a string of ~200MHz bunches at different energies. D: after ϕ-δE rotation and ~80m of cooling; bunches are aligned into nearly equal energies, and transverse emittance has been reduced by a factor of ~2.4. In the plots is momentum (0 to 0.5 GeV/c) ve. longitudinal position with respect to reference particle (-50 to 70m)*

In conclusion, the *ESSμSB Front End* section ensures two main tasks a) the production and collection in excess of $10^{13}$ $\mu^{\pm}$/bunch and b) the muon momentum compression.

According to this proposal, the final r.m.s. energy spread after equalization is about 10 %.

## *10.- Ionization cooling.*

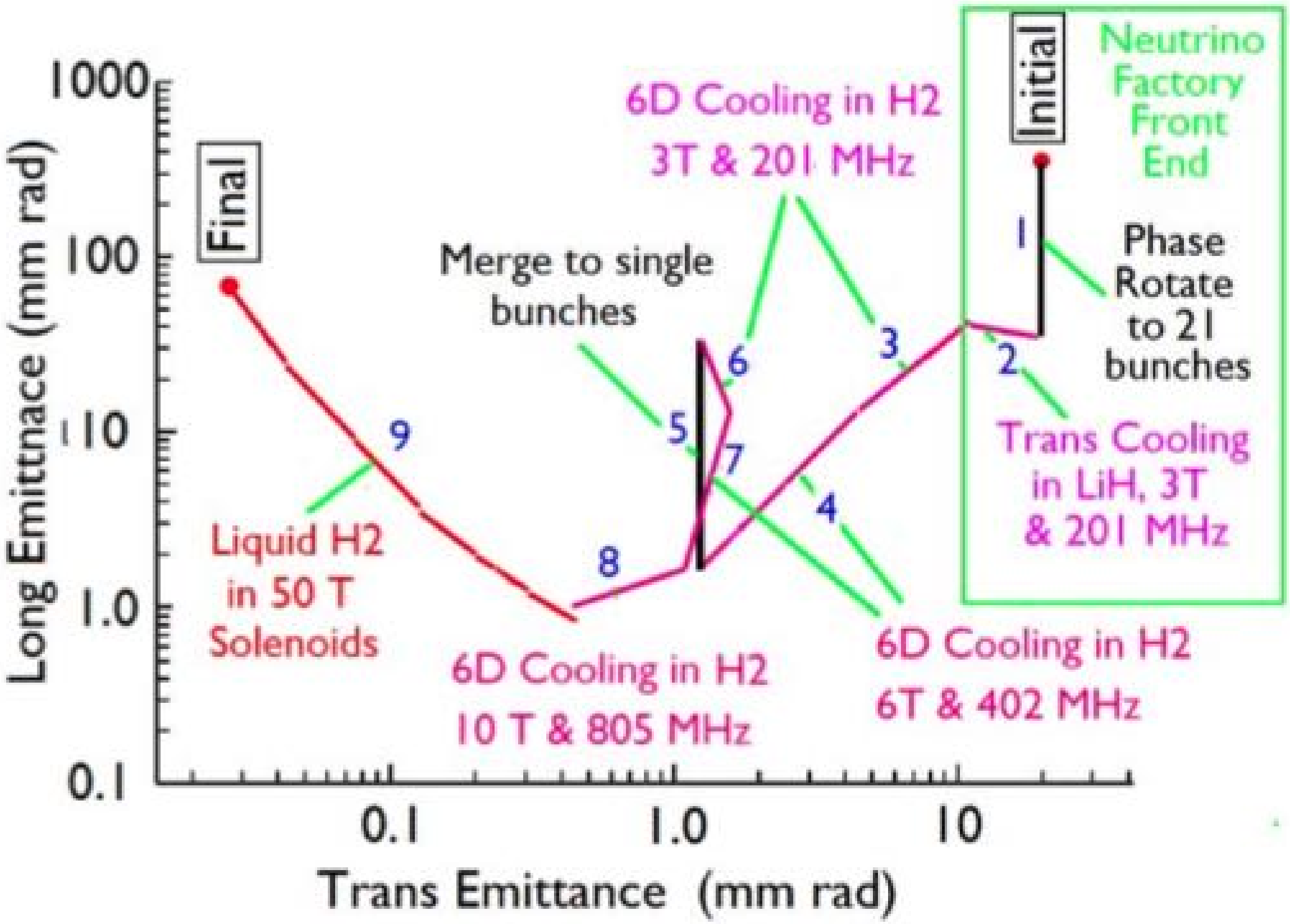

***Figure 25.-*** *Complete scheme for a muon Collider given in COOL 2007 by the US team [64], but adapted to the case of the ESS. This process is there numbered from, (#1) to (#9). Transverse emittances are 6D cooled in* $H_2$ *from 10 to 0.4 (π) mm rad and longitudinal emittance from 30 to 1.0 (π) mm rad.*

After the transport and momentum compression, *Muon cooling* may be initiated at the average momentum of ≈ 250 MeV/c. The muon beam emittance should be reduced with adequate cooling in order to lead to a useful luminosity for the muon Collider. The 6D compression needs to be performed with an overall reduction factor of the order of $10^5$. The muon lifetime may be however long enough to offer ionization cooling with an adequate number of collisions.

As well known, cooling has been already successfully used for the p-pbar Collider program, in which both stochastic and electron cooling have been extensively used. P-pbar Colliders have permitted the discoveries of the $W^{\pm}$ and $Z^{o}$ at CERN and of the Top quark at Fermilab.

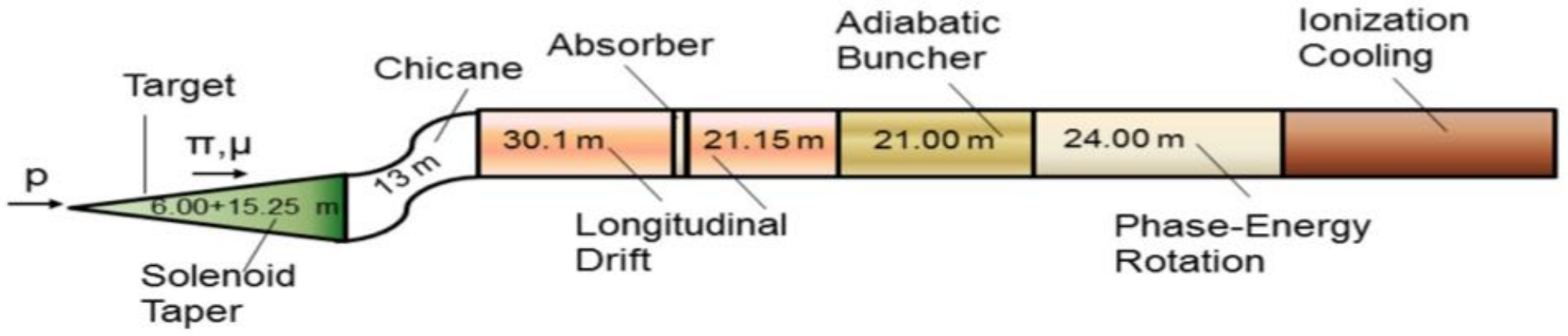

***Figure 26.-*** *After the extraction from the solenoid, secondaries of both signs may be momentum analysed with a magnetic "HFOFO Snake" or "chicane", orbit distortion.*

The muon does not have strong interactions. Its higher mass relative to the electron means that it can pass through matter without significant hadronic or electromagnetic showers. Thus, it is the perfect candidate for Ionization Cooling, in which muons lose energy by passing through a low-Z material with the longitudinal component restored by RF cavities. This technique allows to reduce in a very short time the angular spread of a beam of muons to the limit determined by multiple scattering. However Ionization Cooling, as presently envisioned, will demand large proton intensities in order to cool adequately the initially diffusely produced muon beams.

Key difference between the muon Collider (ESS$\mu$SB) and neutrino factory (ESSvSB) projects is that the muon Collider requires substantially more cooling and that the cooling should apply to the full six dimensional phase space compression of the beam.

At the end of the *ESS$\mu$SB Front End* section (Figure 22), the invariant emittances of the muon beam are assumed to be about $\varepsilon_{V,H}$ = 20 (π) mm rad and $\sigma_L$ = 30 (π) mm rad. For an average $< \beta >$ = 1 m at the cooling ring and $\gamma$ = 2.569 at 250 MeV/c, the r.m.s. radius is $\sigma_{V,H}$ = sqrt($\varepsilon_{V,H} < \beta >/\gamma$) = 8.82 cm, which is too large for a convenient ring geometry. An additional initial phase with a linear transverse pre-cooling is required in order to achieve more adequate dimensions of the ring.

A first purely transverse pre-cooling of several bunches is therefore already initiated in a *linear structure*, Figure 25 (#2), without separating muons according to the sign of their charges and at a ≈ 2 T longitudinal magnetic field with alternate solenoids of an appropriately large diameter. The purpose of this initial cooling is the one of reducing the very large dimensions of the beam before injecting it into a realistic ring structure. A linear system may accumulate strings of µ-bunches of both signs, since half-way between each of the stable RF phases for one sign of µ's there is a stable phase for the opposite sign (Figure 24). The linear muon cooling may be accomplished with cavities at 200 MHz, an average gradient of 16 MeV/m, 400 MWatt peak power for 16 x 80 = 1280 MVolt of total RF acceleration over 80 m. The longitudinal emittances for 250 MeV/c muons may be reduced by about factor six to $\varepsilon_{V,H}$ = 3 (π) mm rad and a r.m.s. radius $\sigma_{V,H}$ = 3.41 cm with $< \beta >$ = 1 m. These reductions in the transverse emittances facilitate the further subsequent injection in a cooling ring. During such pre-cooling, the longitudinal contribution — to be later sequentially recovered — will be growing because of the effect of the straggling.

The hydrogen driven final normalized transverse equilibrium emittances in the rings are expected to be $\sigma_{V,H}$ = 0.4 (π) mm rad, corresponding to the additional ring cooling with a reduction factor sqrt(3.0/0.4) = 2.74 in each of the r.m.s. transverse dimensions.

As already pointed out, initial ideas have been discussed by Budker, Skrinsky, Cline and others in the sixties. A more comprehensive analysis has been given f.i. by Neuffer in the early nineties.

Several ring alternate configurations are shown in Figure 27. A simulation for circular rather than helical geometry was chosen using realistic coil and RF geometries. Preliminary studies suggest that the resulting differences between the choice of the helical and the circular geometries will be negligible.

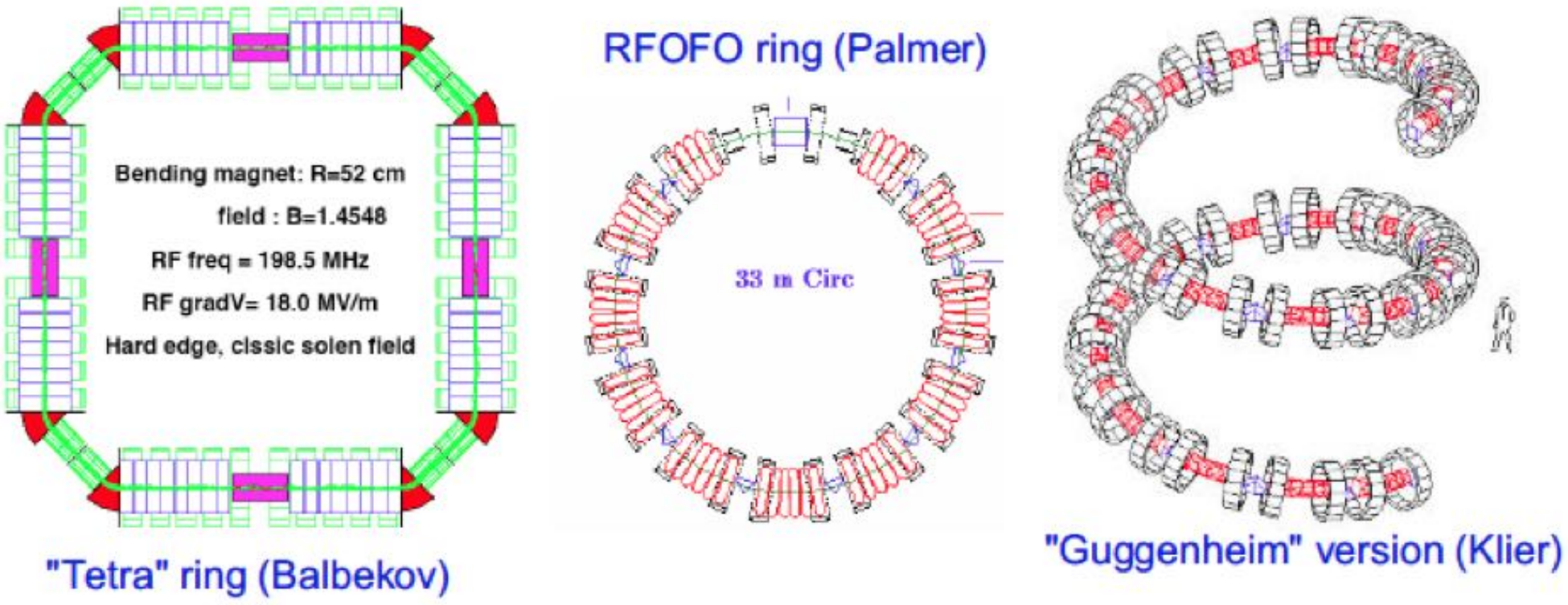

***Figure 27.-*** *Alternative ring configurations: Tetra ring (Balbekov), RFOFO ring (Palmer) and Guggenheim ring (Klier)*

In order to describe in more detail Ionization Cooling, we will look initially at the effects in the transverse plane (perpendicular to the beam trajectory) and then subsequently at longitudinal cooling effects (parallel to the beam trajectory).

However an integrated design including the full complexity of the beam transports, reacceleration and bunching, and including nonlinear beam dynamics coupled with the ionization interactions, are still missing. The later described much cheaper and simpler Initial Cooling Experiment must be the first step in order to verify experimentally the details of cooling.

The simplest case of muon cooling in the *transverse coordinates* is presented first [51]. The emittances $\varepsilon_V$ and $\varepsilon_H$ evolve along the length z of the process, whereby dE/dx losses are balanced by multiple scattering

$$\frac{d\varepsilon_{H,V}}{dz} = \frac{\varepsilon_{H,V}}{\beta^2 E}\frac{dE}{dz} + \frac{\beta^*(13.6\ MeV/c)^2}{2\beta^2 E m_\mu X_o} \to 0 \qquad \textbf{[2]}$$

with β* the value of the betatron function at the crossing point, $m_\mu$ and $\beta_\mu$ the values relative to the muon for the radiation length Xo in cm and dE/dz *the* ionization loss.in MeV/cm. The first term is the ionization loss and the second term is the multiple scattering. The Bethe Bloch parameters for several materials are shown in Table 1.

The cooling process will continue until the equilibrium transverse invariant emittances have been reached:

$$\varepsilon_{V,H} \to \frac{\beta^*(13.6\ MeV/c)^2}{2\beta_\mu m_\mu}\frac{1}{\left(X_o\, dE/dz\right)} \qquad \textbf{[3]}$$

The invariant, normalized equilibrium r.m.s. emittance ($\varepsilon_{V,H}$) with $\varepsilon = \sigma_x\sigma_\theta = sqrt\left(\langle x^2\rangle\langle\theta_X^2\rangle - \langle x\theta_X\rangle^2\right)$ and the corresponding actual transverse emittance ($\varepsilon_{V,H}$ /βγ) are function of the muon momentum. For liquid $H_2$ and cooling at β*= 10 cm, $\varepsilon_{V,H} \leq$ 370 mm mr at 250 MeV/c. In the case of LiH the equilibrium emittances are about a factor 2 larger.

*Table 1. Bethe Bloch parameters for several materials*

| Material | Z | A | Density ($g/cm^3$) | $(dE/dz)_{min}$ (MeV/cm) | $L_R$ (cm) | $L_R(dE/dz)_{min}$ (MeV) |
|---|---|---|---|---|---|---|
| $H_2$ (liquid) | 1 | 1 | 0.071 | 0.292 | 865 | 252.6 |
| Li | 3 | 7 | 0.534 | 0.848 | 155 | 130.8 |
| LiH | 3+ | 7.8 | 0.90 | 1.34 | 102 | 137 |
| Be | 4 | 9 | 1.848 | 2.98 | 35.3 | 105.2 |
| C | 6 | 12 | 2.265 | 4.032 | 18.8 | 75.8 |
| Cu | 29 | 63.5 | 8.96 | 12.90 | 1.43 | 18.45 |

Muons over the chosen momentum spectrum near the ionization minimum have a very small momentum dependence of the dE/dx loss and chromaticity has to be introduced with a wedge shaped “dE/dx foil”, in order to increase the ionization losses for faster particles.

Longitudinal cooling [51] is discussed next, related to producing processes balancing the dE/dx cooling. Balancing heating and cooling for a Gaussian distribution limit gives the expression:

$$\frac{d(\Delta E)^2}{dz} = -2(\Delta E)^2\left[f_A \frac{d}{dE}\left(\frac{dE_o}{ds}\right) + f_A \frac{dE}{ds}\left(\frac{d\delta}{dx}\right)\frac{\eta}{E\delta}\right] + \frac{d(\Delta E)^2_{straggling}}{dz} \quad \textbf{[4]}$$

where the first term is the intrinsic energy loss, the second is the wedge shaped absorber and the third the straggling contribution. There $dE/dz = f_A\, dE/ds$ where $f_A$ is the fraction of the transport length occupied by the absorber, which has an energy absorption coefficient $dE/ds$ ; $\eta$ is the chromatic dispersion at the absorber and $\delta$ and $d\delta/dx$ are the thickness and radial tilt of the absorber. The straggling (for A.Z) is given by

$$\frac{d(\Delta E)^2_{straggling}}{dz} = \frac{\pi (m_e c^2)^2 A(\gamma^2+1)}{4(Z+1)\ln(287)\alpha X_o/\sqrt{Z}} \quad \textbf{[5]}$$

where $X_o$ is the radiation length, $m_e c^2$ the electron mass and $\alpha$ the fine structure constant.

Reaching the equilibrium conditions, the above indicated straggling contribution is exactly balanced by the first two terms. Since this increases as $\gamma^2 + 1$, cooling at low energies is desired. For $p_\mu$ = 200 MeV/c, $T_\mu$ =120 MeV, $\beta$ = 0.88c, we find $L_{decay}$ = 1.2 km and $dE/dx$ = 4.6 MeV/gr cm$^2$.

Balancing heating and cooling terms and for longitudinal cooling of hydrogen [51] we find.

$$\frac{1}{(\Delta E)^2}\frac{d(\Delta E)^2}{dz} = \frac{466\ MeV(1-\gamma^2/12)}{E\gamma^2\beta^4 L_R}\left(1+\frac{1.1(MeV)^2\gamma^3\beta^4(\gamma^2+1)}{(1-\gamma^2/12)(\Delta E)^2}\right) \quad \textbf{[6]}$$

For the typical values $\gamma = 2$ and $\Delta E = 10$ MeV the first term is about twice the second and it can be approximated to the first term as

$$\frac{d(\Delta E)^2}{dz} \approx \frac{(\Delta E)^2}{L_R}$$

Longitudinal and transverse cooling are correlated according to the Robinson's law on sum of damping decrements [52]. Robinson analyzes determinant D of transfer matrix M for one revolution period of a particle in an accelerator $D = \exp(\Sigma\gamma_i)$ where $\gamma 1$, $\gamma 2$, and $\gamma 3$ are damping decrements of the longitudinal, radial, and vertical oscillations. The sum of the radiation damping decrements is equal to $-\left(4P_\gamma(t)/E\right)$, where $P\gamma_\gamma$ is the synchrotron radiation power and E the particle energy for any type of magnetic lattice. It has been remarked [53] that Robinson's proof of the sum rule is invalid as a general proof and it may not hold for non iso-magnetic rings.

Muon bunches of each sign can be cooled in a ring and later, after an intermediate stage, bunch rotated and accumulated to one bunch, which is extracted at the end of the cooling process (Figure 25). In order to ensure the length of the initial bunches and their RF compression in one bunch, the circumference of each cooling ring should be of the order of 30 to 45 m. The rings should permit a final accumulation of a pair of single muon bunches with a final r.m.s. length not longer than about 30 cm.

As already pointed out, the final, normalized emittances after hydrogen cooling at equilibrium at 250 MeV/c are expected to be of $\varepsilon_\perp = 0.4(\pi) mm\,rad$ and $\varepsilon_L = 1.0(\pi)\ mm\ rad$

Figure 5 has already shown a schematic of the main components of the system. A plot of the longitudinal and transverse emittances of the muons is described in Figure 25, #1) to #9) as they progress from production to the requirements for the Collider.

Stage #1) reduces the beam dimension to a size adequate for a linear phase rotation in 21 bunches at 201 MHz, followed by stage #2) before entering the ring with transverse cooling in liquid hydrogen at 3 Tesla and transverse emittances of ≈ 3 (π) mm rad.

Figure 25, #3) of the ring cools simultaneously in all 6 dimensions. The lattice uses ≈ 3 T solenoids for focus, weak dipoles (generated by tilting the solenoids) to generate bending and dispersion, wedge shaped liquid hydrogen filled absorbers where the cooling takes place and 201 MHz RF to replenish the energy lost in the absorbers, followed by a second phase at 402 MHz RF f.i. Figure 25, #4).

Muons are then captured into two single bunches (Figure 25, #5) Capturing the initial muon phase space directly into single bunches does require low frequency RF, and thus low gradients, resulting in slow initial cooling. It is therefore advantageous to capture initially into multiple bunches at 201 MHz and recombine them into a single bunch only after cooling.

This recombination (Figure 25, #5) is done in two stages. The individual bunches are phase rotated to fill the spaces between bunches and lower their energy spread; followed by 5 MHz RF, interspersed along a long drift in order to phase rotate the train into a single bunch that can be captured using 201 MHz.

The design and simulation of a system with the addition of a low frequency RF is separated from a drift in a wiggler system with negative momentum compaction. After bunch merging, the longitudinal emittance of the single bunches can be further reduced (Figure 25, #6 and #7). One more stage of 6D cooling has been designed (Figure 25, #8) with the help of 10 T magnets, hydrogen wedge absorbers, and a 805 MHz RF.

At this point the normalized emittances at equilibrium with hydrogen cooling are $\varepsilon_{\perp} = 0.4(\pi) mm\,rad$ and $\varepsilon_{L} = 1.0(\pi)\ mm\ rad$, for instance appropriate for the case of the s-channel resonance at the √s =125.5 GeV Higgs mass and to observe the many $H^{0}$ decay modes with the remarkable energy resolution R = 0.003 % .

However in order to arrive to an acceptable event rate in excess of L > $10^{34}$ $cm^{-2}$ $s^{-1}$ for much higher energies in the domain of several TeV (Figure 25, #9) the required transverse emittance must be further reduced, at the cost of a Louvillian increase of the longitudinal emittance, This can be obtained in liquid hydrogen with strong solenoids. The results of ICOOL simulation of cooling have been given with seven 50 T solenoids (Figure 25, #9).

Most RF cavities associated with particle accelerators normally operate as closely to vacuum as possible to avoid electrical breakdowns. This is done so that electrons or ions that are accelerated by the high voltages in the RF cavity rarely encounter atoms of the residual gas inhibiting the avalanche process of breakdown [50].

A very attractive development of the cooling channel uses instead a series of high-gradient RF cavities filled with dense hydrogen gas in a magnetic channel composed of solenoidal fields [60]. High-pressure gases in RF cavities facilitate large gradients by suppressing high-voltage breakdown by virtue of Paschen's law. Pressurized RF cavities have been designed to explore the use of hydrogen and helium gases up to more than 100 atm. pressure and at temperature down to 77 K. It has been demonstrated that for instance a 800 MHz cell filled with cold, pressurized $H_2$ gas has reliably achieved 80 MVolt/m. Hydrogen gas at 50 atm and 77 K (0.142 of liquid density) may support gradients of up to 330 MVolt/m. The energy loss, RF regeneration, emittance exchange and transverse cooling occur all simultaneously.

It is useful to define a Merit Factor M = *(Initial 6D)/(final 6D) x transmission*. Several practical examples (Tetra ring (Balbekov), RFOFO ring (Palmer) and Guggenheim ring (Klier)) have been computed in the literature still using vacuum operation however with a much smaller 6D Merit Factor. The use of a compressed gas is a new development that should greatly improve the performance and the Merit Factor M in order to ultimately attain the required reduction factor.

In a first computed example [47], the transverse cooling with a $H_2$ absorber is alternated with emittance exchange in a Lithium Hydride wedge and separate straight sections. The hydrogen absorber is at the center of a 6.67 m straight solenoid focused channel with a 5.1 T field at the absorber. A 200 MHz RF is placed within the solenoid, on either side of the absorber. The circumference of the small ring is 37 m. There is no prevision for injection or extraction of the beam. Dispersion in the following shorter straight is generated by a 45 degree dipole bend. Cooling is observed in all dimensions, with relatively good transmission (48%). The Merit Factor is 38.

A second example [48] is based on the so called *"RFOFO" Cooling Ring.* Focusing is provided by pairs of solenoids with alternating field direction*s* ("FOFO" since "*both focus"* and *"R for 'reverse"). A*ll cells are identical and both transverse and longitudinal cooling are obtained in the same $H_2$ wedge absorbers. A bending field is generated by inclining the focus*sing* coils by alternate 2.6 degree angles and a 200 MHz RF with a gradient of 12 MVolt/m is placed inside the coils. The circumference of the ring is 33 m, including a location for the beam injection and extraction. The three emittances and transmission are plotted on a log scale versus distance. The 6D Merit Factor is 162 as a first approximation of the ring structure (Figure 28).

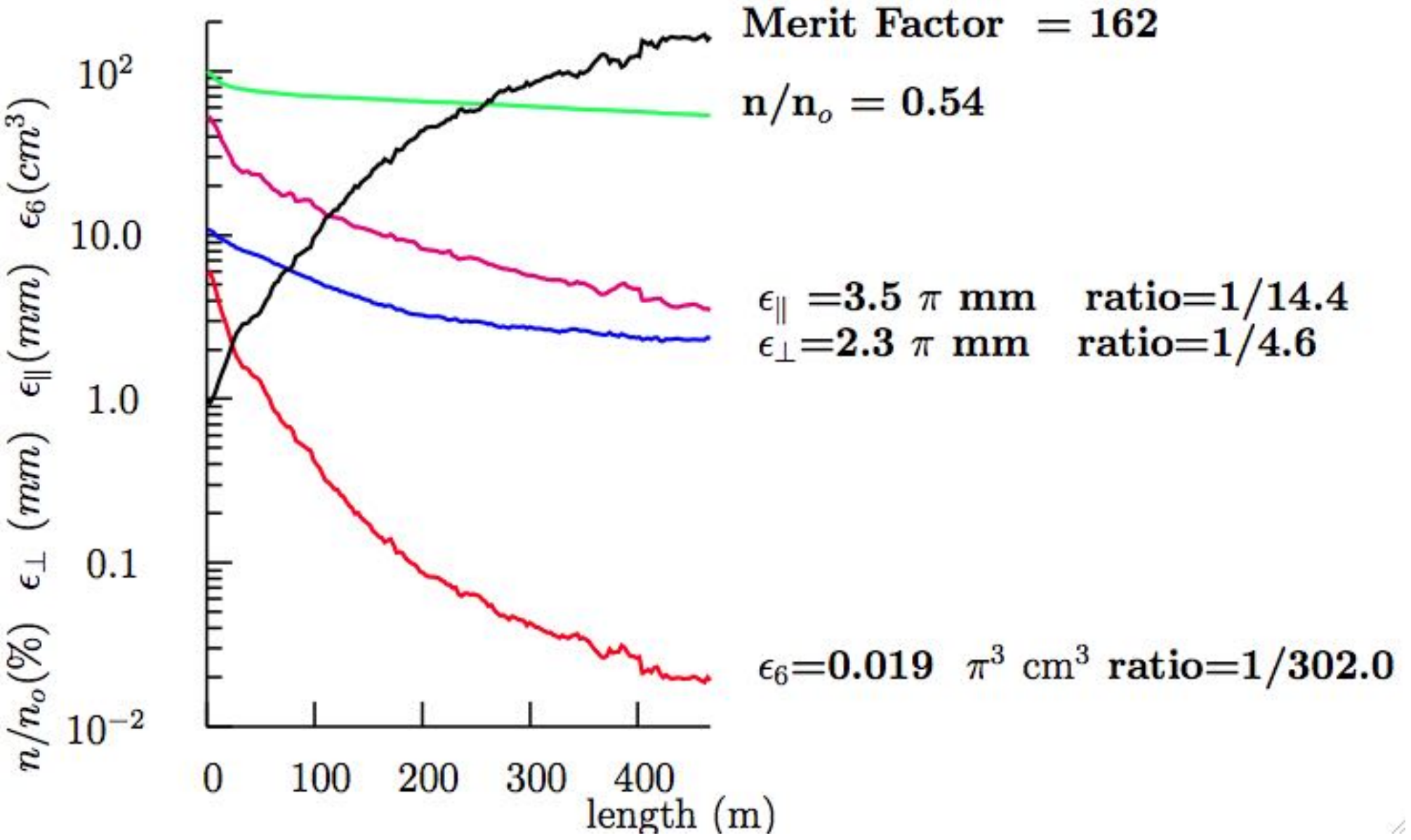

***Figure 28.-*** *Example [48] of a "RFOFO" Cooling Ring ("FOFO" since "both focus" and "R for 'reverse"). The three emittances and transmission are plotted on a log scale versus distance. Focusing is provided by pairs of solenoids with alternating field directions. All cells are identical and both transverse and longitudinal cooling are obtained in the same $H_2$ wedge absorbers. A bending field is generated by inclining the focussing coils by alternate 2.6 degree angles and a 200 MHz RF with a gradient of 12 MVolt/m is placed inside the coils. The circumference includies a location for the beam injection and extraction. The 6D Merit Factor is 162 .*

A third feasibility study has been recently described [49]. The four-sided ring has four 90° arcs with 8 dipoles separated by solenoids. In order to reduce the horizontal and vertical coupling, the fields of successive solenoids are again in alternate direction. Arcs are achromatic both horizontally and vertically. The dispersion is zero in the straight sections between the arcs. For the Injection/extraction very elaborate kickers are used in a straight section and a superconducting flux pipe is used for the injected beam. Liquid hydrogen ($LH_2$) wedge coolers will be inserted into a region with low b and high dispersion. Each $LH_2$ wedge absorber has a length of 19.5 cm, an energy loss rate of 0.303 MeV/cm, at an angle of 23 degrees. Four 201.25 MHz RF accelerating cavities are placed in each super-period. Its accelerating gradient is 15 MVolt/m and RF phase is 30 deg. The RF cavities will restore the energy of the muon beam that is lost in the $LH_2$ absorbers. The compact four-sided ring has a positive dispersion, which means that higher-momentum particles have longer paths around the ring, and lose more energy per turn than low energy particles. Consequently cooling takes place both in the longitudinal and the transverse dimensions. The combined horizontal, vertical and longitudinal invariant emittances of the circulating muon beam, decrease by the factors (5.34, 3.91, 2.56) during 15 revolutions with a transmission efficiency of 65.8 % and a total Merit Factor of 11.8.

These several alternatives of the muon cooling have indicative positive evidence for cooling processes, that however are much smaller than the required merit factor for the presently needed full six dimensional phase space compression. They are initial alternatives for a far more comprehensive future realization of the muon cooling process.

The Initial emitances after phase rotation and before initiating the cooling are $\varepsilon_{V,H}$ = 20 (π) mm rad and $\varepsilon_L$ = 30 (π) mm rad. And a r.m.s. energy spread is about 10 %. After cooling (Figure 25), the ultimate normalized hydrogen driven emittances at equilibrium at 250 MeV/c are expected to be $\varepsilon_{V,H}$ = 0.4 (π) mm rad and $\varepsilon_L$ = 1.0 (π) mm rad. This corresponds to the huge 6D compression factor of 50 x 50 x 30 = 75'000, i.e. a Merit Factor M = *(Initial 6D)/(final 6D) x transmission* of the order of 15'000 = 75'000/5 with a final number of muons including decay losses 1/5 and of the order of 2 x $10^{12}$ for each sign after acceleration to 125.5 GeV (Table 2).

The presently required over all Merit Factor of the order of 15'000 may be combined with the example [48] based on the so called *"RFOFO" Cooling Ring,* where the merit factor of 162 (Figure 28) had been observed after 16 turns of the ring, bringing the 250 MeV/c emittances at equilibrium. Doubling the number of turns of the RFOFO cooling ring will ensure — with the addition of the required phase of linear pre-cooling — the required compression to attain equilibrium of emittances.

In analogy with the equation [1] relative to a proton beam, the Laslett [54] incoherent-space-charge tune shift quantifies the severity of the muon transverse effects. A useful approximation for the space-charge tune shift $\Delta\nu_{sc}$ at the center of a round Gaussian beam is

$$\Delta\nu_{SC} = \frac{N^{+} r_{\mu}}{2\beta\gamma^{2}\varepsilon_{N} b}$$

In the present case, $r_\mu$ = 1.363 x $10^{-17}$ m is the so-called electro-magnetic radius of the muon, β and γ are the usual Lorentz kinematical factors with for instance β = 0.921 and γ = 2.569 at 250 MeV/c, b < 1 is the bunching factor, defined as the ratio of the average beam current to the peak current and $\varepsilon_N$ = 4.0 x $10^{-4}$ (π) m rad are the normalized r.m.s. transverse equilibrium emittances for Hydrogen [54] In the case of LiH, emittances are about a factor 2 larger. As well known, there is no dependence on the radius of the accumulator.

We choose according to Table 2 values $N^+$ = 5.99 x $10^{12}$ and $N^-$ = 3.85 x $10^{12}$ for the two single bunches in the location of after merging and RFOFO cooling. For these values, the larger signal $N^+$ gives $\Delta\nu_{sc}$ = 1.681 x $10^{-3}$/b. However since long-term stability is not needed, the allowable beam-beam tune shift could be significant. Assuming a maximum value $\Delta\nu_{sc}$ = 0.4, the bunching factor is 1/b = 150, corresponding to 23 cm transverse equilibrium r.m.s length for a 35 m long indicative ring.

Further reductions in the number of muons are introduced by the final solenoidal cooling (0.7) and acceleration to 125.5 GeV (0.7), brings the initial number of Collider's muons to $N^+$ = 2.93 x $10^{12}$ and $N^-$ = 1.89 x $10^{12}$. The full evolution of the muon signal is described in Table 2.

*T*able 2. Summary of the estimated event rates at ESS.

| | Positive | Negative |
|---|---|---|
| *ESS dedicated 2.0 GeV proton events at 14/s* | *2.00 x $10^{15}$* | |
| *Selected (4) proton beam pulses at 56/s* | *2.50 x $10^{14}$* | |
| *Pions/p at all angles, including backwards* | *6.72 x $10^{13}$* | *4.15 x $10^{13}$* |
| *Forward ($p_L$>0) pions/p and 50 to 600 MeV/c* | *2.97 x $10^{13}$* | *1.91 x $10^{13}$* |
| *Muon decay length @250 MeV/c (ct =659 m)* | *1.56 km* | |
| *Linear transverse precooling (0.7)* | *2.07 x $10^{13}$* | *1.34 x $10^{12}$* |
| *RFOFO cooling before merging (0.6)* | *1.25 x $10^{13}$* | *8.02 x $10^{13}$* |
| *Merging to single bunches (0.8)* | *9.98 x $10^{12}$* | *6.42 x $10^{12}$* |
| *RFOFO cooling after merging (0.6)* | *5.99 x $10^{12}$* | *3.85 x $10^{12}$* |
| *Final high field solenoidal cooling (0.7)* | *4.19 x $10^{12}$* | *2.70 x $10^{12}$* |
| *$\mu$ after acceleration to 125. GeV (0.7)* | *2.93 x $10^{12}$* | *1.89 x $10^{12}$* |
| *Total survival factor of the process* | ***0.07*** | |

Various approaches to the ring cooler have been discussed. We refer here to a four sided solenoid-dipole configuration of Garren et al. [49], Figure 29, Figure 30 and Figure 31. It has four 90-degree arcs and eight dipoles separated by solenoids in each arc. The arcs are nearly achromatic both horizontally and vertically. The result is that the dispersion is zero in the straight sections between the arcs. In order to reduce the horizontal and vertical coupling the fields of successive solenoids are alternate in direction.

In order to cool the beam, the liquid hydrogen (LH2) wedge coolers will be inserted into a region with low $\beta$ and high dispersion. Each LH2 wedge absorber has a length of 19.5 cm, an energy loss rate of 0.303 MeV/cm, and a wedge angle of 23 degrees. Four 201.25 MHz accelerating cavities (RF) are placed in the superperiod.

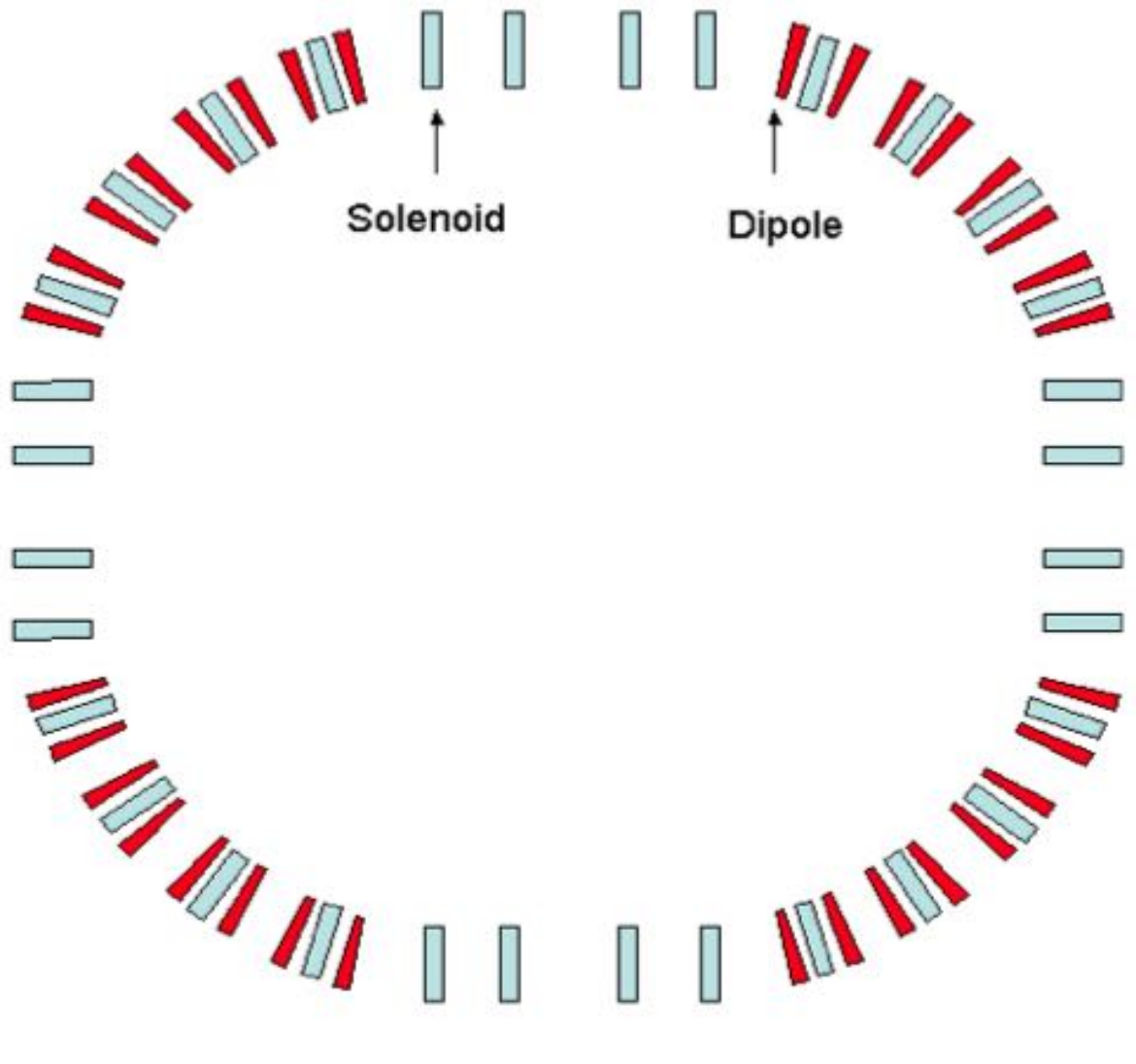

***Figure 29.-*** *Schematic drawing of the four-sided ring cooler using dipoles and solenoids in the configuration of Garren et al. [49]. In order to cool the beam, the liquid hydrogen (LH2) wedge coolers will be inserted into a region with low and high dispersion.*

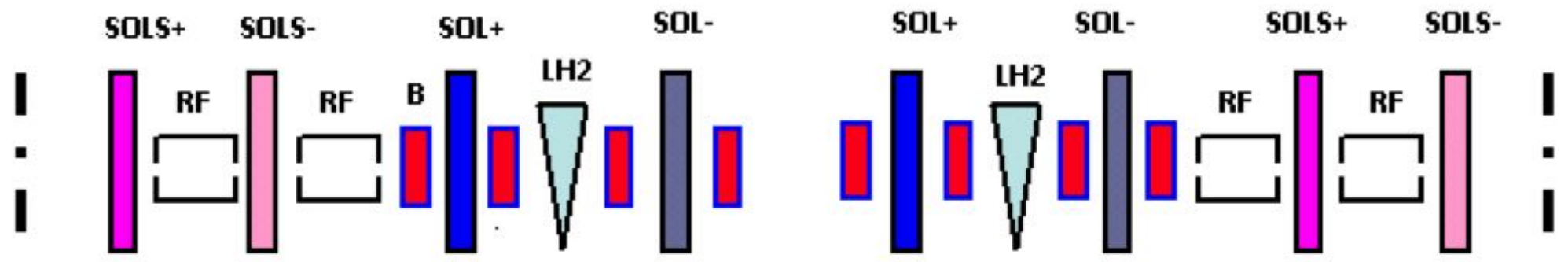

***Figure 30.-*** *Schematic drawing of the ring quadrant in the four-sided and achromatic ring cooler.*

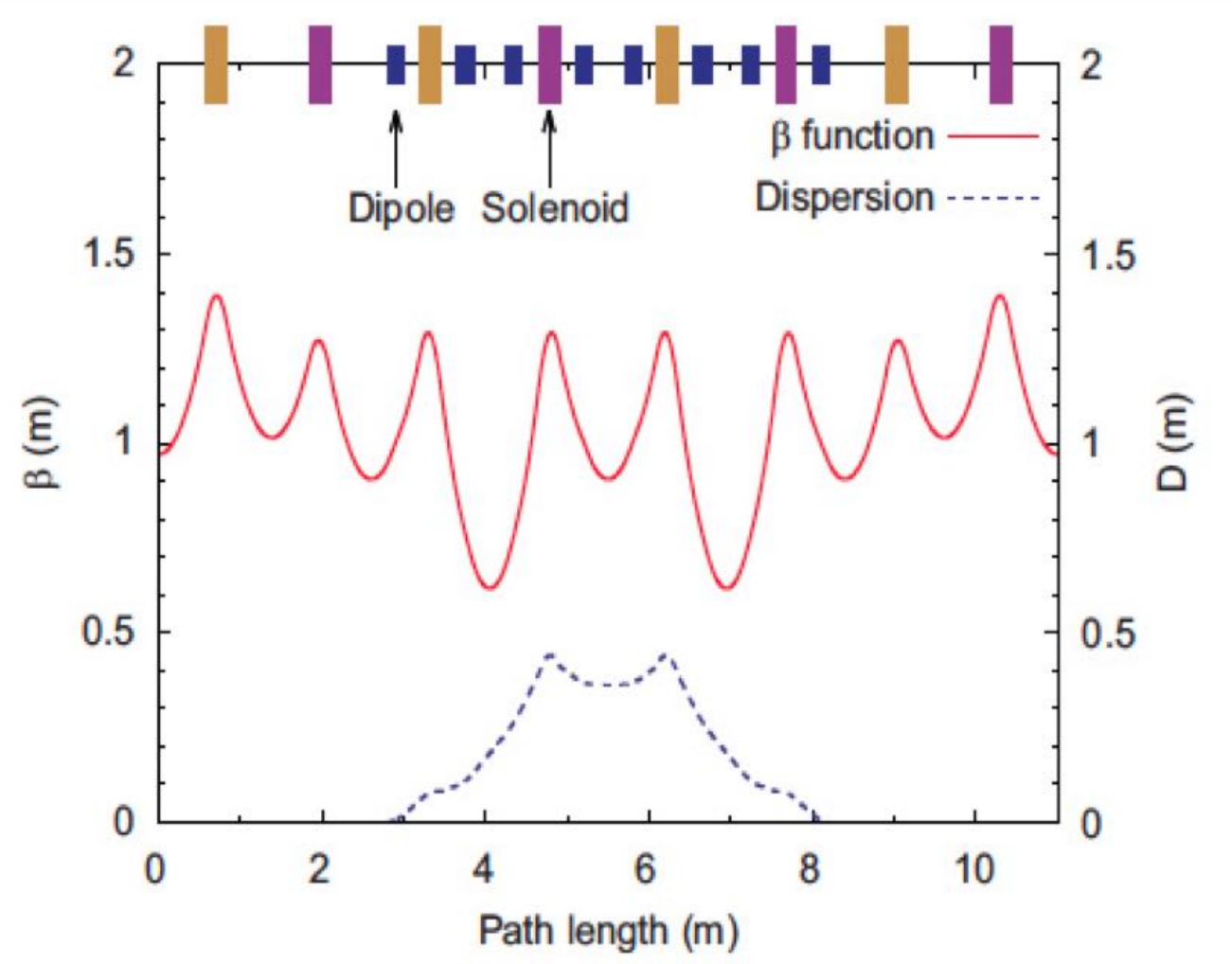

***Figure 31***.- *A tentative design [49] of the "wide band beam". The four sided ring has 90° arcs, each with 8 dipoles separated by solenoids. Arcs are achromatic horizontally and vertically. Dispersion D is zero in the straight sections between the arcs. An elaborate kicker in the straight section and a superconducting flux pipe for the injected beam are used.*

The accelerating gradient is 15 MV/m and the RF phase is 30 deg. Higher-momentum particles have longer paths around the ring, and thus lose more energy per turn than low energy particles do. Consequently cooling takes place in the longitudinal as well as the transverse dimensions.

## *11.- Muon acceleration after cooling.*

Short, intense bunches of protons of several GeV and with a beam power of a few MWatt are focused onto a target to produce pions that decay into muons, which are cooled and accelerated, offering $\mu^+ - \mu^-$collisions of adequate luminosity. The $\mu^+$ and $\mu^-$ bunches are finally counter-rotating in a single Collider ring and focused to collide in two interaction regions.

In order to realize a Higgs Factory at the known energy of 125.5 GeV, an acceleration system increases progressively the energy of captured muons to $m_{Ho}/2$ with a longitudinal Liouvillian acceleration. Assuming a rather large average acceleration of the order of 20 MVolt/m and a single, common LINAC, the transport of muons of both signs to 62.5 GeV would require a length of the order of 3 km (9 $\mu$s), far too large to fit in the ESS site.

Therefore in order to reduce linear dimensions, the narrow, few ns long accelerating muons bunches of both signs should perform several successive passages through a common LINAC structure with small RF adjustments. The Liouvillian acceleration system increases progressively the energy of captured muons to $m_{Ho}/2$. After a pre-accelerator to an indicative energy of 2.5 GeV, the transport merges and accelerates $\mu^+$ and $\mu^-$.

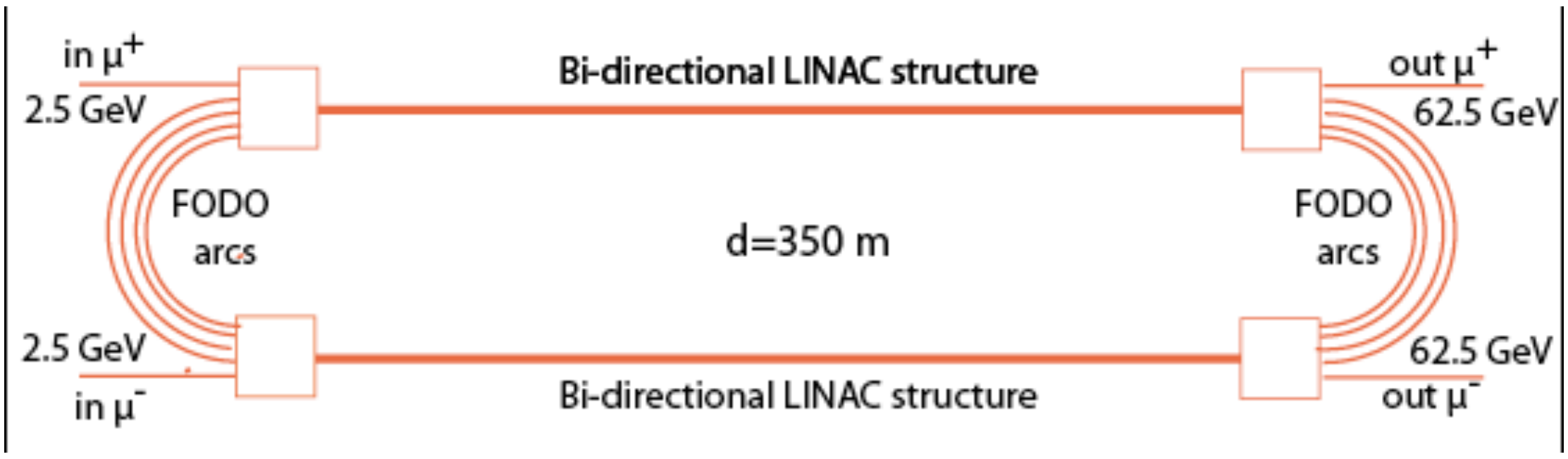

***Figure 32.-*** *Aa acceleration system is progressively rising the energy of muons to $m_{Ho}/2$ with the help of re-circulating linear accelerators (LRA). For instance with a eight turn arrangement the energy increase of the LRA is 7.8 GeV/pass, corresponding to a length of ≈ 350 m for an average gradient of 20 MeV/m. A pre-accelerator is increasing the energy up to 2.5 GeV.*

The setup with two linear LINAC bi-directional structures traversed in opposite directions by the muon charges is described in Figure 32. The option corresponds to N = 4 passages and to the linear LINAC structure traversed nine times (4 x 2 +1) in opposite directions by the muon charges . The beam energy gain for each passage is then 60,0/9 = 6.66 GeV, corresponding to a length of about 330 m for a particle RF gain of about 20 MeV/m. The nominal energies of the FODO 180° arcs are then respectively 9.17, 22,5, 35.8, 49.16 GeV for the right arc and 15.8, 29.16, 42.5, 55.8 GeV for the left arc. Stray collisions between $\mu^+$ and $\mu^-$ occur once at every mid crossing of the FODO 180° arcs. An additional passage before the muon exit adds another 6.66 GeV, reaching the final output energies of 62.5 GeV. At the largest right arc with 55.8 GeV, the average magnetic bending radius at 5.0 Tesla gives $\rho$ = 37.2 m, corresponding to a FODO bending radius equals to 1.4 x 37.2 = 52.08 m. The 180° arcs have similar concentric structure located in the same holding support, for correspondingly lower fields with diameters about 3 meters apart.

The total length traversing the 9 LINAC segments is 2970 m corresponding to an energy integrated $\mu^+$ or $\mu^-$ decay loss of 2.44 %. Therefore the loss over the full path of Figure 32 due to decays from 2.5 GeV to the final Higgs mass is 6.4 %. The corresponding $\mu^+ + \mu^-$ decay losses of the pre-accelerator LINAC from 250 MeV/c to 2.5 GeV is another 3.62 %. Therefore the full $\mu^+ + \mu^-$ decay losses up to the Higgs mass are ≈ 10 %.

The higher energy option at $\sqrt{s}$ = 0.5 TeV will extend the lengths. The same option with N = 4 x 2 multiple passages to the linear LINAC structure traversed nine times would extend the LINAC lengths to 1.4 km, requiring the FODO multiple magnetic bending with higher magnetic fields and/or larger radii. Of course different values are also possible, as a balance and optimization between number of arcs and lengths of LINACS.

### *12.- The ESS Collider.*

The circular beam Collider for the previously described option (1) is a relatively small SC ring at 62.5 GeV with a typical radius of ≈ 60 m and two low β sections with a free length of about 10 m, where the two detectors are located (Figure 33). cm be $\beta_x = \beta_y = \beta^* = 5$ cm. The main parameters are given in Table 3.

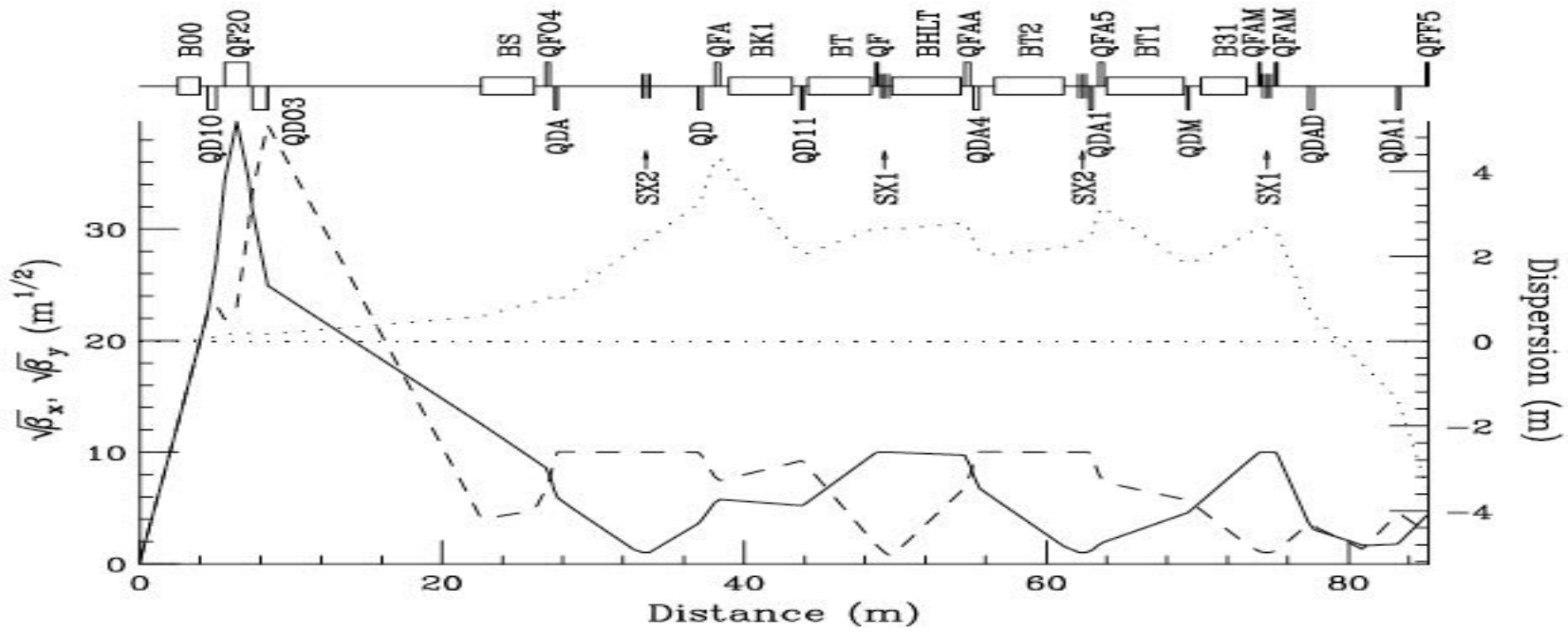

***Figure 33.-*** *Lattice structure at crossing point of the 60 m radiua ESS Collider,including local chromaticity corrections with $\beta_x = \beta_y = \beta^* = 5$ cm*

An important background which has to be carefully analyzed is coming from the ≈ 2 x $10^{12}$ muons/bunch of each sign from $\mu \to e\nu_\mu\nu_e$decays, emitting a total of 2 x $10^{11}$ $e^{\pm}/s/meter$ in a narrow average cone of $\langle\theta\rangle \approx$ 1.6 mr from the beam axis, producing an average electron showers power of 1.6 kW/m. This amount of power, due to electron showering is comparable to the one in the case of synchrotron radiation from TLEP [57] or SuperTRISTAN [58]) and it must be handled with an appropriate geometry of the locations of the shower stoppers (Figure 34).

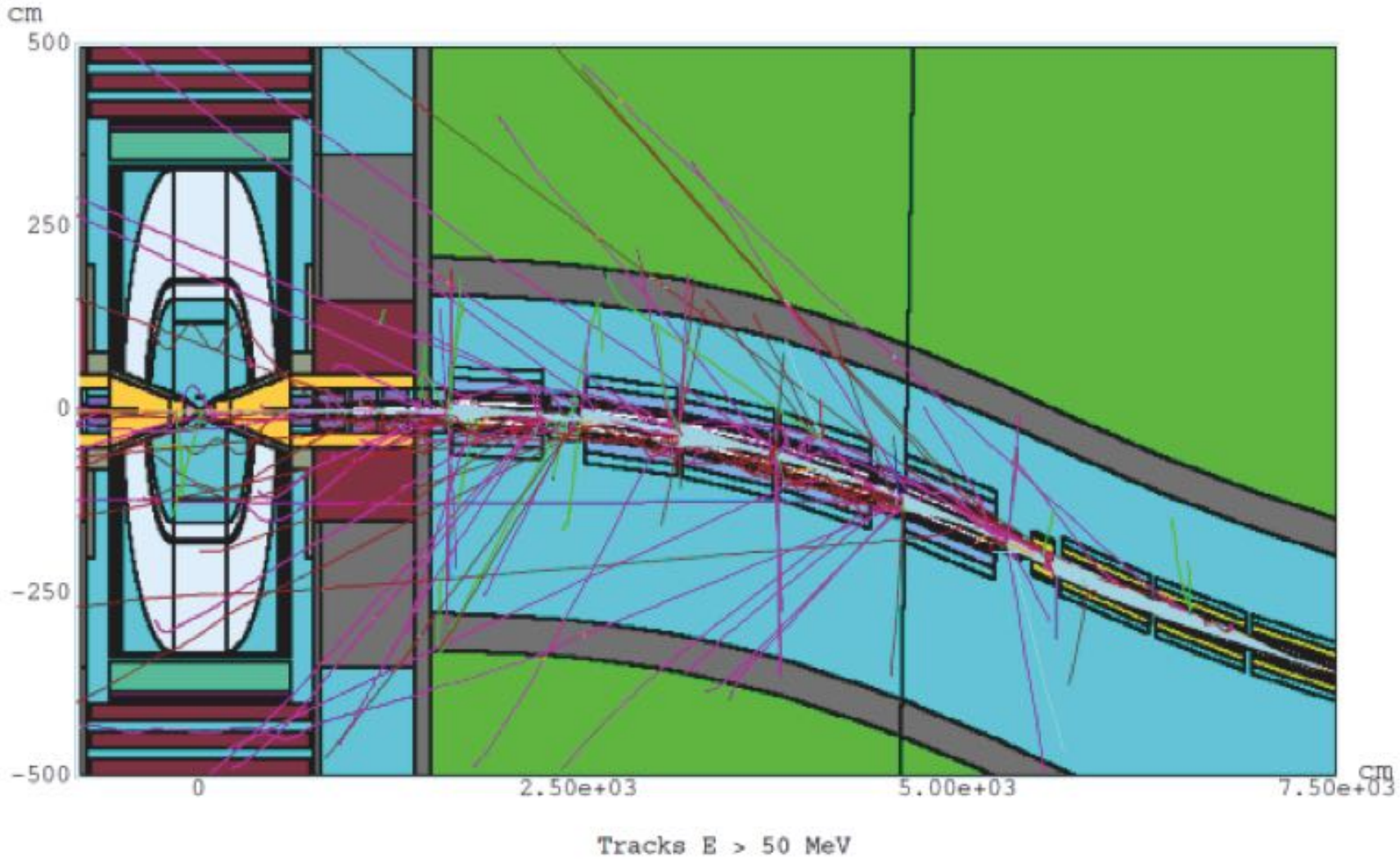

***Figure 34.-*** *The amount of power due to 1.9 x$10^7$ decays per m of electron showering coming from 6 x$10^{12}$ $\mu \to e\nu_\mu\nu_e$decays is very large and it must be handled with an appropriate geometry of the locations of the shower stoppers.*

The angular distribution of the decay electrons and positrons should be used in order to measure and control as a function of time the individual muon bunches observing the g-2 precession frequency of polarized decays.

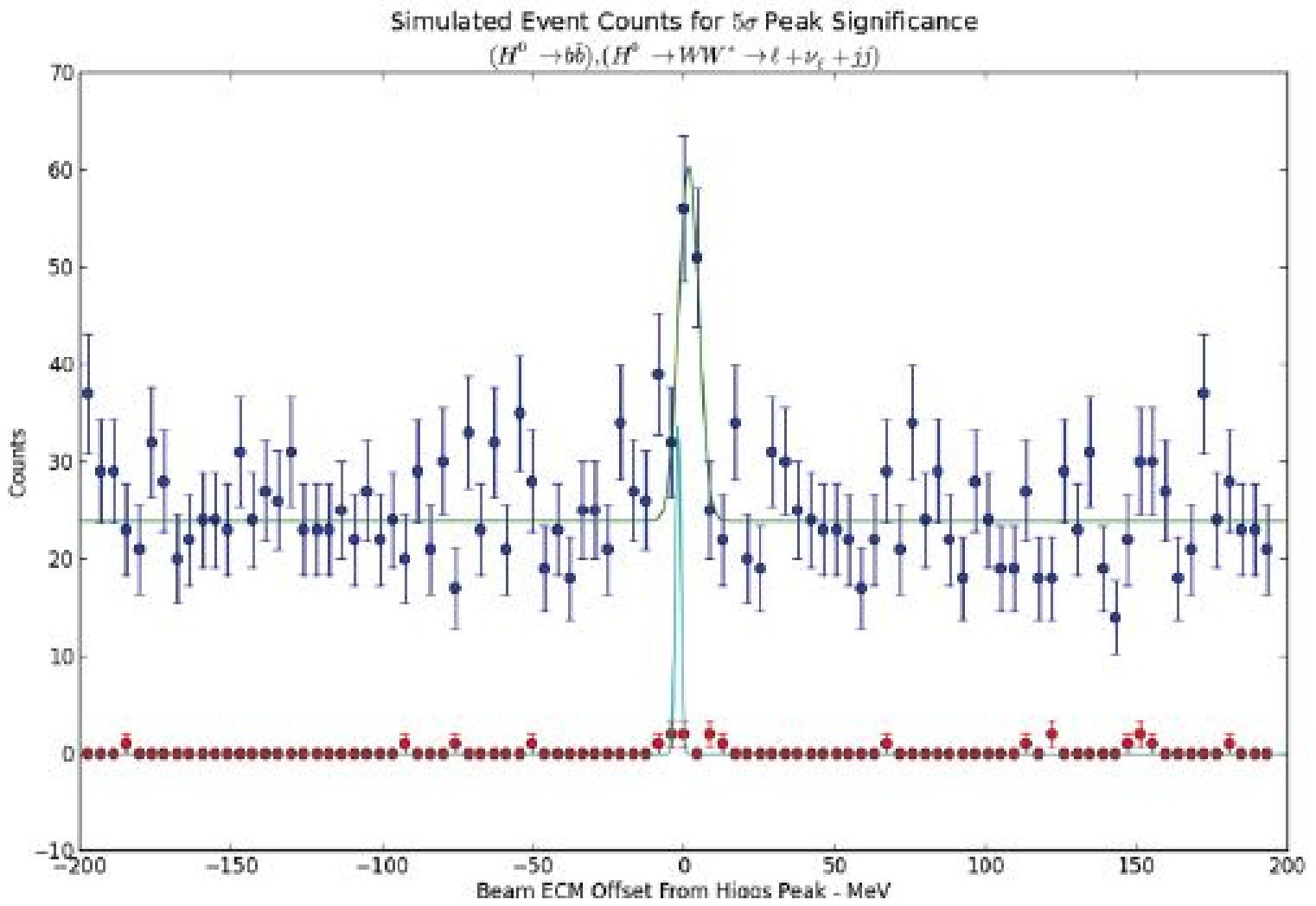

***Figure 35.-*** *Presently the Higgs mass is known to some 600 MeV., At the muon collider we need to locate $M_H$ to ~ 4 MeV. The location of the resonance can be performed initially with a few months running at the relatively modest initial 1.7 x $10^{31}$ luminosity.*

The present uncertainty with which the mass of the $H^o$ has been so far measured by ATLAS [1] and CMS [2] has been of the order of about 1 GeV. No doubt future measurements will improve this result and a window of the order of ± 200 MeV should become reasonable in the forthcoming future.

The first step of a future $\mu^+ - \mu^-$ Collider should therefore be the one of finding the position of its signal in view of its actual very narrow resonance width. This is particularly favorable for the (WW) channel that has a substantial branching ratio of 22% according to the SM.

The angular distribution of the decay electrons and positrons should be used in order to measure and control as a function of time the individual muon bunches observing the g-2 precession frequency of polarized decays.

The first step of a future $\mu^+ - \mu^-$ Collider should therefore be the one of finding the position of its signal in view of its actual very narrow resonance width (Figure 35). This is particularly favorable for the (WW) channel that has a substantial branching ratio of 22% according to the SM.

The competing non-resonant background signal in the vicinity of the Ho is extremely small, of the order of ≈ 7 fb. In the bin closest to the signal for a Gaussian convoluted beam spread of 7 MeV and for an energy scan in steps of 10 MeV (40 steps) the presence of the Higgs will record a minimum $H^o$ signal of 750 fb. An initial exposure of 10 effective days at 3 x $10^{31}$ $cm^{-2}s^{-1}$ luminosity corresponds to 2.6 x $10^{37}$ $cm^2$ and > 20 events for the maximum of the (WW) resonant signal with a negligible background in each of the two experimental detectors.

Once the actual location of the H° resonance has been identified with the $\mu$+$\mu$- Collider, the actual value of its mass can be determined with a remarkable precision observing the g-2 precession frequency of polarized muon decays. Raia and Tollestrup [20] have investigated the feasibility of determining the precise energy of the circulating muon beams (once accelerated to the final energy of 62.5 GeV) with the turn by turn variation of the electron energy spectrum produced by the decay $\mu \rightarrow e\nu\nu$.

This variation based on a non-zero value of (g – 2) for the muon and of a finite polarization of the beam. This angular frequency/turn in the energy spectrum of the decay electrons $\omega = 2\pi\gamma(g-2)/2 \approx -0.7 \times 2\pi$ can be measured with high precision. Because of the narrow width of the Higgs boson, it is mandatory control the energy of the individual muon bunches to a precision of a few parts in a million. With adequate statistics it may then be feasible to determine the mass of the H° particle to the order of 100 keV, i.e. $\delta E/E \approx 10^{-6}$.

After having reached the nominal energy of 62.5 GeV for each of the two buckets of muons of opposite signs with $\gamma_\mu$ = 589 and $\mu$ lifetimes of 1.295 ms, collisions are recorded at two low beta points of a storage superconducting ring of an appropriate radius of about 60 m.

Further reductions in the number of muons are introduced by the final solenoidal cooling (0.7) and acceleration to 125.5 GeV (0.7), brings the initial number of Collider's muons to $N^+ = 2.93 \times 10^{12}$ and $N^- = 1.89 \times 10^{12}$.

Table 3. Collider ring properties at 125.5 GeV

| | | |
|---|---|---|
| *Proton kinetic energy* | *2.0* | *GeV* |
| *Number accelerated protons/pulse* | *1.1 x $10^{15}$* | *p/p* |
| *Proton power* | *5.0* | *MWatt* |
| *Proton current* | *62.5* | *mA* |
| *Proton event rate* | *14* | *ev/s* |
| *Duration of proton events* | *2.86* | *Ms* |
| *Number accelerated protons/event* | *1.1 x $10^{15}$* | *p/ev* |
| *Proton driven target collision rate* | *56 =14 x 4* | *ev/s* |
| *Time between proton driven collisions* | *17.86* | *Ms* |
| *Number accelerated protons/collision* | *2.5 x $10^{14}$* | *p/coll* |
| *Number $\mu$+ after final cooling at 250 MeV/c* | *4.19 x $10^{12}$* | *$\mu$+/coll* |
| *Number $\mu$– after final cooling at 250 MeV/c* | *2.70 x $10^{12}$* | *$\mu$–/coll* |
| *Final muon momentum* | *62.5* | *GeV/c* |
| *Final $\beta\gamma$ of muons* | *594* | |
| *Final muon lifetime* | *1.295* | *Ms* |
| *Initial number of $\mu$+ at collider ring* | *2.93 x $10^{12}$* | *$\mu$+/coll* |
| *Initial number of $\mu$+ at collider ring* | *1.89 x $10^{12}$* | *$\mu$–/coll* |
| *Equalized number of muone in each ring* | *2.41 x $10^{12}$* | *$\mu$/coll* |
| *Transverse invariant emittances, $\varepsilon_N$ (no PIC)* | *0.37* | *π mm rad* |
| *Longitudinal invariant emittance* | *1.9* | *π mm rad* |
| *Betatron function at collision point, $\beta_x = \beta_y = \beta$* | *5* | *cm* |
| *Tranverse r.m.s. bunch radius* | *140 (176)* | *$\mu$m* |
| *R.m.s. transverse emittance, $\varepsilon_{rms} = \varepsilon_N/(\beta\gamma)$* | *0.62 x$10^{-4}$* | *π cm rad* |
| *Circumference of collider ring* | *350* | *m* |
| *Number of effective luminosity turns* | *555* | |
| *Effective repetition rate/crossing, f = 54 x 555* | *29'970* | *s-1* |
| *Higgs luminosity at each collision point* | *4.0 x $10^{31}$* | *$cm^{-2}$ $s^{-1}$* |
| *Nominal Higgs cross section at 125.5 GeV/c* | *3.0 x $10^{-35}$* | *$cm^2$* |
| *Nominal Higgs events, $10^7$ s/y, each crossing* | *12'000* | *ev/y* |

The luminosity at √s = 125.5 GeV is given by

$$L = \frac{fN^+N^-}{4\pi\varepsilon_{rms}\beta^*}$$

where fo $\varepsilon_{V,H}$ = 0.4 (π) mm rad (0.04 (π) cm rad) without PIC, we have $\varepsilon_{rms} = \varepsilon_{V,H}/\gamma = \varepsilon_{V,H}/594$ = 0.673 x $10^{-4}$ (π) cm rad, a collision betatron function β* = 5 cm (equal in both planes) and a number of effective luminosity crossings in each of the two collision points f = 54 x 555 = 29'970/s, resulting in L = 4.0 x $10^{31}$ $cm^{-2}$ $s^{-1}$at each of the two interaction points. With a Higgs resonant cross section $H^o$ = 3.0 x $10^{-35}$ $cm^2$ and an effective live time of $10^7$ s/year, the Higgs related event rate is expected to be L x $H^o$ x $10^7$ = 12'000 ev/y.

The value β* = 5 cm is not a critical parameter and values of β* as small as ≈ 1.0 cm may be possible, with the corresponding luminosity increases, but only provided that the longitudinal dimension of the bunch is of comparable size, which requires an appropriate longitudinal emittance.

The 126 GeV Higgs scalar is quite near to the 91.2 GeV $Z^o$ which has a production cross section almost 1000 times larger (3 x $10^4$ pb), namely 3000 $Z^o$/day at L = $10^{30}$ $cm^{-2}$ $s^{-1}$. Its larger cross section may offer a large number of events at the $Z^o$ peak during the earlier operation. Separation of signal from backgrounds can be initiated with relatively easy parameters. The $Z^o$ energy measurement would be an early part of the Higgs program.

### *13.- The Parametric Ionization Cooling (PIC).*

The proposed next-generation Collider may encourage new major technical advances in order to further reduce the transverse emittances beyond the standard equilibrium cooling. The *Parametric-resonance Ionization Cooling (PIC)* is therefore investigated for the final 6D cooling stage of the high-luminosity muon Collider.

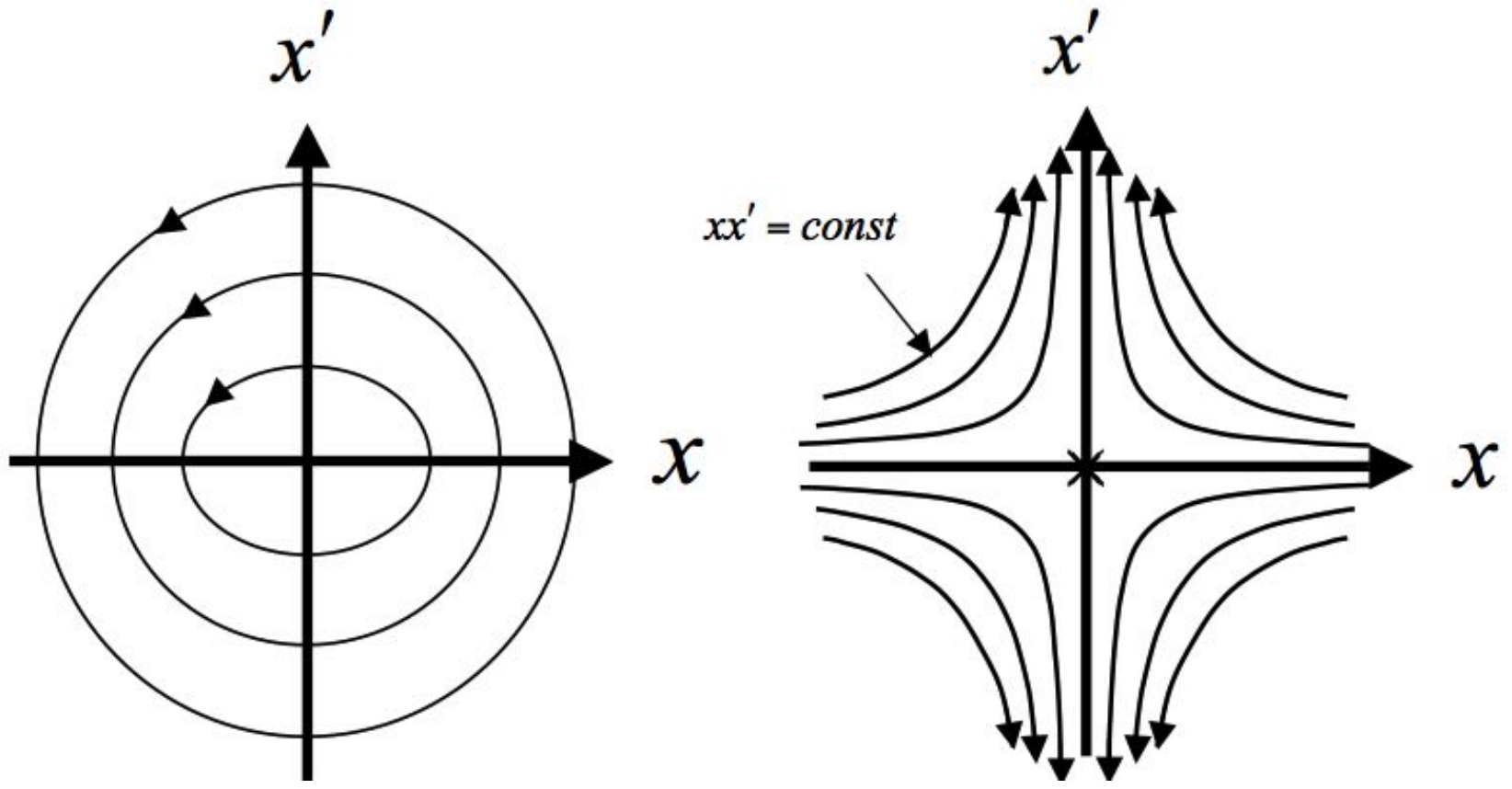

***Figure 36.-*** *LEFT ordinary oscillations; RIGHT hyperbolic motion induced by perturbations near an (one half integer) resonance of the betatron frequency.*

PIC dynamics is expected to produce a smaller final transverse beam emittances and may offer the potentials either for a higher luminosity and/or a smaller number of muons.

A second linear magnetic transport channel with different and appropriate ring magnetic field parameters [55] has to be added. A linear magnetic transport channel has been designed by Ya.S. Derbenev et al. where a half integer resonance is induced such that the normal elliptical motion of particles in x-x' phase space becomes hyperbolic, with particles moving to smaller x and larger x' at the channel focal points. (Figure 36). The half-integer parametric resonance introduces strong *transverse focusing* while ionization cooling limits the beam's *angular spread* with appropriately placed energy absorbers.

*Table 4. Indicative parameters [55] of a preliminary PIC Cooling (PIC).*

| Parameter | Initial | Final | Unit |
|---|---|---|---|
| Muon beam momentum, p | 250 | 250 | MeV/c |
| Number of particles/bunch, $N_b$ | 1 | 1 | 10^10 |
| Normalized transverse emittances (rms) | 230 | 23 | μm |
| Beam size at absorbers (rms) | 0.7 | 0.1 | Mm |
| Angular spread at absorbers (rms) | 130 | 130 | Mrad |
| Momentum spread (rms), $\Delta p/p$ | 2 | 2 | % |
| Bunch length (rms) | 10 | 10 | Mm |

The parameters of the original proposal [55] are illustrated in Table 4. At the chosen value of 250 MeV/c,the muon momentum is limited to 2 % r.m.s. and to $10^{10}$ particles, much smaller than what required for the ESS alternative.

Assume a periodic focusing lattice of period $\lambda$ along the beam path coordinate z. The transformation matrix **M** along the transverse particle transverse coordinate and angle x and x' between two points at $z_o$ and $z_o + \lambda$ has the general form

$$M = \begin{bmatrix} e^{-\lambda\Lambda d}\cos\psi & g\sin\psi \\ -\sin\psi / g & e^{\lambda\Lambda d}\cos\psi \end{bmatrix}$$

In particular, the optical period can be designed in a way that sin $\psi$ = 0, (i.e. $\psi = \pi$ or $\psi = 2\pi$). Then the evolving particle coordinate and angle (or momentum) appear uncoupled: $[x] = \pm\, e^{-\lambda\Lambda_d}[x]$ and $[x'] = \pm\, e^{-\lambda\Lambda_d}[x']$. Thus, if the particle angle at point $z_o$ grows ($\Lambda_d > 0$), the transverse position experiences damping, and vice versa. Liouville's theorem is not violated, but particle trajectories in phase space are hyperbolic (x x' = const). Exactly between the two resonance focal points the opposite situation occurs where the transverse particle position grows from period to period, while the angle damps. If we now introduce an energy absorber plate of thickness *w* at each of the resonance focal points, ionization cooling damps the angle spread at a rate $\Lambda_c$.

More generally, assuming a balanced 6D ionization cooling, the three partial cooling decrements have been equalized as follows, using emittance exchange techniques:

$$\Delta_c = \frac{1}{3}\Delta \quad \Delta = 2\frac{\langle \gamma'_{abs} \rangle}{\gamma} = 2\frac{\langle \gamma'_{aee} \rangle}{\gamma} \quad \langle \gamma'_{abs} \rangle = 2\frac{w}{\lambda}\gamma'_{abs}$$

where $\gamma'_{abs}$ and $\gamma'_{acc}$ are the intrinsic absorber energy loss and RF acceleration rate. For $\Lambda_d = \Lambda_c / 2$, the angle spread and beam size are damped with a decrement $\Lambda_c / 2$:

$$\begin{bmatrix} x \\ x' \end{bmatrix}_{zo+\lambda} = e^{-\lambda\Lambda c/2}\begin{bmatrix} x \\ x' \end{bmatrix}_{zo}$$

The emittance with PIC is improved compared to a conventional cooling channel by the factor $(\pi w)/(\lambda sqrt(3)) = (\pi\gamma'_{aee})/(sqrt(12)\gamma'_{abs})$ Thin absorbers placed at the focal points of the channel cool the angular divergence with the usual ionization cooling. Without damping, the beam dynamics is not stable because the beam envelope grows with every period. Energy absorbers at the focal points stabilize the beam through ionization cooling. The longitudinal emittance is maintained constant tapering the absorbers and placing them at points of appropriate dispersion.

In chapter 10, "Ionization Cooling" and Table 2, the expected number of muons after RFOFO cooling and merging in a single bunch are $N^+ = 6.0 \times 10^{12}$ and $N^- = 3.8 \times 10^{12}$. The normalized r.m.s. transverse equilibrium emittance for Hydrogen [54] is $\varepsilon_N = 4.0 \times 10^{-4}$ ($\pi$) m rad. Introducing the PIC cooling will then bring the emittances to exceedingly large tune shift $\Delta\nu$ values. Therefore muons should be accelerated substantially — for instance to 1.0 GeV/c — before entering a second “PIC” ring . Finally the momentum spread should be some five rimes larger.

PIC cooling for the present project requires therefore some very bold extrapolations which have to be valuated. The realization of a PIC Ionization Cooling requires a delicate compensation for tunes spreads caused by numerous aberrations: chromatic, spherical, and non-linear field effects. Techniques are needed to reduce the detuning. One of the PIC challenges is compensation of beam aberrations over a sufficiently wide parameter range while maintaining the dynamical stability with correlated behavior of the horizontal and vertical betatron motion and dispersion.

Only after PIC has been fully confirmed with the much cheaper and simpler *Initial Cooling Experiment* — as described in the present paper on chapter 14 — the final setup can be realized. For the previously indicated $N^+ = 2.93 \times 10^{12}$ and $N^- = 1.89 \times 10^{12}$ the value of $\Delta\nu_{sc}$ largely is exceeded since any decrease of the transverse emittance will correspondingly augment the Laslett tune shift, for two bunches already largely at the limit in the previous conventional cooling conditions. The momentum spread is also now avout 10% r.m.s.

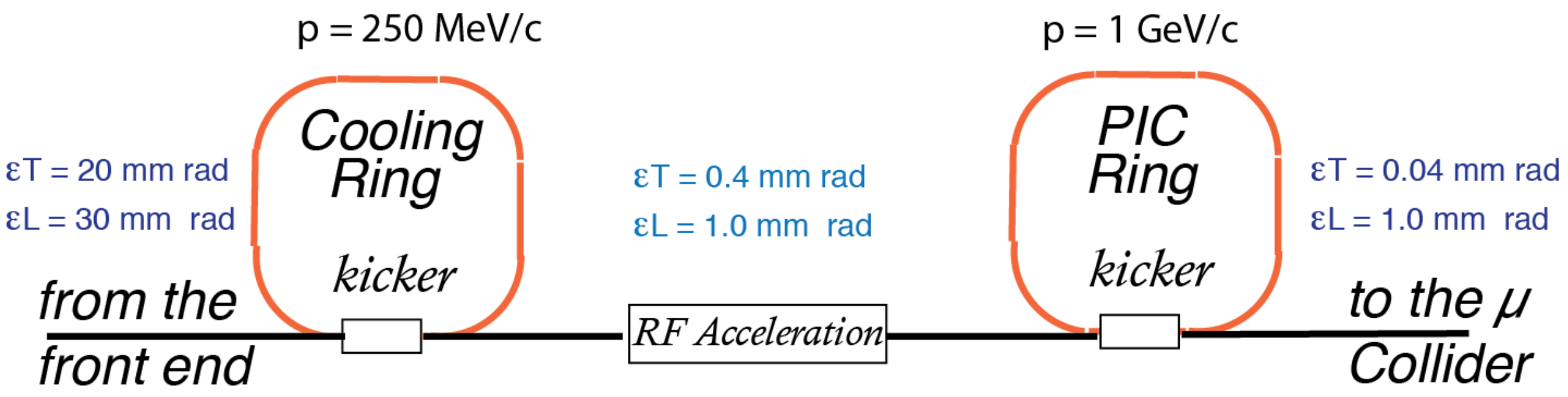

***Figure 37.-*** *Schematic description of the PIC cooling. Muons from the front end are entering n the main cooling ring where transverse emittances rreach the cooling equilibrium. After further acceleration to about 1 GeV/c a second larger ring at the PIC configuration near a resonance arrangement is further reducing the emittances of a factor about ten in the traverse radius of the beam, prior to final acceleration.*

Comparison of cooling factors (ratio of initial to final 6D emittance) with and without the PIC condition vs. number of cells — if fully successful — indicates an ultimate gain of about 10 times for a final invariant emittances at 1000 MeV/c.

Thin absorbers placed at the focal points of the channel then cool the angular divergence. The longitudinal momentum spread is maintained constant tapering the absorbers and placing them at points of appropriate dispersion, vertical $\beta$ and horizontal $\beta$ (Figure 37). Combining muon ionization cooling with parametric resonant dynamics is expected to allow final transverse beam emittances smaller than with the conventional ionization cooling as described in the previous paragraphs.

The PIC accumulator at 4 times the momentum will have a four times the circumference (120 m) reducing correspondingly the b factor to 1/b = 600 and the transverse beam radius by about a factor ten for an acceptable tune shift. Because of the acceleration, the longitudinal emittance will be also appreciably reduced to $0.25^{1/4} = 0.7$.

To conclude, the implementation of such a vastly different configuration requires serious experimental confirmations *before being considered as a realistic alternative.* One of the PIC open challenges is compensation of beam aberrations over a sufficiently wide momentum range while maintaining the dynamical stability with correlated behavior of the horizontal and vertical betatron motion and dispersion. We explore use of transverse coupling to reduce the dimensionality of the problem and to shift the dynamics away from non-linear resonances. On the other hand in order to profit of the 10 x increased luminosity the numbers of muons must be maintained.

## *14.- The Initial Cooling Experiment*

As an initial part of the program, cooling should be experimentally studied in an appropriate $\mu^+\mu^-$ ring configuration and the much cheaper and simpler *Initial Cooling Experiment* (Figure 38).

With its ring conference of about 45 meters, the Initial Cooling Experiment will require a much less intense muon beam and it can be realized at a far modest cost. It consists essentially of ten LiH energy absorbers and two fast kickers (in and out) with 48 RF cavities.

Several European laboratories, like for instance in the UK, CERN or Sweden (Lund) could be a priori considered for a possible location for this Initial Cooling Experiment. The ESS related alternative at its presently approved stage is described here. Other locations are of course possible, like for instance at the CERN-PS or SPS.

The main goal of this experimental R&D program is to develop the muon ionization cooling hardware-wise to the point where a complete ionization cooling channel can be confidently designed for the ESS$\mu$SB Muon Collider. Provided muon cooling has been experimentally verified in all its main aspects, the subsequent realization of the full scale $\mu^+\mu^-$ Collider program may follow for instance at the ESS at higher intensity with the addition of several conventional accelerator technologies of reasonable dimensions.

The Initial Cooling Experiment will require a proton target with a much more modest intensity than the one needed for the final cooling. Out of a standard ESS pulse (2.86 ms and 62.5 mA) one may select a separate ≈ 100 ns long proton pre-pulse of about 3 x $10^{11}$ protons at 2.0 GeV/c (≈ 70 bunches at 704 Mhz, 1.42 ns apart) and at a repetition rate for instance of a few seconds.

The experimental succession of operations at the Initial Cooling Experiment will closely follow the previous description of points # 2 to # 8 of Figure 25 as given in COOL 2007 [64] adapted to the ESS. At the end of this process and ejection by the second kicker, outgoing particles are analysed in momentum and direction by an adequate external comprehensive setup and wire chambers.

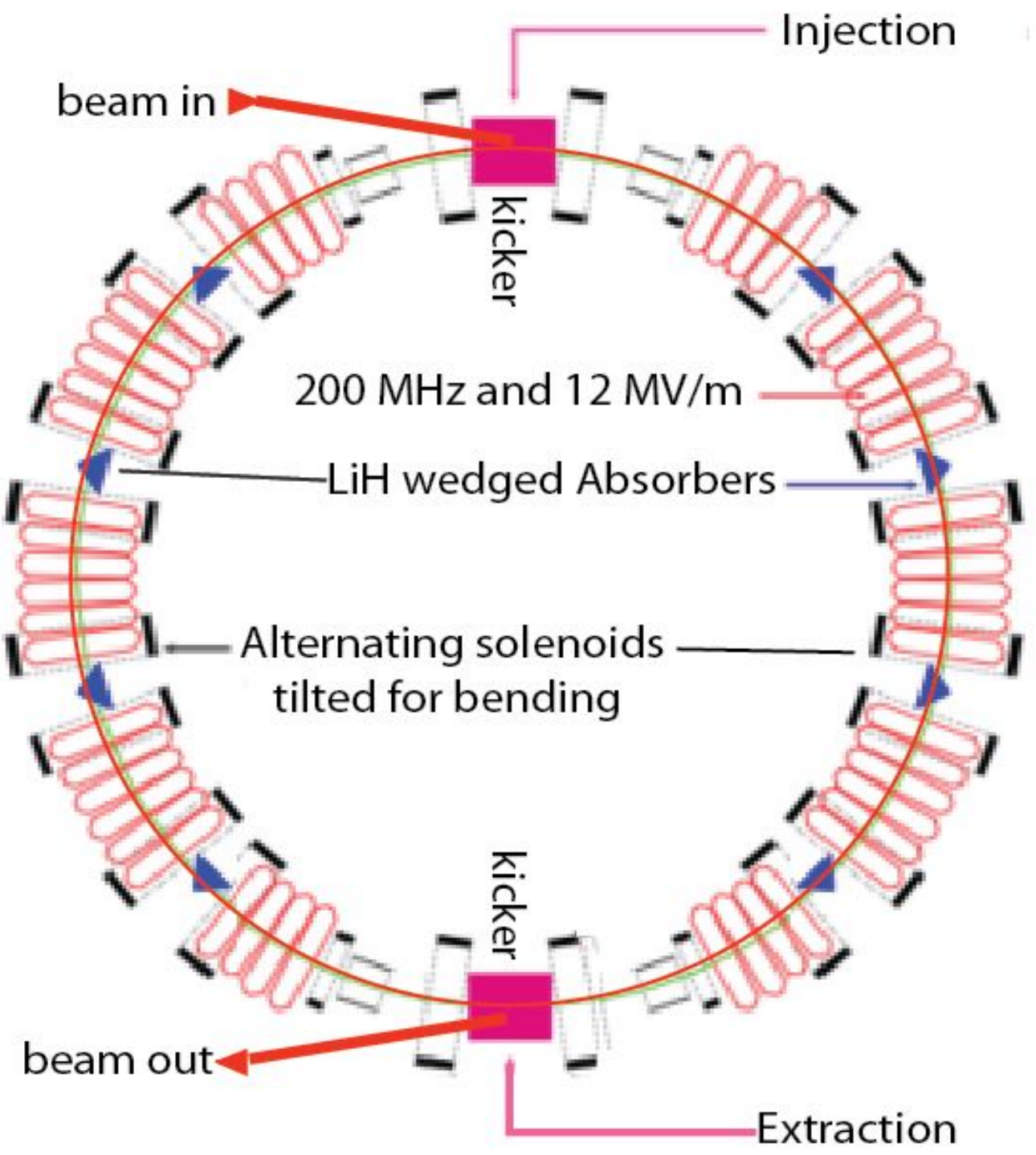

***Figure 38.***- *The "RFOFO" ring of 45 m circumference] is made of 12 identical cells with about 3 degrees tilted solenoids of opposite orientations providing focusing and bending. RF cavities may be either under vacuum or pressurized with hydrogen and helium gas up to more than 100 atm pressure at temperature down to 77 K.*

The particle producing horn can be made with a solid mass of high Z material, for instance a 15 cm thick Tungsten rod (m.p. 3695 K), and about two hadron interaction lengths. Assuming one half of the black body temperature, the Stefan Boltzmann radiation will bring a 18 $cm^2$ rod to slightly less than 2000 K.

Alternatively one might use the CERN-PS at 24 GeV with one bunch of 2 x $10^{12}$ protons and a pion intensity at 250 MeV/c about a factor 20 larger. The ESS configuration has been chosen.

The horn will be followed by a kicker that will send products to the ring (Figure 36). Before reaching the kicker a linear transport of adequate length (not shown) will build up muons from pion decays. At 250 MeV/c, pions have ct = 7.8 m, a half-life of 14 m. Muons have instead ct = 659 m, corresponding to the much longer decay length of 1.56 km. The peak pion momentum at 2 GeV/c protons of ESS is about 100 MeV/c, as shown in the spectrum of Figure 20. Therefore a lower but still acceptable muon yield is expected for 250 MeV/c muons.

Secondary particles of a nominal momentum of about 250 MeV/c and of a given sign are then injected in the ring with the help of a kicker magnet. The ring may have a circumference of the order of 45 m, (Figure 38), corresponding to about 100 ns long injected pulse and allowing about additional 60 ns for the kickers.

At the ESS, the proton beam of the LINAC is tightly bunched at 704.4 MHz. For the muon cooling ring a 1/3 or 1/4 synchronous sub-harmonic may be selected, either of 234.8 or 176.1 MHz. A frequency for the RF cavities of the ring of 234.8 MHz has been chosen. The exact length of the ring has been accurately adjusted to 40 RF cycles at nominal momentum and to a length of 162.51 ns.

For an average $\beta_{x,y} \approx 50$ cm and a beam radius of 5 cm the resulting accumulated $\mu^-$ rate at 250 MeV/c within ± 25 MeV/c ( ± 10%) window is estimated in an initial population of 3 x $10^7$ $\mu^-$/pulse. The $\mu^+$/pulse rate is about twice this value. .

Two magnetic configurations are in principle possible for the ring: either (1) a conventional structure of bending magnets and quadrupoles or (2) a more sophisticated set of slightly bent solenoidal fields with alternate polarities, the so called "RFOFO" arrangement (Figure 27) that will offer a significantly wider momentum spread and be closer to the final cooling setup. The "RFOFO" arrangement (Figure 37) and the ESS alternative at Lund have been chosen.

RFOFO focusing is provided by pairs of solenoids with alternating field direction ("FOFO" since both focus, "R" is for 'reverse'. All solenoids are identical and both transverse and longitudinal cooling are obtained in the same LiH wedge absorbers. The solenoids are not evenly spaced: those on either side of the absorbers are closer. This increases the focusing. A bending field is generated by tilting the focus coils by alternate 3° angles.

The "RFOFO" ring of 45 m circumference [48] is made of 12 elements with 3 degrees tilted solenoids of opposite orientations providing both focusing and bending. The two kind of solenoids are 50 cm long and have an inner radius of 77 cm and an outer radius of 88 cm. One solenoid is tilted by +3° with respect to the horizontal plane. The next solenoid is tilted by -3°. Both solenoids are centered 10 cm "outside" the ring central line, in such a way that the circular trajectory along the central line follows more or less the magnetic field lines.

Each ring unit of Figure 36 covers an arc length of 36 degrees along the central line. Each of the RF cavity elements is a cylinder, 28.75 cm long and 25 cm in aperture, 33 cm apart on the central line. As already pointed out, RF cavities may be either under vacuum or pressurized with hydrogen and helium gas up to more than 100 atm pressure at temperature down to 77 K [50]. At 50 atm and 77 K, hydrogen gas density is about 21.5% of liquid hydrogen and may support gradients of up to 330 MV/m.

The resulting RF gain/turn is (2.5 MeV gain) x 6 cavities x 10 cells =150 MeV balanced with the absorbers. The individual sets are empty inside and have at the edges two Beryllium walls, each of 0.5 mm. The loss/turn of the 2.x.10 = 20 windows for muons is 1.34 MeV namely 0.8 % of the RF gain/turn corresponding to 1.0 x $10^{-2}$ radiation lengths. As an alternative the RF cavities may have no inside windows.

The absorber thickness is increasing with growing radius since it is a wedge of 110° opening angle. The wedge is 40 cm wide and its tip is 9.5 cm above the central line.

The upper injection kicker (Figure 38) is storing at 250 MeV/c also other stable particles and in particular protons and electron-positrons at the same momentum. A nominal RF gain of 150 MeV/turn has been assumed. According to the ring of Figure 38 there are four units with 6 RF cells and four units of 4 RF cells, for a total of 40 cells. Therefore a total of 48 windows are required — assuming that two contiguous cells have one common window. Each of the windows is made *of* 0.5 mm thick Beryllium foil, adding to a total thickness of 48 x 0.05 = 2.4 cm and an energy loss of 2.4 cm x 2.98 MeV/cm = 7.152 MeV.

The eight wedged absorbers are made with LiH (1.34 MeV/cm) and a total length at the nominal beam trajectory of 150 MeV/ (1.34 MeV/cm) = 112 cm corresponding to 14 cm long wedges. The 112 cm of LiH represent 112/102 = 1.10 radiation lengths for each turn. The absorbers are wedged, namely the thickness is growing linearly with the radius of the particle beam. In order to "cool" also longitudinally, chromaticity has to be introduced with a wedge shaped "dE/dx foil", to increase the ionisation losses for faster particles at a location where there is dispersion. This results in an exchange of longitudinal and transverse emittances. With transverse cooling, this allows cooling in all dimensions. The wedge has radially an opening angle of $100^{o.}$

Secondary protons have a kinetic energy of 32.73 MeV, $\beta = 0.257$ and a range of 810 mg/cm$^2$ in a low Z absorber and the wrong RF frequency. They are quickly absorbed already at the first turn.

Electrons and positrons initially at 250 MeV/c mostly from $\pi_o$ decays will differ for a constant, higher $\Delta\beta = 7.7\%$ affecting the synchronization of the RF phases. They will undergo strong radiation losses in the absorbers of the ring. The 112 cm of LiH represent 112/102 = 1.10 radiation lengths for each turn. A substantial background at lower energies will persist only for a few turns at the exit of the outgoing kicker and should be practically absent at the external measuring setup.

As in the previously described final setup (Figure 25, #5), muons should be captured to two single bunches. Capturing the initial muon phase space into single bunches does require low frequency RF and thus low gradients. This recombination is done in two stages. The individual bunches are phase rotated after initial cooling to fill the spaces between bunches and to lower their momentum spreads to the minimal cooling equilibrium, followed by a 5 MHz RF interspersed along a long drift in order to phase rotate the train into a single bunch that then can be re-captured using the high frequency RF.

The Initial Cooling Experiment may expand with a more elaborate setup the already approved MICE experiment [59] under development at the Rutherford Appleton Laboratory, UK, a single lattice cell cooling by about 10% and expected to be operational by 2020.

After a specified number of turns in the cooling ring the muon beam will be extracted with the help of a second pulse of the kicker and momenta and directions will be fully analyzed in an elaborate external spectrometer channel similar to the one described for the MICE experiment.

An important development that should be tested is related to the study of Parametric Ionization Cooling (PIC) that should eventually allow in the future important increases of the luminosity with a further decrease of the transverse emittances.

As already pointed out in paragraph 13, angles for the RFOFO configuration between the solenoids of opposite orientations providing both the focusing and the bending should be modified in order to achieve a half integer resonance such that the normal elliptical motion of particles in x-x' phase space becomes hyperbolic. The half-integer parametric resonances should permit a stronger *transverse focusing* while ionization cooling limits the beam's *angular spread* with appropriately placed energy absorbers.

However the performances as PIC, although already at the low intensity of 3 x $10^7$ $\mu^-$/pulse of the Initial Cooling Experiment require serious developments that are so far hard to predict. The largest momentum spread that can be useful for the PIC is a especially relevant question.

### *15.- Concluding remarks.*

The experimental realization of an appropriate $\mu^+\mu^-$ Ring Collider may represent a very attractive addition for the future programs at the ESS. in synergy with the realization of long distance neutrino beams, requiring however a substantial amount of prior R&D on the cooling, which must be initially and promptly tested with the help of a simpler and much cheaper *Initial Muon Cooling Experiment* program, as described in the present pap

The number of circulating muons of each sign will be determined accurately as a function of time and compared with the theoretical expectations for transverse and longitudinal 3D cooling and for a variety of incoming muon momenta over a window variable between 100 and 300 MeV/c.

All described components for both options (1) at the $H^o$ mass and (2) at $\sqrt{s}$ = 700 GeV may easily fit within the existing ESS site, doubling the repetition rate of the already approved initial LINAC design dedicated to the spallation neutron production and that may be summarized as a number of main elements:

> *a* pair of proton rings, the *Accumulator and the Compressor*, both with a radius of ≈ 35 m, splitting the main LINAC beam from 14 to 56 bunch/s and compressing the bunches to ≤ 2 ns and 2.5 x $10^{14}$ protons/bunch;

> a linear *Front End Channel* of about 100 m length converting the muons to narrow $\Delta p$ bunches at about 250 MeV/c;

> a linear 60 m long *Pre-cooler* to initially reduce the beam dimensions as adequate for a ring;

> a pair of robust $\mu^+$ and $\mu^+$ ionization cooling rings each with ≈ 30-45 m circumference, compressing in 6D with a number of narrow bunches;

> a rotation with the help of RF of the beams to two single bunches with an acceptable tune shift $\Delta\nu_{sc}$ and invariant equilibrium transverse $\varepsilon_{H,V}$ = 0.4 (π) mm rad and $\varepsilon_t$ = 1.0 (π) mm rad longitudinal emittances.

> a fast recirculating LINAC acceleration system of about few hundred m to bring ≈ 2 x $10^{12}$ muons/bunch of both signs (i.e. 0.8 percent of protons) to the required Collider energies;

> a Collider ring at 7 Tesla and ≈ 60 m radius for option (1) at the $H^o$ mass with L = 4.0 x $10^{31}$ $cm^{-2}$ $s^{-1}$ and ≈ 330 m for option (2) at $\sqrt{s}$ = 700 GeV with with two interaction points where detectors are located at $\beta_x = \beta_y = \beta_o$ = 5 cm.

For a Higgs resonant cross section $H^o$ = 3.0 x $10^{-35}$ $cm^2$ and an effective live time of $10^7$ s/year, the resulting event rate for option (1) is L x $H^o$ x $10^7$ = 12'000 ev/y.

Presently the Higgs mass is known.at the LHC only within some 600 MeV. In order to initiate the operation of the muon Collider the position of the resonance at the $H^o$ mass (Figure 35) should be more accurately located with a few months exploratory scan of a relatively modest initial luminosity of 1.7 x $10^{31}$ $cm^{-2}$ $s^{-1}$.

The proposed next-generation Collider may encourage new major technical advances. The *Parametric-resonance Ionization Cooling (PIC)* has been investigated. Combining the muon ionization cooling with the parametric resonant dynamics is expected to allow final transverse beam emittances smaller than with the conventional ionization cooling alone and may offer the potentials either for a higher luminosity and/or a smaller number of muons, However the value of $\Delta\nu_{sc}$ may be easily exceeded since any decrease of the transverse emittance may augment the Laslett tune shift already largely at the limit in the previous conventional cooling conditions. To overcome the space charge impact, one must recombine bunches only after a further fast acceleration to a sufficiently high relativistic energy. Such a vastly different configuration requires serious additional experimental confirmations.

In conclusion, compared to other alternatives, the advantages of such an appropriate $\mu^+\mu^-$ Ring Collider, even without PIC, are mainly the following:

- Large cross section σ ($\mu^+\mu^- \rightarrow h$) = 22 pb in s-channel resonance.
- Small size footprint: it may fit in the ESS site.
- Cost so far unknown but far smaller than the ILC.
- No synchrotron radiation and beamstrahlung problems
- Precise measurements of line shape and decay width $\Gamma$
- Exquisite measurements of all channels and tests of SM.
- A low cost demonstration of muon cooling to be done first.

However several important challenges must be overcome:

- Foil or LASER stripping at the ESS require future developments and conclusive proof.
- The concentration of 2.5 x $10^{14}$ protons to few nanoseconds at the compressor requires a substantial proton beam diameter.
- Muon 2D and 3D cooling needs to be experimentally demonstrated.
- A very small relative energy spread (0.003%) is required
- Backgrounds from muon decays must be removed.
- A major R&D towards end-to-end design is necessary

To conclude, the ESS may become the highlight project for the further studies of the Higgs sector in Europe beyond the LHC, since in other existing locations the development of an adequately intense proton beam is presently no longer under active consideration.

At the Nominal Higgs cross section of 3.0 x $10^{-35}$ cm$^2$ at 125.5 GeV/c and$10^7$ s/y, at each crossing we expect the rate of 12’000 ev/y even without PIC cooling, starting with 1.1 x $10^{15}$ dedicated protons/cycle and 14 cycles/s. PIC cooling, if successful, may either increase the luminosity or reduce the number of dedicated proton cycles up to about one order of magnitude.

## 16- References.